\documentclass[letterpaper,11pt]{article}

\usepackage[margin=1in]{geometry}

\usepackage[T1]{fontenc}
\usepackage{mathtools, amssymb, amsthm, physics, graphicx, placeins, mleftright, color}
\usepackage{colortbl}
\usepackage{amsmath}
\usepackage{enumitem}
\usepackage[ruled,vlined,linesnumbered]{algorithm2e}
\usepackage[dvipsnames]{xcolor}
\usepackage{todonotes}
\usepackage{dsfont}
\usepackage{authblk}
\usepackage{longdivision}

\usepackage{hyperref}
\hypersetup{
    colorlinks = true,
    linkcolor  = blue,
    citecolor  = green!60!black,
    urlcolor   = blue
}

\usepackage[capitalize,noabbrev]{cleveref}
\makeatletter
\@ifundefined{RemoveFromHook}{}{%
    \RemoveFromHook{label}[firstaid/cleveref]%
}
\makeatother

\theoremstyle{plain}

\newtheorem{theorem}{Theorem}[section]

\newtheorem{lemma}[theorem]{Lemma}
\newtheorem{corollary}[theorem]{Corollary}

\newtheorem{fact}[theorem]{Fact}

\theoremstyle{definition}

\theoremstyle{remark}

\newcommand{\p}[1]{\mleft( #1 \mright)}

\newcommand{\curly}[1]{\mleft\{ #1 \mright\}}

\renewcommand{\square}[1]{\mleft[ #1 \mright]}

\newcommand{\ceil}[1]{\mleft\lceil #1 \mright\rceil}
\newcommand{\floor}[1]{\mleft\lfloor #1 \mright\rfloor}

\renewcommand{\P}[1]{\mathbb{P} \square{ #1 } }
\newcommand{\Ex}[1]{\mathbb{E} \square{ #1 } }
\newcommand{\Vari}[1]{\text{Var} \square{ #1 } }
\def\E{\mathrm{E}}
\def\Var{\mathrm{Var}}

\newcommand{\indeg}{\textup{\textsc{indeg}}\xspace}
\newcommand{\innbr}{\textup{\textsc{in}}\xspace}
\newcommand{\outdeg}{\textup{\textsc{outdeg}}\xspace}
\newcommand{\outnbr}{\textup{\textsc{out}}\xspace}
\newcommand{\jump}{\textup{\textsc{Jump}}\xspace}
\newcommand{\din}{d_\mathrm{in}}
\newcommand{\dout}{d_\mathrm{out}}
\newcommand{\Nin}{\mathcal{N}_{\mathrm{in}}}
\newcommand{\Nout}{\mathcal{N}_{\mathrm{out}}}
\newcommand{\Hin}{{H}_{\mathrm{in}}}
\newcommand{\Hout}{{H}_{\mathrm{out}}}

\newcommand{\Td}{T^*(\delta)}
\newcommand{\Tstar}{T^*}
\newcommand{\TA}{T_f(r)}
\newcommand{\TAd}{T_f(\delta,r)}
\newcommand{\TB}{T_b(r)}

\newcommand{\totalactivatef}{U_f}
\newcommand{\totalactivateb}{U_b}
\newcommand{\cntf}{\operatorname{cnt}_f}
\newcommand{\cntb}{\operatorname{cnt}_b}

\newcommand{\nr}{q}
\newcommand{\rmax}{\bar{r}}

\newcommand{\rmaxb}{\bar{r}_b}
\newcommand{\rsumf}{R_f}
\newcommand{\rsum}{R}

\newcommand{\frozenf}{F_f}
\newcommand{\frozenb}{F_b}

\newcommand{\candf}{C_f}
\newcommand{\candb}{C_b}

\newcommand{\vpi}{\pi}
\newcommand{\epi}{\hat{\pi}}

\newcommand{\edgesub}{E'}
\newcommand{\notedgesub}{\overline{E'}}

\def\rb{r_b}
\def\pb{p_b}
\def\rf{r_f}
\def\pf{p_f}

\def\pr{P}
\def\qr{Q}

\def\cG{\mathcal{G}^*}

\def\tp{\pi}
\def\budget{\kappa}
\def\budgetf{\kappa_f}
\def\budgetb{\kappa_b}

\newcommand{\indicator}{\mathds{1}}

\def\ifrozenf{L_f}
\def\ifrozenb{L_b}

\def\rela{\varepsilon_r}

\def\pushback{\textup{\texttt{BackwardPush}}\xspace}
\def\pushforward{\textup{\texttt{ForwardPush}}\xspace}
\def\frozenpushback{\textup{\texttt{BackwardPushWithFreezing}}\xspace}
\def\frozenpushforward{\textup{\texttt{ForwardPushWithFreezing}}\xspace}

\def\threephase{\textup{\texttt{MonteCarlo}}\xspace}

\def\TriPPR{\texttt{TriPPR}}

\def\mP{\mathbf{P}}

\usepackage{tikz}
\usetikzlibrary{arrows.meta}

\newcommand{\ignore}[1]{}

\title{Instance-Optimality of Single-Pair Personalized PageRank\footnote{This work was supported by the VILLUM Foundation grant 54451.}}

\author[1]{Christian Bertram}
\author[1]{Mads Vestergaard Jensen}
\author[2]{Mikkel Thorup}
\author[3]{
\and \vspace{-1em} Hanzhi Wang}
\author[1]{Shuyi Yan}

\vspace{0.3em}
\affil[1,2]{BARC, University of Copenhagen \vspace{0.2em}}
\affil[3]{The University of Melbourne \vspace{1.3em}}
\affil[1]{\texttt{\{chbe, mvje, shya\}@di.ku.dk}\vspace{0.2em}}
\affil[2,3]{\texttt{\{mikkel2thorup, hanzhi.hzwang\}@gmail.com}}

\date{}

\begin{document}

\maketitle
\begin{abstract}

We consider discounted random walks in a directed simple graph $G=(V,E)$. Such a walk terminates at each step with a constant probability $\alpha\in (0,1)$ and otherwise continues to a uniform random out-neighbor. Given a pair $(s,t)\in V^2$, we denote by $\pi(s,t)$ the probability of such a discounted random walk starting at $s$ terminates at $t$. This probability is referred to as the Personalized PageRank (PPR) of $t$ with respect to $s$. The problem of estimating $\pi(s,t)$ within a constant multiplicative factor with constant success probability for a given vertex pair $(s,t)$ is called the single-pair PPR estimation problem. 

We investigate the instance-optimality of single-pair PPR estimation in directed graphs. We present a new algorithm that aims at instance-optimal running time up to polylogarithmic factors. In other words, no correct algorithm can run faster than our algorithm on any valid graph by more than a polylogarithmic factor. We show that our algorithm is instance-optimal in the standard adjacency-list query model if $n$ is not known (as is often the case for most of the large networks to which PPR is applied). When $n$ is known, our algorithm is still instance-optimal for sparse graphs. For denser graphs, when $n$ is known, we show that no algorithm can achieve general instance-optimality.

For the multiplicative approximation of PPR, we also improve the worst-case complexity from $O(mn\log n)$ to the near-optimal  $O(m\log n)$\footnote{This improvement was originally presented as part of our unpublished arXiv paper \cite[Version 4]{bertram2025estimating}, but we move it to the current paper, where it fits better and will take it out in the next version of \cite{bertram2025estimating}.}. This is not only for the single-pair discussed above, but also for the single-source and single-target variants of the PPR estimation problem. 
    
\end{abstract}


{
  \hypersetup{linkcolor=black}
}

\section{Introduction}
Random walks in directed graphs is an algorithmic building block in modern network analysis. One important type is the discounted random walk, where at each step, the walk either terminates with some probability $\alpha \in (0,1)$, or moves to a uniformly random out-neighbor with probability $1-\alpha$. For simplicity, we assume that every vertex has at least one out-neighbor. This assumption can be easily satisfied by adding a self-loop to each dangling vertex. Following previous works~\cite{lofgren2014fast, wang2017fora, bressan2023sublinear, wang2024revisiting, thorup2025instance}, we also assume that $\alpha$ is a constant throughout the paper. 

Let $\vpi(s,t)$ denote the probability that a discounted random walk starting from vertex $s$ terminates at vertex $t$. This probability is referred to as the \emph{Personalized PageRank (PPR)} of $t$ with respect to $s$. The notion of PPR was introduced by the co-founders of Google in~\cite{brin1998anatomy,page1998pagerank} for personalized web search and has since become a widely used measure of node proximity, finding various applications in graph analysis and learning~\cite{andersen2006local, recsys_LeeSKLL11, wsdm_BackstromL11, klicpera2018predict, bojchevski2020scaling, epasto2022differentially, nips_WeiC024}. The ``Google'' papers~\cite{brin1998anatomy,page1998pagerank} also introduced PageRank for global webpage ranking. The PageRank $\pi(t)$ for a vertex $t$ is defined as the average of the PPR values $\pi(s,t)$ over all vertices $s\in V$, i.e., $\pi(t)=\sum_{s\in V}\pi(s,t)/n$, where $n$ is the number of vertices in the graph. Thus, by definition, an algorithm for estimating PPR $\pi(s,t)$ can be extended to estimate the PageRank $\pi(t)$ by adding a pseudo vertex $s^*$ with outgoing edges 
to all vertices in the graph and estimating $\pi(s^*,t)$.

In this paper, we study the \emph{single-pair PPR problem}, which aims to estimate $\pi(s,t)$ for a given source vertex $s$ and a given target vertex $t$ in a directed graph $G=(V,E)$ within a constant multiplicative factor with constant success probability. 
This problem and its variants have received considerable attention over the past decades~\cite{lofgren2014fast, lofgren2016personalized, wang2017fora, bressan2018sublinear, bressan2023sublinear, wang2024revisiting, soda_Thorup0W026, thorup2025instance}, with applications in diverse areas, including web search~\cite{page1998pagerank, gleich2007approximating}, recommender systems~\cite{wtf_GuptaGLSWZ13}, node embedding~\cite{STRAP_YinW19}, graph neural networks~\cite{PPRGo_BojchevskiKPKBR20}, and many others~\cite{gleich2015pagerank, yang2024efficient}. There is also a line of work that studies this problem and its variants on undirected graphs~\cite{lofgren2015bidirectional, wang_kdd2024_revisiting, bertram2026undirected, cui2026accelerating, li2026nearly}, but in this paper, we consider only directed graphs. 

The focus of this paper is to find an instance-optimal algorithm for the single-pair PPR problem, that is, an algorithm whose running time is close to optimal on every input graph among all correct algorithms, or rather probabilistically correct since most algorithms for PPR are randomized.

\subsection{Instance-Optimality}

Traditionally, theory of computing has focused on worst-case complexity, but algorithms that are worst-case efficient may perform very poorly on easy instances encountered in practice, and it is their performance on practical instances that determines actual computational costs and energy consumption.
This has motivated research on designing algorithms that are average-case efficient or perform well on specific classes of instances, e.g., power-law graphs or sparse graphs. However, it is often difficult to determine which classes of instances or graph parameters are most relevant (e.g., whether sparse or dense graphs are more important), as the answer depends on the applications, which may evolve over time. An ideal solution would be to design an algorithm that achieves optimal running time on \emph{every} possible instance, thereby covering all instances encountered in practice. Such an algorithm is called an \emph{instance-optimal} algorithm. 

A beautiful illustration of the advantage of an instance-optimal algorithm over a worst-case optimal algorithm is the shortest-path problem: What is the fastest way to get from $A$ to $B$? Dijkstra's algorithm~\cite[Problem 2, stated for single-pair, not single-source]{Dij59} with a priority queue establishes the $\tilde\Theta(m)$ worst-case complexity of this problem assuming non-negative weights, with many great works devoted to understanding the logarithmic factor hidden by the $\tilde\Theta$ notation (see~\cite{DuanMMSY25, icalp_DuanMSY26} and the references therin). 
However, individual instances can be much easier. 
As an example, imagine that you want to go to Rome.
This is easy because all roads lead to Rome. More  mathematically, think of an undirected star graph with Rome in the center: Dijkstra will find the center from any node in constant time.  However, when you want to go back home from Rome after the party, Dijkstra may lead you to explore the whole graph. Bidirectional Dijkstra will be fast in both directions, and it is, in fact, instance-optimal within a logarithmic factor for positively weighted multigraphs \cite{sosa_bidirectional_HaeuplerHRTT25}. 

Instance-optimality is much stronger than worst-case optimality or average-case optimality, since it requires the algorithm to be the best, or close to best, among all good algorithms on every instance, rather than merely excel in the worst case or on average. It is therefore regarded as the ultimate notion of optimality~\cite{colt_ChenL16_openproblem, instnce-optimal_AfshaniBC17}. Here, a \emph{good} (randomized) algorithm refers to one that is (probabilistically) \emph{correct on every valid instance}. When we say that an algorithm is instance-optimal, we mean that its running time on every instance is at most a small factor larger than that of the fastest good algorithm on that instance, e.g., a polylogarithmic factor when the running times are polynomial. 
A good algorithms may be \emph{instance-smart} in the sense that it is tuned to be fast on a particular instance (but may be slow on other instances). This subsume algorithms that receive good but not fully trusted advice (e.g., advice from AI~\cite{MitzenmacherV22}). But the instance-smart algorithm must verify that its answer is probabilistically correct, so that it is not fooled when given an input that is not the instance for which it was tuned. 
Thus, an algorithm that simply outputs an
answer without verifying its correctness is not considered a competitor of instance-optimal algorithm. 

For most problems, instance-optimal algorithms do not exist~\cite{instnce-optimal_AfshaniBC17, FOCS_best_HaeuplerHRTT24}. 
Some positive examples are known, e.g., in quantity estimation~\cite{siamcomp_DagumKLR00}, database aggregation~\cite{FaginLN03_instance_optimal}, sequential estimation~\cite{stoc_ValiantV16_instance_optimal, dang_neurips_23, focs_NarayananRTT24_thorup}, computational geometry~\cite{instnce-optimal_AfshaniBC17, BergHRRW25_esa_instanceopt_convexhull}, distribution testing and learning~\cite{siamcomp/ValiantV17, books_ValiantV20, focs_BlancCW25_clement}, contextual bandits~\cite{nips_LiRNJJ22}, and the above-mentioned shortest-path problem with positive weights~\cite{sosa_bidirectional_HaeuplerHRTT25, our_bi-dij_shortestpath}. 
Additionally, \cite{FOCS_best_HaeuplerHRTT24} establishes \emph{universal optimality} for a new priority-queue in Dijkstra's single-source shortest paths problem. The notion of universal optimality is strictly weaker than instance-optimality but still substantially stronger than worst-case optimality. 
The most relevant work is a recent paper that investigates the instance-optimality of estimating a vertex’s PageRank~\cite{thorup2025instance}. We will return to this work and its connection to our paper later.

\subsection{Personalized PageRank (PPR)}\label{subsec:ppr_intro}
As mentioned earlier, in this paper, we study the single-pair Personalized PageRank (PPR) problem and investigate its instance-optimality. Recall that the problem is to estimate $\pi(s,t)$ with constant relative error and constant success probability, given a source vertex $s$ and a target vertex $t$ in a directed graph $G$. Some previous works also introduce an approximation threshold parameter $\delta \in [0,1]$ and only require an estimate of $\pi(s,t)$ when $\pi(s,t)>\delta$. 
We note that $\pi(s,t)$ can be as small as $1/n^{\Omega(n)}$ (we shall see example in \Cref{fig:small_estimate}).
For our own algorithms, we will at first not  parameterize the problem by $\delta$, that is, we consider $\delta=0$, but we will later show that our instance-optimality results extend to the setting where such an approximation threshold $\delta$ is given.

We will now discuss some of the previous work on this problem, most of which involves $\delta$.

\paragraph{Worst-case complexity.}
Let $n$ and $m$ denote the number of vertices and edges in $G$, respectively. 
The best-known worst-case complexity is $O(\min\{1/\delta, m\log(1/\delta)\})$, obtained simply by combining Monte Carlo sampling and the power method. Specifically, Monte Carlo sampling estimates $\pi(s,t)$ in $O(1/\delta)$ time by generating $O(1/\delta)$ random walks from the source vertex $s$ and using the fraction of walks that terminate at $t$ as the estimate of $\pi(s,t)$. The \emph{power method}, introduced in the seminal paper that proposes PPR and PageRank~\cite{page1998pagerank}, 
iteratively applies the definition equation of PPR: $\boldsymbol{\pi}_s=(1-\alpha)\mP\boldsymbol{\pi}_s+\alpha \boldsymbol{1}_s$, where $\boldsymbol{\pi}_s$ is an $n$-dimensional PPR vector with $\boldsymbol{\pi}_s[t]=\pi(s,t)$, $\mP$ is the random-walk transition matrix, and $\boldsymbol{1}_s$ is an $n$-dimensional one-hot vector with only $\boldsymbol{1}[s]=1$. Each iteration takes $O(m)$ time for computing $\mP\boldsymbol{\pi}_s$ as it touches all edges in the graph. The geometric convergence rate of $(1-\alpha)$ ensures that after $\log_{(1-\alpha)}\delta=O(\log(1/\delta))$ iterations, the absolute error falls below $\delta$, thus yielding the $O(m\log{(1/\delta)})$ runtime.\footnote{
The power method is described in the seminal paper introducing PPR and PageRank, although the description is somewhat sketchy. The $O(m\log(1/\delta))$ bound is not explicitly stated in the paper, but it has since become a folklore result widely used in subsequent work, such as~\cite{maehara2014computing, lofgren2014fast, lofgren2015efficient, wang2017fora, wei2018topppr}. We refer interested readers to the survey paper~\cite{tnse_Chung14_survey} and the doctoral thesis~\cite[Section 2.2]{lofgren2015efficient} for a more detailed discussion. }

The $O(\min\{1/\delta, m\log(1/\delta)\})$ upper bound also holds for the single-source PPR problem, that is, estimating $\pi(s,t)$ for all possible targets $t\in V$, since both Monte Carlo sampling and the power method can estimate $\pi(s,t)$ simultaneously for all targets $t\in V$. One might hope that the single-pair complexity should be smaller than that of the single-source. However, our unpublished manuscript \cite{bertram2025estimating} establishes a tight $\Omega(\min\{1/\delta, m\})$ lower bound for both the single-pair and single-source PPR problems. 
Therefore, in the worst case, solving the single-pair PPR problem requires as much time as solving the single-source problem. 

\paragraph{Better algorithms for some instances.}
Analogous to the aforementioned shortest path problem where Dijkstra’s algorithm is worst-case optimal yet other algorithms, such as bidirectional Dijkstra, can be much faster on particular instances, 
previous work on single-pair PPR has also sought to improve performance on individual instances despite the worst-case lower bound.
This includes \cite{siamcomp_DagumKLR00}, which proposes an adaptive Monte Carlo sampling method that estimates $\pi(s,t)$ in $O(\min\{1/\delta,1/\pi(s,t)\})$ time, which is faster on instances where $\pi(s,t)>\delta$. 

\def\BiPPR{\texttt{BiPPR}}
\def\push{\texttt{push}}

Another line of work~\cite{lofgren2014fast,lofgren2016personalized} seeks to go beyond worst-case complexity by considering the average-case complexity: complexity of estimating $\pi(s,t)$ for an arbitrary source $s$ and a uniformly random target $t\in V$. \cite{lofgren2016personalized} proposes a bidirectional search algorithm, which we refer to as \BiPPR. The $\BiPPR$ algorithm combines Monte Carlo sampling with a so-called \emph{backward push} method~\cite{andersen2008local}, which deterministically propagates random-walk probability mass backward from the target vertex $t$ along incoming edges, step by step.  
For this bidirectional $\BiPPR$, the worst-case running time is still  $O(\min\{1/\delta, m\log(1/\delta)\})$, but
~\cite{lofgren2016personalized} proves that it achieves an average-case complexity of $O(\sqrt{(m/n)/\delta})$~\cite{lofgren2016personalized}, which is better than the aforementioned worst-case complexity of $O(\min\{1/\delta,m\log(1/\delta)\})$ when $\delta>n/m$.

Another set of research~\cite{bar2008local, bressan2023sublinear, wang2024revisiting, soda_Thorup0W026, thorup2025instance} studies the PageRank estimation problem: estimating the PageRank $\pi(t)$ of a given target vertex $t$. Recall that an algorithm for estimating single-pair PPR $\pi(s,t)$ for a given source $s$ and a given target $t$ can be trivially extended to estimate the PageRank $\pi(t)=\sum_{s\in V}\pi(s,t)/n$ for a given target $t$: adding a pseudo vertex $s^*$ with outgoing edges to all vertices in the graph and estimating $\pi(s^*,t)$ for the given target $t$. 
\cite{wang2024revisiting} revisits the above-mentioned $\BiPPR$ method and shows that it solves the PageRank estimation problem in $O(n^{1/2}m^{1/4})$ time and that this bound is already worst-case optimal. Recently, \cite{thorup2025instance} showed that the adaptive variant of this \BiPPR\ method is even instance-optimal for PageRank estimation on a class of graphs $\cG$ in which each graph has at most a polylogarithmic number of vertices with in- or out-degree at least $(1-o(1))n$. The graph class $\cG$ contains almost all graphs including all sparse graphs.

For all that is known, it could be that $\BiPPR$ was always instance-optimal for PPR, but we shall see that it can be off by a factor $\Omega(m/\log n)$ even when $\pi(s,t)=\Omega(1/m)$.

\subsection{Main contributions} 

In this paper, we design a simple algorithm called $\TriPPR$ for the single-pair PPR problem: estimating $\pi(s,t)$ for a given $s$ and $t$ with constant relative error and constant success probability. Like in many previous works, we can also benefit from an approximation threshold parameter $\delta \in [0,1]$ where we only require an estimate of $\pi(s,t)$ when $\pi(s,t)>\delta$.  

We show that our new
algorithm instance-optimal if $n$ is not known. When $n$ is known, our algorithm is still instance-optimal for sparse graphs with $m=\tilde O(n)$. For denser graphs, when $n$ is known, we show that no algorithm can achieve general instance-optimality.

As we discuss below in Section \ref{sec:model}, not knowing $n$ is a common case for some of the large networks for which we want to compute PPR values.  Having an algorithm not knowing $n$ is thus very relevant.

For the multiplicative approximation of PPR, that is, $\delta=0$, we also improve the worst-case complexity from $O(mn\log n)$ \cite{page1998pagerank} to the near-optimal  $O(m\log n)$. This improvement is not only for the single-pair discussed above, but also for single-source and single-target.

\subsection{Model}\label{sec:model}
For our algorithms and running-time upper bounds we assume the standard unit-cost RAM model with word-size $\Theta(\log n)$. This includes arithmetic operations on floating-point numbers with $O(\log n)$-bit exponent and 
mantissa. 

Our lower bounds are proved in the adjacency-list query model for sublinear graph algorithms. 
The latter supports in- and out-degree queries $\indeg(v)$ and $\outdeg(v)$, which return the in- and out-degree of a vertex $v$, respectively, as well as in- and out-neighbor queries $\innbr(v,i)$ and $\outnbr(v,i)$, which return the $i$-th in- and out-neighbor of $v$, respectively. Each query incurs unit cost.
Basically, we are only given direct access to $s$ and $t$ in a large unknown network. The only way to access other nodes is to follow in- and out-edges from vertices already visited. We do not have any other global information on the network. In particular, we do not necessarily know the number $n$ of vertices in the graph. Our $\TriPPR$ algorithm does not use knowledge of $n$. We consider its instance-optimality both when $n$ is known and when $n$ is not known.

The above kind of limited local access without knowledge of $n$ aligns with the graph queries provided by application programming interfaces (APIs) for large real-world networks. For example, consider the World Wide Web. No single entity owns or maintains the entire web, and there is no universal API that provides the exact number of existing web pages. Similarly, large-scale social networks such as Facebook do not provide an API operation for querying the total number of vertices $n$~\cite{TAO_BronsonACCDDFGKLMPPSV13}.

This also reflects the capabilities of APIs for large real-world networks. For example, Google has explicitly acknowledged that it does not know the exact number of unique pages on the web~\cite{google_web_2008}, noting that the web is too large for its crawlers to discover every publicly accessible URL~\cite{google_crawl_budget}. Similarly, consider large-scale social networks such as Facebook. Maintaining an exact, globally consistent vertex count in Facebook's massive, continuously evolving social graph could require substantial communication and synchronization overhead~\cite{osdi_LyerlyPDRBH25}.

We remark that prior work on PageRank estimation, such as~\cite{avrachenkov2007monte, borgs2012sublinear, borgs2014multiscale, bressan2023sublinear, wang2024revisiting, thorup2025instance}, also introduces a query called $\jump()$, which returns a uniformly random vertex in the graph. This $\jump$ query is important for PageRank estimation as it because, by definition, the PageRank score $\pi(t)$ of a vertex $t$ is the average of $\pi(s,t)$ over all source vertices $s$ in the graph, i.e., $\pi(t)=\E[\pi(\jump(),t)]$. \cite{bar2008local, bressan2013power} show that without the $\jump$ query, no sublinear-time algorithm exists for PageRank estimation. In contrast, the $\jump$ query is not necessary for PPR estimation. Prior work on PPR estimation, whether for the single-pair, single-source, or single-target problem, e.g., \cite{lofgren2016personalized, wang2017fora, andersen2008local}, does not assume access to such a query, and neither do we. Indeed, a $\jump$-query that is guaranteed to finds \emph{any} vertex with probability close to $1/n$ has issues similar to a query on $n$, e.g., it is not available in an an API like~\cite{TAO_BronsonACCDDFGKLMPPSV13}. Thus, when we prove instance optimality  not knowing $n$, we also assume that there is no $\jump$ query. Observe that when we reduce PageRank to PPR, we introduced a node $s^*$ with out-edges to all other nodes, so going to a random out-neighbor from $s^*$ is the same as a $\jump$, except that we may not have the knowledge that $\dout(s^*)=n$. 

Note that our lower-bounds are information theoretic based on query complexity. In particular, they can never go above $\Omega(m)$ if we read the whole graph. For contrast, our upper bounds are time bounds, and our algorithms may do multiple operations on the same vertices and edges. 

\subsection{Technical Perspective}
The design of $\TriPPR$ is motivated by two challenges.
First, existing PPR algorithms can spend substantial time exploring parts of a graph that an instance-smart algorithm can handle much more efficiently.
We explain how $\TriPPR$ addresses this by combining forward push, backward push, and Monte Carlo sampling.
Second, existing pushing algorithms can become very expensive when estimating extremely small PPR values.
We address this problem with a vertex-freezing technique that limits the amount of work spent on each vertex.
Finally, we consider what changes when the number of vertices $n$ is known.
We explain why this additional information makes instance-optimality impossible, and then provide a more refined characterization of the achievable instance-optimality ratio, showing in particular that $\TriPPR$ remains instance-optimal for sparse graphs and a much broader class of locally solvable instances.

\subsubsection{Issues with Existing Algorithms}
Before inventing a new algorithm for single-pair PPR estimation, let us first ask if the currently best algorithm $\BiPPR$ could be instance optimal. We recall that it is worst-case optimal within a factor $O(\log 1/\delta)$, and we know from \cite{thorup2025instance} that it is instance-optimal for the special case of PageRank estimation. 

\begin{figure}[h]
\centering
\includegraphics[width=0.78\linewidth]{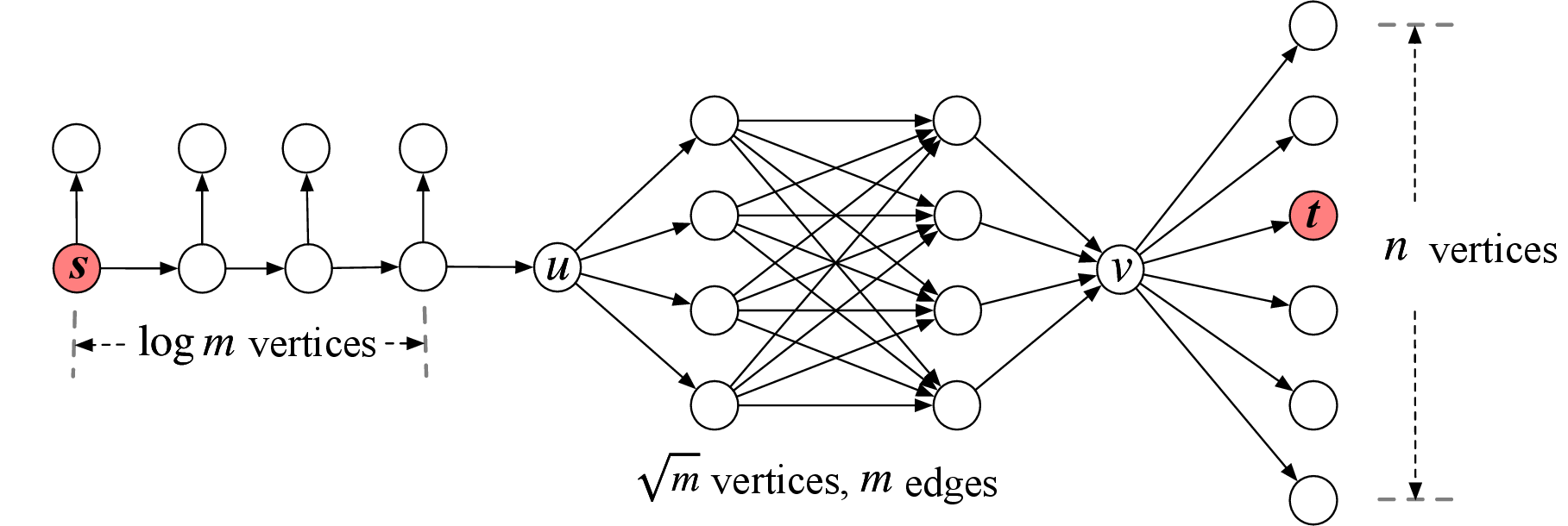}
\caption{Hard instances for single-pair PPR estimation, on which an instance-smart algorithm takes $O(\log n)$ time, while the previously best $\BiPPR$ algorithm~\cite{lofgren2016personalized} takes $O(m)$ time, and pure Monte Carlo sampling takes $\omega(mn)$ time.}
\label{fig:high-level-idea}
\end{figure}

The example instance of \Cref{fig:high-level-idea} shows that  $\BiPPR$ may be very far from instance-optimal. In this example, $\BiPPR$ spends 
$\Omega(m)$ time while the instance  can be solved by an instance-smart algorithm in $O(\log n)$ time. 

The toy graph in \Cref{fig:high-level-idea} consists of a binary tree of depth $\log m$ connecting the source vertex $s$ to a vertex $u$, a complete bipartite graph with $\sqrt{m}$ vertices on each side connecting $u$ to another vertex $v$, and a depth-one tree rooted at $v$ with $n$ leaves (the getting-home-from-Rome example mentioned earlier), one of which is the target vertex $t$. 
We note that in this graph, 
\[\pi(s,t) = \pi(s,u)\pi(u,v)\pi(v,t)/\alpha^2 =((1-\alpha)/2)^{\log{m}}(1-\alpha)^3((1-\alpha)/n)\alpha =o(1/(nm))\] 

To estimate $\pi(s,t)$, a pure forward Monte Carlo sampling of discounted random walks from $s$ needs $\omega(nm)$ samples since $\pi(s,t)=o(1/(mn))$;
otherwise it will not  reach $t$ with constant probability.

The $\BiPPR$ method requires $\Omega(m)$ time as it combines Monte Carlo sampling from $s$ and backward push from $t$.
One of them, or both, must reach $u$. To reach $u$ with constant probability, the Monte Carlo sampling needs $\Omega(1/\pi(s,u))=\omega(m)$ samples.
The backward-push method~\cite{andersen2008local} deterministically propagates random-walk probability mass backward from the target vertex $t$ along incoming edges, step by step. To reach $u$ from $t$, it will explore the $\Omega(m)$ edges in the subgraph between $u$ and $v$. It follows that $\BiPPR$ takes $\Omega(m)$ time on this instance.

In contrast, an instance-smart algorithm can first completely explore the $O(\log m)$-sized subgraph between $s$ and $u$ and ascertain that any path from $s$ leaving this subgraph must pass through $u$. Having explored this subgraph completely, it will know $\pi(s,u)$, but only costing $O(\log m)$ time. Likewise, since $(v,t)$ is the only edge entering $t$, it knows that every $s$-$t$ path must pass through $v$ and that $\pi(v,t)=(1-\alpha)/\dout(v)=(1-\alpha)/n$. It remains to estimate $\pi(u,v)$, which can be done by Monte Carlo sampling in constant time; for all paths in the middle subgraph lead from $u$ to $v$ in three hops, so $\pi(u,v)=(1-\alpha)^3=\Omega(1)$.

\subsection{Basic Idea for the $\TriPPR$ Algorithm}\label{subsec:basic_idea}

Our new $\TriPPR$ algorithm is motivated by examples such as the one in \Cref{fig:high-level-idea}.
It overcomes the limitations of \Cref{alg:bippr} by using a three-component algorithm, a bit like the instance-smart algorithm mentioned above: forward push from $s$, backward push from $t$, and Monte Carlo sampling.

Forward push is the counterpart to backward push and was proposed even earlier~\cite{andersen2006local}.
Like backward push, it deterministically propagates random-walk probability mass step by step through the graph, but along outgoing edges rather than incoming edges.
Looking again at~\Cref{fig:high-level-idea}, forward push estimates $\pi(s,u)$ in only $O(\log n)$ time, and backward push estimates $\pi(v,t)$ in constant time.
Monte Carlo sampling then estimates the probability of the intermediate walk from $u$ to $v$, by sampling random walks from $u$ until one of them successfully terminates at $v$, spending only constant time on the example.
These three probability estimates then combine to estimate $\pi(s,t)$.

We show how this three-part combination extends to general instances.
Our more surprising main result is that this combination is instance-optimal for all instances (at least if $n$ is not known).
We invite the reader to consider an alternative example where two copies of~\Cref{fig:high-level-idea} are placed in series (the second copy with source $s'$ and target $t'$ identifies its source $s'$ with the current target $t$, and the goal is to compute $\pi(s,t')$).
Is it clear that \TriPPR{} is instance-optimal in this instance, or would it be better with a 6-part  ``\texttt{HexaPPR}'' algorithm?
Our proof says that $\TriPPR$ is close to optimal, also on this instance.

\begin{figure}[b]
    \centering
    \includegraphics[width=0.5\linewidth]{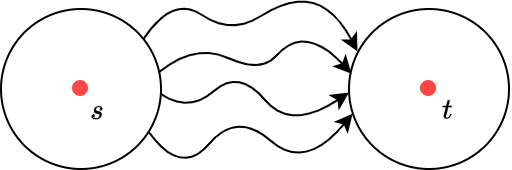}
    \caption{The three components of \TriPPR{}: Forward push, backward push and Monte Carlo.}
    \label{fig:trippr-ball-squiggle}
\end{figure}

We now provide a high-level sketch of $\TriPPR$ for the sake of intuition.
Let us temporarily imagine that we can guess a parameter $r \in (0,1]$; note that in our technical section, we do not explicitly use $r$, but what the algorithm finally does will be similar to what we describe below with the optimal $r$.

See~\Cref{fig:trippr-ball-squiggle} for an illustration of the three components in general.
First, we perform backward pushes from $t$, deterministically finding all vertices $u$ with $\pi(u,t)\ge r$.
The procedure starts with probability mass 1 at $t$ and repeatedly pushes this mass backward along all in-edges of vertices with more than $r$ mass, so it runs in asymptotic time $T_b(r) = \sum_{u\in V,\pi(u,t)\ge r}(1+\din(u))$.

Next, we perform forward pushes from $s$, by starting with probability mass 1 at $s$ and repeatedly pushing this mass forward along all out-edges of certain vertices.
This set of vertices vaguely expands like a ball around $s$ with some frontier $F \subseteq V$.
We will shortly describe more precisely which vertices we push from.

The last part of the algorithm is Monte Carlo sampling.
Recall that we can estimate $\pi(s,t)$ in asymptotic time $1/\pi(s,t)$ by Monte Carlo sampling (i.e.\ simulating) random walks from $s$, each of which runs in expected time $1/\alpha=O(1)$.
We can give our random walks a head start by starting them from a vertex $v\in F$ sampled proportionately to $\pi(s,v)$, instead of starting them from $s$.
Instead of sampling $1/\pi(s,t)$ walks from $s$, it suffices to sample $\pi(s,u)/\pi(s,t)$ walks from every $u \in F$.
Furthermore, our random walks do not have to hit $t$ to succeed.
If they hit the explored backward set $B = \{u \in V\mid \pi(u,t)\ge r\}$, we can consider them as having achieved an $r$ fraction of success in our estimator.
To get sufficiently low error, we now only need to sample $\pi(s,u)r/\pi(s,t)$ random walks from every $u \in F$.

We are now ready to tell which vertices we perform forward push from.
We forward push from vertices where pushing is cheaper than Monte Carlo sampling, i.e.\ vertices $u$ where $\dout(u) < \pi(s,u)r/\pi(s,t)$.
Of course, we do not know $\pi(s,t)$ and only have estimates of $\pi(s,u)$; we also do not really know $r$.
So, the real algorithm will actually try to guess the optimal running time, use it as a budget, and spend equal time in the three parts. Then it decides whether the estimate of $\pi(s,t)$ is good enough. If not, it doubles the budget and repeat.

This gives us a total asymptotic running time of
\[\Tstar = \inf_{r\in(0,1]}\max\{\TA,\TB\} = \max_{r\in(0,1]}\min\{\TA,\TB\},\]
where
\begin{equation*}
\begin{aligned}
\TA=\sum_{u\in V,\pi(s,u)>0}\min\left\{\dout(u),\frac{\pi(s,u)r}{\pi(s,t)}\right\}, \quad \text{ and }\quad &\TB=\sum_{v\in V,\pi(v,t)\ge r}(1+\din(v)).
\end{aligned}
\end{equation*}
For this sketch to go through, existing pushing algorithms are not quite sufficient, so we strengthen them as explained below.

\subsubsection{The Difficulty of Small PPR Values}

\begin{figure}[h]
\centering
\includegraphics[width=0.35\linewidth]{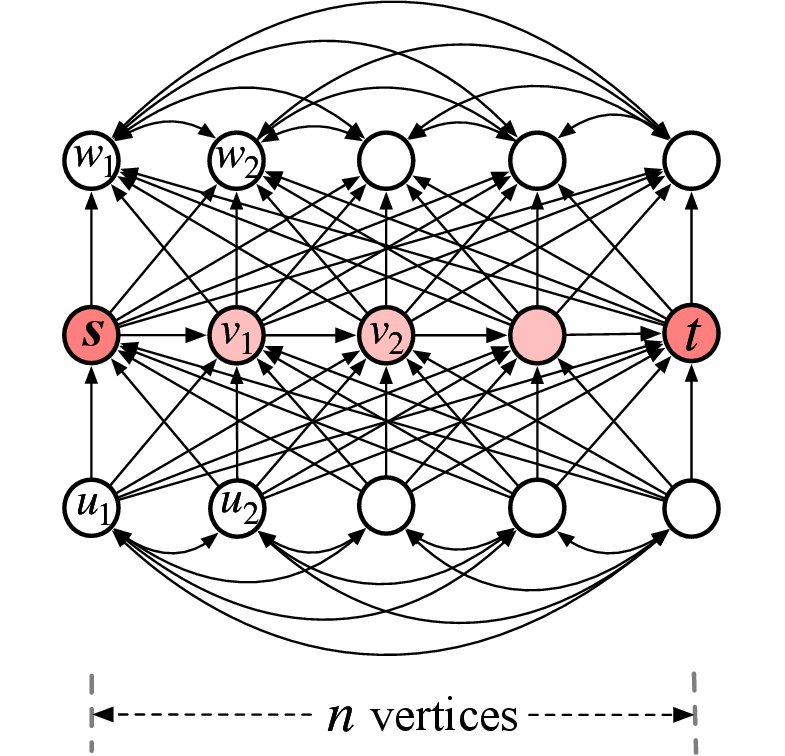}
\caption{Illustrative example where $\pi(s,t)$ is extremely small. }
\label{fig:small_estimate}
\end{figure}

Another challenge in achieving instance-optimality is that prior work cannot efficiently estimate $\pi(s,t)$ when it is very small.  Recall from~\Cref{subsec:ppr_intro} that the previous best worst-case complexity is $O(\min\{1/\delta, m\log(1/\delta)\})$, while the lower bound is $\Omega(\min\{1/\delta, m\})$. So there is a $O(\log{1/\delta})$ gap. 
This gap can be as large as $\Omega(n\log{n})$ because $\pi(s,t)$, and hence $\delta$, can be as small as $1/n^{\Theta(n)}$. An example of this is illustrated in~\Cref{fig:small_estimate}. 

Closing such gap may be of lesser interest from a practical worst-case perspective since
we may only care about $\pi(s,t)$ values above a not so small threshold $\delta$. A devil's advocate could argue that this is a case of sour grapes~\cite{aesop600fox}; that $\delta$ was introduced in practice because small probabilities were too expensive to estimate. Theoretically, it is clearly nicer not to be restricted to larger probabilities.. 

The point in instance-optimality is that we do not need to speculate on what instances are most important in practice. Ambitiously, we want an  algorithm that is best, or close to best, on every possible instance, and this includes those with small $\pi(s,t)$.

The issue is illustrated in the  toy graph shown in~\Cref{fig:small_estimate}. The target vertex $t$ can be reached from the source vertex $s$ only along the path $sv_1v_2\cdots t$ of length $n$. Each vertex on this path also has outgoing edges to each of another $n$ vertices, which we refer to as the $w$-vertices, and incoming edges from each of another $n$ vertices, which we refer to as the $u$-vertices. Every pair of $u$-vertices and every pair of $w$-vertices are connected by bidirectional edges.
At every node on the path $sv_1v_2\cdots t$, there is a $1-1/(n+1)$ probability of going to the $w$-vertices, and then it would never come back to the path, so $\pi(s,t)<1/n^{n-1}$ in this example. 

In this example, Monte Carlo sampling is useless, since a random walk starting from almost any vertex in the graph has an extremely low probability of reaching $t$. The previously best-known upper bound of $O(\min\{1/\delta, m\log(1/\delta)\})$ on this example becomes $O(mn\log n)$. It is  achieved by both the power method~\cite{page1998pagerank} and the $\BiPPR$ method~\cite{lofgren2016personalized} and the bound is tight for this example. 
Basically forward pushing will keep pushing probability mass between the $w$-vertices, and backwards pushing will keep pushing probability mass between the $u$-vertices.
For contrast, an instance-smart algorithm would take at most $O(m)$ time to verify the above structure of the instance.

\subsubsection{Vertex Freezing}

Motivated by this example, our $\TriPPR$ algorithm introduces a \emph{vertex-freezing} strategy, to improve the pushing strategies in forward and backward push. We will be far more conservative about when we push from a vertex, not pushing if we can deduce that it is not needed for the final approximation. A  vertex \emph{freezes} if no more pushing is needed from it, and we will show that at at most $O(\log n)$ pushes are needed from each vertex. This enables $\TriPPR$ to run in only $O(m\log n)$ time, improving upon the previous best $O(mn\log n)$ upper bound and almost matching the $\Omega(m)$ lower bound from \cite{bertram2025estimating} up to a logarithmic factor.
We get this upper-bound, not only for single-pair, but also for the single-source and single-target version of PPR. In combination with the pure Monte Carlo based bound of $O(1/\delta)$, we now get that all these problem have a worst-case complexity of
\begin{equation}\label{eq:worst-case}
    O(\min\{1/\delta,m\log n\})\textnormal,
\end{equation}
almost matching the 
$\Omega(\min\{1/\delta,m\})$ lower-bound from \cite{bertram2025estimating}.

Note that bounding the number of pushes to $O(\log n)$ is crucial for getting instance-optimality within factors of $(\log n)$ for our lower bounds are all based on query complexity, which would only count for the first push on the vertex. 

Another example of work aimed at estimating small (not PPR) values more efficiently is the
 recent work \cite{stoc26_sddm} on  computing entrywise approximate solutions for SDDM systems in 
$O(mn^{o(1)})$ time. Here, SDDM stands for any invertible symmetric diagonally dominant M-matrix. In our setting, entries correspond to our individual $\pi(s,t)$ values in the $\pi(s,\cdot)$ vector. The problems are of course very different. The SDDM systems are more general, but symmetric, while our directed graphs are asymmetric. Also, we have a factor $\log n$ instead of $n^{o(1)}$, and our vertex-freezing technique is very simple.

\subsubsection{Understanding of PPR Instance Optimality When $n$ is Known}
\def\eps{\varepsilon}

As stated earlier, our main contribution  when $n$ is unknown is  that $\TriPPR$ is instance-optimal.
We now consider what changes when $n$ is known.

The fundamental issue is that, while knowing $n$ may not be useful for a natural general-purpose PPR algorithm, there exist instances of denser graphs where an instance-smart algorithm can exploit the knowledge of $n$ in ways that are unavailable to a general PPR algorithm. 
For example, an instance-smart algorithm may deliberately query the degrees of a particular vertex $v$ whose in- or out-degree is $n-1$, thereby inferring that it has edges to all other vertices in the graph. In contrast, a general algorithm may struggle to find the vertex $v$.

We show that $\TriPPR$ is still instance-optimal for all sparse graphs, and complement this by showing that 
there cannot be a general instance-optimal algorithm for  denser graphs.
More precisely, we will prove that the best possible instance-optimality factor is $\tilde \Theta(d)$ where $d=\ceil{m/n}$ and $\tilde\Theta$ hides polylog-factors. 

Technically, it makes a big difference whether or not an instance-smart algorithm 
can solve an instance in $o(n)$ time. When $o(n)$ time is possible, we call the instance \emph{locally solvable}.

If an instance is not locally solvable, then an instance-factor of $O(d\log n)$ is trivial; for  by definition, an instance-smart algorithm needs $\Omega(n)$ time, and our worst-case bound is $O(m\log n)$. The hard part here is to show a lower-bound  of $\tilde \Omega(d)$ on the instance-optimality factor. 

For any density $d=\ceil{m/n}$, we need to find a class of graph instances where each instance can be solved by an  instance-smart algorithm in $\tilde O(n)$ time but where for any  general algorithm, some instances takes $\tilde\Omega(m)$ time. For each of our instances, an instance-smart algorithm can first visit all vertices (since $n$ is known, it knows when it has seen them all), check their degrees, and use this information to estimate $\pi(s,t)$. For contrast, a general algorithm may have no efficient way to find all the vertices. 
We note that an instance-smart algorithm has time to check all vertices only because the instance is not locally solvable.

We now consider the locally-solvable instances, noting that for very large graphs, locally solvable instances may be the only ones we can realistically solve in practice. 
For these locally solvable instances, the $\tilde O(d)$-instance-optimality bound is far from obvious.
We resolve this case by showing that $\TriPPR$ obtains the following instance factor (in general).

We prove that if there are at most $h$ vertices with degree above $(1-\eps)n$, then $\TriPPR$ is instance-optimal within a factor of
\begin{equation}\label{eq:h-mess-factor}
    \tilde O\p{\frac1\varepsilon\max\curly{h+1,\frac{1}{n}\min\curly{m,\frac{1}{\delta}}}}.
\end{equation} 
We emphasize that this bound holds for any chosen $\varepsilon$ and matching $h$.

As a first consequence, choosing $\eps=1/2$ gives $h=O(m/n)$, so \Cref{eq:h-mess-factor} implies that $\TriPPR$ achieves an instance-optimality factor of $\tilde O(d)$ on all instances, including the non-trivial locally solvable ones, where $d=\ceil{m/n}$.
In particular, $\TriPPR$ is instance-optimal up to polylogarithmic factors on all sparse graphs with $m=\tilde O(n)$.

We now consider instead cases where $1/\delta=\tilde O(n)$, noting that this may include very dense graphs. 
In that case \Cref{eq:h-mess-factor} simplifies to,
\begin{equation}\label{eq:h-factor}
    \tilde O\p{{(h+1)}/{\varepsilon}}.
\end{equation}

Define the graph class $\cG$ to include all graphs where the factor in  \Cref{eq:h-mess-factor} is polylogarithmic, that is, $(h+1)/\eps=\tilde O(1)$ for some $\eps$ and $h$.
Then $\cG$ is much more than just sparse graphs, e.g., it includes almost all graphs; for  a random graph has max-degree below $2n/3$ with very high probability, so with $\eps=1/3$ we get $h=0$. 
The definition of the graph class $\cG$ is taken from \cite{thorup2025instance} and we shall further discuss the relation below.

\subsubsection{Relation to the Instance-Optimality for PageRank in {\cite{thorup2025instance}}}

We will now explain how our result for PPR generalizes and  improves the 
result on PageRank from \cite{thorup2025instance}. 

First recall that PageRank reduces to PPR with an artificial source $s^*$ with out-edges to all other nodes including $t$. With $n$ known, we can recognize this instance, so we know that taking a random out-neighbor from $s^*$ corresponds to a $\jump$ in the original 
graph. We call this the PageRank-instance of PPR.

We have $\pi(s^*,t)\geq \alpha(1-\alpha)/n$, so we can set $\delta=\alpha(1-\alpha)/n$, implying that $\TriPPR$ solves the PageRank-instance within the instance-optimality factor in \Cref{eq:h-factor}. In fact, due to the high out-degree from $s^*$, $\TriPPR$ may never push forward from $s^*$, so it reduces to $\BiPPR$ on the PageRank-instance, and this is exactly the algorithm analyzed in \cite{thorup2025instance} for PageRank.

However, in 
\cite{thorup2025instance}, the instance-optimality factor they proved was 
$((h+1)^2/\eps)\log^{O(1)} n$, that is, they had a quadratic dependence on $(h+1)$ whereas our dependence is linear. For the definition of $\cG$ is does not matter if the dependence is linear or quadratic, but as described above, we need the linear dependence to prove an instance-optimality factor of $\Tilde O(\ceil{m/n})$ for the general PPR problem.

Summing up the major differences, 
\cite{thorup2025instance} shows that within the class $\cG$, the classic $\BiPPR$ is instance-optimal for PageRank. However, PageRank is only a very special case of PPR, and we demonstrate that $\BiPPR$ has almost worst possible performance on some instances of PPR. We propose the new algorithm $\TriPPR$ for PPR that we show is instance-optimal on all sparse graphs. To strictly generalize the result from \cite{thorup2025instance}, we also
show that $\TriPPR$ is instance-optimal within a factor $\tilde O((h+1)/\eps)$ on any PPR instance graph from $\cG$ with $\pi(s,t)=\tilde\Omega(1/n)$, and this subsumes the case of PageRank. We note that for such graphs, we do not have any issues with the $(\log 1/\delta)$-factor since it is $O(\log n)$.

We also show that for the PPR problem, there cannot exist an instance-optimal algorithm for denser graphs. More precisely, we show that the best general instance-optimality factor is $\tilde \Theta(\ceil{m/n})$. In \cite{thorup2025instance} there is no proof that instance-optimality for all graphs is impossible for PageRank.

Finally, we show that $\TriPPR$ is \emph{instance-optimal on every PPR instance} if the number of nodes \emph{$n$ is not known}. 
As pointed out in Section \ref{sec:model}, not knowing $n$ is a common case for large networks. We believe that we are the first work to make this point and use it to get general instance-optimality which is not possible when $n$ is known.

\subsection{Paper Organization} In the body the paper, we first describe our $\TriPPR$ algorithm and establish its running-time upper bound in~\Cref{sec:upperbound}. Then, in~\Cref{sec:lowerbound_withoutn}, we establish an instance-wise lower bound, showing that when $n$ is unknown, no good algorithm can outperform $\TriPPR$ by more than a polylogarithmic factor. We next consider the case where $n$ is known. In~\Cref{sec:instance-non-optimality}, we first show that no instance-optimal algorithm exists for dense graphs. Finally, in~\Cref{sec:lowerbound_withn}, we establish an instance-wise lower bound for the case where $n$ is known.

\subsection{The use of generative AI} The paper is entirely human-made. The authors wrote this paper and used ChatGPT Pro only to assist with working out some technical details and with proofreading, including comments for the exposition.




\section{Review of Push Operations}\label{sec:push_review}
As mentioned before, our $\TriPPR$ algorithm consists of three components: forward push from $s$, backward push from $t$, and Monte Carlo sampling. 

In this section, we briefly review forward and backward push, along with their properties that will be used later in our analysis. At the end of this section, we also discuss the $\BiPPR$ method~\cite{lofgren2016personalized} and the power method in more detail. The former has been proven instance-optimal for PageRank estimation, while the latter achieves the best-known worst-case complexity. However, neither is instance-optimal for single-pair PPR estimation. Readers primarily interested in our $\TriPPR$ algorithm may skip this final part.

\subsection{Forward Push and Its Invariant}

The forward push operation was introduced in~\cite{andersen2006local} for estimating single-source PPR and identifying local clusters around the given source vertex. It deterministically propagates random-walk probability mass along outgoing edges step by step through the graph. Each forward push operation maintains two variables for each vertex $v\in V$: the {\em forward reserve} $\pf(v)$ and the {\em forward residue} $\rf(v)$. Initially, $\pf(v)$ and $\rf(v)$ are set to $0$ for all $v$, except $\rf(s)=1$. The following invariant is maintained for every $v \in V$:

\begin{align}\label{eq:forward-inv}
\pi(s,v)=\pf(v)+\sum_{u\in V}\rf(u)\pi(u,v). 
\end{align}

The forward push operation is atomic and can be applied to any vertex $u$ in the graph, following the steps outlined in~\Cref{alg:forwardpush}. \cite[Lemma 1]{andersen2006local} proves that the invariant~\eqref{eq:forward-inv} is maintained before and after every forward push operation, regardless of the order in which vertices are processed.

\begin{algorithm}[H]
    \caption{$\pushforward(u)$~\cite{andersen2006local}}
    \label{alg:forwardpush}
    \KwIn{vertex $u$}
    \KwOut{updated $\pf()$ and $\rf()$}
        $r \gets \rf(u)$\;
        $\rf(u)\gets 0$\;
        $\pf(u)\gets \pf(u) + \alpha r$\;
        \For{\textup{each} $v\in \Nout(u)$}{
            $\rf(v) \gets \rf(v) + (1-\alpha)r / \dout(u)$\;
        }
\Return $\pf()$ and $\rf()$\;
\end{algorithm}

\subsection{Backward Push and Its Invariant}

As a counterpart of forward push, the backward push operation deterministically propagates random-walk probability mass backward from the target vertex $t$ along incoming edges, step by step.  It was originally introduced in~\cite{andersen2008local} for the estimating single-target PPR. 

Analogous to forward push operations, each backward push operation maintains two variables for each vertex $v\in V$: the {\em backward reserve} $\pb(v)$ and the {\em backward residue} $\rb(v)$. Initially, $\pb(v)$ and $\rb(v)$ are set to $0$ for all $v$, except $\rb(t)=1$ for the target vertex $t$. The following invariant is maintained for every $u \in V$:

\begin{align}\label{eq:back-inv}
\pi(u,t)=\pb(u)+\sum_{v\in V}\pi(u,v)\rb(v). 
\end{align}  

A backward push operation is also atomic. It consists of the steps shown in~\Cref{alg:backwardpush}. A backward push operation can be applied to any vertex. As proven in~\cite[Lemma 1]{andersen2008local}, the invariant~\eqref{eq:back-inv} is maintained throughout the backward push process, regardless of the push order. 

\begin{algorithm}[h]
\DontPrintSemicolon
\caption{$\pushback(v)$~\cite{andersen2008local}}
\label{alg:backwardpush}
\KwIn{vertex $v$}
\KwOut{updated $\rb()$ and $\pb()$}
$r\gets \rb(v)$\; 
$\rb(v)\gets 0$\; 
$\pb(v)\gets \pb(v)+\alpha r$\; 
\For{\textup{each} $u \in \Nin(v)$}{
    $\rb(u)\gets \rb(u)+(1-\alpha)r/\dout(u)$\; 
}
\Return $\rb()$ and $\pb()$ \;
\end{algorithm}

\subsection{The $\BiPPR$ Method}

Recall that~\cite{lofgren2016personalized} proposes a method called $\BiPPR$, which combines backward push and Monte Carlo sampling to estimate single-pair PPR. Specifically, it sets a push threshold $\rmax$, and repeatedly performs backward push operations until no vertex has a backward residue $\rb(v)\geq \rmax$. Then it performs Monte Carlo sampling by generating discounted random walks from $s$. The algorithm balances the running times of the two phases, and the final estimate $\epi(s,t)$ of $\pi(s,t)$ is obtained by combining their results, as illustrated in~\Cref{alg:bippr}.

\begin{algorithm}[h]
\DontPrintSemicolon
\caption{$\BiPPR (t, \rmax, \nr)$~\cite{lofgren2016personalized}} 
\label{alg:bippr}
\KwIn{target vertex $t$, threshold $\rmax$, total number of random walks $\nr$}
\KwOut{estimate for $\vpi(t)$}
$\pb(),\rb()\gets$ arrays with all entries $0$ \;
$\rb(t)\gets 1$ \;
\While{$\exists v \in V$ \textup{with} $\rb(v)\ge \rmax$}{
        $\pushback(v)$ $\quad$ \textcolor{gray}{// invoke \Cref{alg:backwardpush}}\;
}
$\nr(v) \gets 0$ for each $v\in V$ $\quad$ \textcolor{gray}{// recording the number of random walks terminating at $v$} \; 
\For{$i$ \textup{from} $1$ \textup{to} $\nr$}{
Simulate a discounted random walk from $s$ and use $v$ to denote the ending vertex\;
$\nr(v)\gets \nr(v)+1$\;
}
\Return $\epi(s,t)\gets \pb(s)+\sum_{v\in V}(\nr(v)/\nr)\rb(v)$ \;
\end{algorithm}

The recent work~\cite{thorup2025instance} shows that the adaptive variant of $\BiPPR$ is instance-optimal for PageRank estimation on all graphs with at most polylogarithmically many vertices whose in- or out-degree is $(1-o(1))n$, i.e., in the graph class $\cG$. However, it is easy to see that no reasonable variant of $\BiPPR$ is instance-optimal for single-pair PPR estimation, as it can be much slower than an instance-smart algorithm on the examples in~\Cref{fig:high-level-idea}. In the example in~\Cref{fig:high-level-idea}, Monte Carlo sampling requires $\Omega(m)$ time to reach $u$ from $s$, and backward push requires $\Omega(m)$ time to propagate probability mass from $t$ to $u$. The time cost of $\BiPPR$ is the summation of that of Monte Carlo sampling and backward push, and achieves the minimum when balancing the time cost of the two phases, thus also requiring $\Omega(m)$ time on the example. In contrast, the instance-smart algorithm costs only $O(\log{n})$ time on the example, as explained in~\Cref{subsec:basic_idea}.

\subsection{The Power Method}

Recall that the power method was proposed in the seminar paper~\cite{page1998pagerank, brin1998anatomy} that introduces PPR and PageRank. We introduced the power method earlier in \Cref{subsec:ppr_intro} from the perspective of iteratively applying the PPR recurrence, but it is actually equivalent to performing forward push on all vertices synchronously and repeatedly until the probability mass accumulated at $t$ converges, as illustrated in~\Cref{alg:push_by_level}.

\begin{algorithm}[h]
    \caption{\texttt{ForwardPowerMethod}$(s,\delta)$}
    \label{alg:push_by_level}
\KwIn{source vertex $s$, approximation threshold $\delta$}
\KwOut{estimate for $\vpi(s, t)$}
        Initialize $\pf(v)\gets 0$ and $\rf(v)\gets 0$ for all $v\in V$\; 
        $\rf(s)\gets 1$\;
        \For{$\ell=0,1,2,\dots, \lceil \log_{1-\alpha} \delta \rceil$}{
            Run $\pushforward(v)$ synchronously on every $v\in V$\;
        }
        \Return $\pf(t)$\;
\end{algorithm}

The running time of the power method is $O(m\log(1/\delta))$ for estimating $\pi(s,t)\ge \delta$. However, as illustrated in~\Cref{fig:small_estimate}, this can be as large as $O(mn\log n)$ when estimating extremely small values of $\pi(s,t)$. Specifically, in the example in~\Cref{fig:small_estimate}, starting from the source vertex $s$, forward push first propagates probability mass to $v_1$ and all the $w$-vertices (i.e., the outgoing neighbors of $s$). In expectation (depending on the ordering of vertices in the adjacency lists), the algorithm may perform forward push on almost all the $w$-vertices before pushing from $v_1$. Only after pushing from $v_1$ does $v_2$ receive probability mass. Similarly, before pushing from $v_2$, the algorithm may again perform forward push on almost all the $w$-vertices. This pattern repeats as the algorithm progresses along the path $s v_1 v_2 \cdots t$. Since each forward push on a $w$-vertex costs $\Theta(n)$ time to propagate probability mass to its $\Theta(n)$ outgoing neighbors, and $\Theta(n)$ such $w$-vertices are processed before advancing one step along the path, the total running time before probability mass reaches $t$ is $\Omega(mn)$, i.e., $\Omega(n^3)$ in this example. 

The power method can also be implemented in a backward fashion by repeatedly performing backward push synchronously on all vertices, starting from $t$. For completeness, we provide its pseudocode in~\Cref{alg:powermethod_backward} below. 

\begin{algorithm}[h]
    \caption{\texttt{BackwardPowerMethod}$(s,\delta)$}
    \label{alg:powermethod_backward}
\KwIn{target vertex $t$, approximation threshold $\delta$}
\KwOut{estimate for $\vpi(s, t)$}
        Initialize $\pb(v)\gets 0$ and $\rb(v)\gets 0$ for all $v\in V$\; 
        $\rb(t)\gets 1$\;
        \For{$i=0,1,2,\dots, \lceil \log_{1-\alpha} \delta \rceil$}{
            Run $\pushback(v)$ synchronously on every $v\in V$\;
        }
        \Return $\pb(s)$\;
\end{algorithm}

On this example, this backward version behaves symmetrically: starting from $t$, it spends substantial time processing the $v$-vertices before advancing backward along the path $t \cdots v_2 v_1 s$, thus also incurring a running time of $O(mn\log n)$.

\section{Push with Vertex Freezing}

Recall that our $\TriPPR$ algorithm introduces a vertex-freezing technique to efficiently estimate $\pi(s,t)$ when its value is extremely small. Before presenting the full $\TriPPR$ algorithm, we first describe this technique in this section. The technique is simple and can be readily incorporated into both forward and backward push processes. Below, we describe the technique in the context of forward push and defer its backward counterpart to the end of this section for completeness.

\subsection{Basic Idea}
The high-level idea behind the vertex-freezing technique is that if the probability mass a vertex $v$ will receive in the future is sufficiently small compared to the mass it has already received, then further pushes from $v$ will contribute only negligibly to $\pi(s,t)$. We can therefore \emph{freeze} $v$ in this case, meaning that no further pushes will be performed from $v$.

More specifically, given a source vertex $s$, we initialize $\rf(u)=\pf(u)=0$ for all $u\neq s$, and set $\rf(s)=1$ and $\pf(s)=1$, as in the standard forward-push algorithm. The pseudocode for initialization is given in~\Cref{alg:init_f}. 

After initialization, we proceed iteratively for $\ell=0,1,2,\ldots$, performing forward push synchronously in each iteration, as in the power method described in~\Cref{alg:push_by_level}. However, instead of pushing from every vertex in each iteration, we start pushing from a vertex $u$ only when its forward residue satisfies: 
\begin{align*}
\rf(u)/\dout(u)\ge \rsumf/\budgetf. 
\end{align*}
We refer to this as \emph{activating} a vertex. Once activated, a vertex continues to be pushed in subsequent iterations until it is frozen.

Specifically, we maintain a candidate set $\candf$, which contains all activated vertices that have not yet been frozen. Whenever a vertex is activated, it is added to the candidate set. In each iteration, we perform forward push only on vertices in $\candf$, and only when their total forward residual mass satisfies: 
\begin{align*}
\sum_{u\in \candf^{(\ell)}}\rf^{(\ell)}(u)\ge \rsumf^{(\ell)}/2. 
\end{align*}

Moreover, we introduce a freezing threshold $\ifrozenf$ to limit the number of forward-push operations performed on each vertex. Specifically, we perform forward push on each vertex $u$ at most $\ifrozenf$ times, after which the vertex is \emph{frozen}, meaning that no further forward-push operations will be performed on $u$. 
We maintain a set $\frozenf$ of frozen vertices. Once a vertex is frozen, it is added to $\frozenf$ and permanently removed from the candidate set $\candf$. Frozen vertices would never be unfrozen, so the size of $\frozenf$ is non-decreasing.

The pseudocode for the forward-push process with the vertex-freezing technique is provided in~\Cref{alg:pushforward_frozen}. The algorithm takes a running-time budget $\budgetf\ge 0$ as input. Before each iteration $\ell$, we compute its anticipated running time, given by $\sum_{u\in \candf}\dout(u)$, which is the cost of performing forward push on all vertices in $\candf$ in the current iteration. If the anticipated running time exceeds the remaining budget, we do not proceed with the iteration.

\begin{algorithm}[H]
\DontPrintSemicolon
\caption{$\textup{\texttt{ForwardPush-Init}}(s)$}
\KwIn{source vertex $s$}
\label{alg:init_f}
$\frozenf \gets \emptyset$ \textcolor{gray!60}{\tcp*[r]{set of frozen vertices}}
$\candf\gets \{s\}$\textcolor{gray!60}{\tcp*[r]{set of active vertices}}
$\cntf(u)\gets 0$ for each $u\in V$\textcolor{gray!60}{\tcp*[r]{count the number of forward pushes on $u$}} 
$\rf(u)\gets 0$ for each $u\neq s$, and  $\rf(s)\gets 1$\; 
$\pf(u)\gets 0$ for each $u\in V$\;
\end{algorithm}

\begin{algorithm}[h]
\DontPrintSemicolon
\caption{$\frozenpushforward(s,\budgetf, \ifrozenf)$}
\label{alg:pushforward_frozen}
\KwIn{source vertex $s$, budget $\budgetf$, freezing threshold $\ifrozenf$}
$\budgetf'\gets \budgetf$ \textcolor{gray!60}{\tcp*[r]{remaining budget}}
\For{$\ell=1,2,\ldots$}{
    $\rsumf \gets \sum_{u\in (V \setminus \frozenf)}\rf(u)$  \textcolor{gray!60}{\tcp*[r]{total residue of all non-frozen vertices}}
    $\candf \gets \left\{u\in (V \setminus \frozenf) \mid \rf(u)/\dout(u)\ge \rsumf/\budgetf \right\} \bigcup \left\{\candf \setminus \frozenf \right\}$\;  
    \textbf{if} 
    $\candf$ is empty
    \textbf{or} $\sum_{u\in \candf}\rf(u)<\rsumf/2$, 
    \textbf{or} $\sum_{u\in \candf}\dout(u)>\budgetf'$ \textbf{then break}\;
    Run $\pushforward(u)$ synchronously on every $u\in \candf$ \textcolor{gray!60}{\tcp*[r]{invoke \Cref{alg:forwardpush}}}
$\budgetf'\gets \budgetf'-\sum_{u\in \candf}\dout(u)$\;
Increase $\cntf(u)$ for each vertex $u\in \candf$ by $1$ and add all $u$ with $\cntf(u)\ge \ifrozenf$ into $\frozenf$\;
}
\end{algorithm}

\subsection{Approximation Error}
We now analyze the approximation error of \Cref{alg:pushforward_frozen}.
For each iteration $\ell\ge 0$, we use $\candf^{(\ell)}$ and $\frozenf^{(\ell)}$ to denote the set of vertices that have been activated and have been frozen at the beginning of iteration $\ell$, respectively.  We also define $\rsumf^{(\ell)}=\sum_{u\in V\setminus \frozenf^{(\ell)}}\rf^{(\ell)}(u)$, that is, the total forward residue over all non-frozen vertices at the beginning of iteration $\ell$. 
Additionally, we use $\rf^{(\ell)}(u)$ and $\pf^{(\ell)}(u)$ to denote the forward residue $\rf(u)$ and reserve $\pf(u)$, respectively, at the beginning of iteration $\ell$.

\begin{lemma}\label{lem:decay_forward}
For each integer $\ell\ge 0$, performing forward push synchronously on all vertices in $\candf^{(\ell)}$ only when $\sum_{u\in \candf^{(\ell)}}\rf^{(\ell)}(u)\ge \rsumf^{(\ell)}/2$ ensures that
\begin{align*}
\rsumf^{(\ell)} \le (1-\alpha/2)^\ell, \quad \text{ and } \quad \rsumf^{(\ell+1)}\le (1-\alpha/2)\rsumf^{(\ell)}. 
\end{align*}
\end{lemma}

\begin{proof}
    We prove this lemma by induction on $\ell$. For $\ell=0$, the frozen vertices' set $\frozenf^{(0)}$ is an empty set and $\rf^{(0)}(u)=0$ for all $u$ except $\rf^{(0)}(s)=1$, so 
    \begin{align*}
    \rsum^{(0)}=\sum_{u\in \big(V\setminus \frozenf^{(0)}\big)}\rf^{(0)}(u)=\rf^{(0)}(s)=1=(1-\alpha/2)^0. 
    \end{align*}
    So \Cref{lem:decay_forward} holds when $\ell=0$. 

    We now fix an $\ell$ and assume the following holds for $\ell$: 
    \begin{align}\label{eqn:rsum_assumption}
    \rsumf^{(\ell)}=\sum_{u\in \big(V\setminus \frozenf^{(\ell)}\big)}\rf^{(\ell)}(u)\le (1-\alpha/2)^\ell. 
    \end{align}
    We know that $\frozenf^{(\ell)}\subseteq \frozenf^{(\ell+1)}$, since once a vertex is frozen, it will never be unfrozen. 
    This implies 
    \begin{equation}\label{eqn:induction_nextstep}
    \begin{aligned}
    \rsumf^{(\ell+1)}=\sum_{u\in \big(V\setminus \frozenf^{(\ell+1)}\big)}\rf^{(\ell+1)}(u) 
\le \sum_{u\in \big(V\setminus \frozenf^{(\ell)}\big)}\rf^{(\ell+1)}(u)
\le \sum_{u\in \big(V\setminus \frozenf^{(\ell)}\big)}\rf^{(\ell)}(u)-\sum_{u\in \candf^{(\ell)}}\alpha \rf^{(\ell)}(u),
    \end{aligned}
    \end{equation}
    where the last inequality is not necessarily an equality, since some residue may be pushed to frozen vertices, whereas $\rsum$ only accounts for the total residue over non-frozen vertices. 
    
    By definition $\sum_{u\in \big(V\setminus \frozenf^{(\ell)}\big)}\rf^{(\ell)}(u)= \rsumf^{(\ell)}$. 
    We also note that we only push on vertices in $\candf^{(\ell)}$ when $\sum_{u\in \candf^{(\ell)}}\rf^{(\ell)}(u)\ge \rsumf^{(\ell)}/2$. 
    Substituting these into~\Cref{eqn:induction_nextstep}, we obtain 
    \begin{align*}
    \rsumf^{(\ell+1)}=\sum_{u\in \big(V\setminus \frozenf^{(\ell+1)}\big)}\rf^{(\ell+1)}(u)\le \rsumf^{(\ell)}-\alpha \rsumf^{(\ell)} /2 =(1-\alpha/2) \rsumf^{(\ell)}. 
    \end{align*}
    Since we assume $\rsumf^{(\ell)}\le (1-\alpha/2)^{\ell}$ in~\Cref{eqn:rsum_assumption}, we then have $\rsumf^{(\ell)}\le (1-\alpha/2)^\ell$. Therefore, $\rsumf^{(\ell+1)}\le (1-\alpha/2) (1-\alpha/2)^\ell = (1-\alpha/2)^{\ell+1}$, completing the inductive step. 
\end{proof}

\begin{lemma}\label{lem:mid_freezing_f}
For a vertex $u$, activating it for the first time when $\rf(u)/\dout(u)\ge \rsumf/\budgetf$ and then freezing it after $\ifrozenf$ forward-push operations ensure that 
\begin{align*}
\rf(u)\pi(u,t)\le (1-\alpha/2)^{\ifrozenf}\budgetf\pi(s,t) 
\end{align*}
after it is frozen. 
\end{lemma}

\begin{proof}
Suppose that $u$ is activated in iteration $\ell_u$. This implies
\begin{align}\label{eqn:upper_pi_t}
\rf^{(\ell_u)}(u)\ge \frac{\dout(u)\rsumf^{(\ell_u)}}{\budgetf}.
\end{align}
By the forward push invariant~\eqref{eq:forward-inv}, 
at the beginning of iteration $\ell_u$, we have
\begin{align}\label{eqn:invariant_oneresidue}
\pi(s,t)
= \pf^{(\ell_u)}(t)+\sum_{v\in V}\rf^{(\ell_u)}(v)\pi(v,t) 
\ge \rf^{(\ell_u)}(u)\pi(u,t). 
\end{align}
Combining~\Cref{eqn:upper_pi_t} and~\Cref{eqn:invariant_oneresidue} yields
\begin{align}\label{eqn:upper_pi_ut}
\pi(u,t)
\le \frac{\pi(s,t)}{\rf^{(\ell_u)}(u)}
\le \frac{\pi(s,t)\budgetf}{\dout(u)\rsumf^{(\ell_u)}}.
\end{align}

Since vertex $u$ is activated in iteration $\ell_u$, it will be frozen at the end of iteration $\ell_u+\ifrozenf-1$. 
Let $\rf(u)$ denote the forward residue of $u$ after it is frozen. We have
\begin{align}\label{eqn:forward_sum}
\rf(u)
\le \sum_{x\in V\setminus \frozenf^{(\ell_u+\ifrozenf)}}\rf^{(\ell_u+\ifrozenf)}(x)=\rsumf^{(\ell_u+\ifrozenf)}.
\end{align}
This is straightforward because, once $u$ is frozen, it can no longer push residual mass to other vertices, and can only receive residual mass propagated from other vertices that have not been frozen by the end of iteration $\ell_u+\ifrozenf-1$. Therefore,
\begin{align*}
\rf(u)\le
\sum_{x\in V\setminus \frozenf^{(\ell_u+\ifrozenf)}}\rf^{(\ell_u+\ifrozenf)}(x)\Pr{x\to u},
\end{align*}
where $\Pr{x\to u}$ denotes the fraction of residue mass that can be transferred from $x$ to $u$. Since $\Pr{x\to u}\le 1$ by definition, we obtain \Cref{eqn:forward_sum}.

Recall from~\Cref{lem:decay_forward} that $\rsumf^{(\ell+1)}\le (1-\alpha/2)\rsumf^{(\ell)}$ holds for every iteration $\ell$. Applying this bound repeatedly and combining it with~\Cref{eqn:forward_sum}, we obtain
\begin{align}\label{eqn:upper_rf}
\rf(u)\le \rsumf^{(\ell_u)}(1-\alpha/2)^{\ifrozenf}.
\end{align}
Further combining~\Cref{eqn:upper_pi_ut} with~\Cref{eqn:upper_rf} yields
\begin{align*}
\rf(u)\pi(u,t)
\le \rsumf^{(\ell_u)}(1-\alpha/2)^{\ifrozenf}
\cdot \frac{\pi(s,t)\budgetf}{\dout(u)\rsumf^{(\ell_u)}} 
= \frac{(1-\alpha/2)^{\ifrozenf}\budgetf}{\dout(u)}\pi(s,t) 
\le (1-\alpha/2)^{\ifrozenf}\budgetf\pi(s,t),
\end{align*}
as claimed. 
\end{proof}

\begin{lemma}\label{lem:freeze_f}
Setting $\ifrozenf=\lceil \log_{1-\alpha/2} \frac{\rela}{2^i \budgetf^2} \rceil$ ensures that
\begin{align*}
\sum_{u\in \frozenf}\rf(u)\pi(u,t)\le (\rela/2^i) \pi(s,t)
\end{align*}
at the end of \Cref{alg:pushforward_frozen}. 
\end{lemma}

\begin{proof}
Let $\frozenf$ denote the set of frozen vertices at the end of~\Cref{alg:pushforward_frozen}. 
Since each vertex must be pushed at least once before being frozen, and each push incurs at least unit cost, the number of frozen vertices satisfies $|\frozenf|\le \budgetf$. By \Cref{lem:mid_freezing_f}, for each $u\in \frozenf$, 
$\rf(u)\pi(u,t)
\le (1-\alpha/2)^{\ifrozenf}\budgetf\pi(s,t)$. 
Summing over all frozen vertices, we obtain 
\begin{align*}
\sum_{u\in \frozenf}\rf(u)\pi(u,t)
\le (1-\alpha/2)^{\ifrozenf}\budgetf^2\pi(s,t).
\end{align*}
Therefore, setting $\ifrozenf=\left\lceil \log_{1-\alpha/2}\frac{\rela}{2^i \budgetf^2}\right\rceil$
ensures that $\sum_{u\in \frozenf}\rf(u)\pi(u,t)
\le \frac{\rela}{2^i}\pi(s,t)$.
\end{proof}

\subsection{Running Time}

We now upper bound the running time of~\Cref{alg:pushforward_frozen}.

\begin{lemma}\label{lem:runtime_freezing}
The running time of~\Cref{alg:pushforward_frozen} is $O(\budgetf)$. 
\end{lemma}

\begin{proof}
The running time of~\Cref{alg:pushforward_frozen} is dominated by two components. 
The first is the cost of performing forward push in each iteration. Recall that in~\Cref{alg:pushforward_frozen}, before starting an iteration, we compute the anticipated cost of the forward-push operations and check whether it exceeds the remaining budget. If so, we terminate the algorithm without executing that iteration. Therefore, the total number of arithmetic operations performed throughout the forward-push process is $O(\budgetf)$.

\paragraph{Maintaining $\boldsymbol{\rsumf}$.}

We maintain $\rsumf=\sum_{u\in V\setminus\frozenf}\rf(u)$ in each iteration using only addition, and claim that the total number of addition operations is $O(\budgetf)$.

For active vertices, the cost of summing their residues is covered by the forward-push operations performed on them in that iteration. For inactive, unfrozen vertices with nonzero residues, we claim that the cost of summing their residues in each iteration is also covered by the forward-push operations performed in that iteration.

To see this, consider an inactive, unfrozen vertex $v$ with nonzero residue in iteration $\ell$. Suppose that $v$ first receives residue from an in-neighbor $u$ in iteration $\ell_u\le \ell$. By the vertex activation rule in~\Cref{alg:pushforward_frozen}, we have $\rf^{(\ell_u)}(u)/\dout(u) \ge \rsumf^{(\ell_u)}/\budgetf$. 
The forward-push operation performed on $u$ in iteration $\ell_u$ increases $\rf(v)$ by $(1-\alpha)\rf(u)/\dout(u)$, thus implies
\begin{align}\label{eqn:residue_lowerbound}
\rf^{(\ell_u)}(v)
\ge \frac{(1-\alpha)\rf^{(\ell_u)}(u)}{\dout(u)}
\ge \frac{(1-\alpha)\rsumf^{(\ell_u)}}{\budgetf},
\end{align}

Note that $v$ first receives residue from $u$ in iteration $\ell_u$. This means that $u$ is activated in iteration $\ell_u$, since once $u$ is activated, forward push starts being performed on $u$, and $v$ starts receiving residue. 
At the end of iteration $\ell_u+\ifrozenf-1$, $u$ will be frozen. By~\Cref{lem:decay_forward,lem:freeze_f}, we have
\begin{align*}
\rsumf^{(\ell_u+\ifrozenf)}
\le (1-\alpha/2)^{\ifrozenf}\rsumf^{(\ell_u)}\le \frac{\rela}{8\budgetf^2}\rsumf^{(\ell_u)},
\end{align*}
where the second inequality follows from
$\ifrozenf=\lceil \log_{1-\alpha/2} \frac{\rela}{8\budgetf^2} \rceil$.
Consequently,
\begin{align*}
\rf^{(\ell_u+\ifrozenf)}(v)
\le \rsumf^{(\ell_u+\ifrozenf)} \le \frac{\rela}{8(1-\alpha)\budgetf}
\cdot\frac{(1-\alpha)\rsumf^{(\ell_u)}}{\budgetf}.
\end{align*}
As a result, when $\budgetf>\rela/(8(1-\alpha))$, $v$ must have been activated by iteration $\ell_u+\ifrozenf$. Otherwise, this would contradict~\Cref{eqn:residue_lowerbound}, since the residue of an inactive vertex never decreases before it becomes active. Since $\rela/(8(1-\alpha))=O(1)$, we conclude that, for $\budgetf$ larger than this constant, $v$ must become active before $u$ is frozen. More generally, when $\budgetf>\rela/(8(1-\alpha))$, every inactive vertex with nonzero residue continues to receive residue from at least one of its active in-neighbors until it becomes active. Therefore, the cost of summing the residue of $v$ in each iteration can be charged to the forward-push operations that propagate residue to $v$ in that iteration.

When $\budgetf\le \rela/(8(1-\alpha))=O(1)$, the analysis is simpler. In this case, the number of inactive vertices in each iteration is at most $\budgetf$ and is therefore constant. Thus, we can simply scan all inactive vertices in each iteration to compute their total residue. Since each such vertex remains inactive for at most $\budgetf$ iterations, the total cost is $O(\budgetf)=O(1)$.

Consequently, the total cost of maintaining $\rsumf$ is covered by the forward-push cost and is therefore $O(\budgetf)$.

\paragraph{Finding newly activated vertices.} We now analyze the cost of identifying the vertices to be activated in each iteration, i.e., computing $\candf^{(\ell)}$ for each iteration $\ell \ge 1$. Recall that
\begin{align*}
\candf^{(\ell)} = \left\{u\in (V \setminus \frozenf^{(\ell)}) \mid \rf^{(\ell)}(u)/\dout(u)\ge \rsumf^{(\ell)}/\budgetf \right\} \bigcup \left\{\candf^{(\ell-1)}\setminus \frozenf^{(\ell)} \right\}.
\end{align*}
Given $\candf^{(\ell-1)}$ and $\frozenf^{(\ell)}$, the key is to compute $\candf^{(\ell)}\setminus \candf^{(\ell-1)}$ for each $\ell$, i.e., to identify all newly activated vertices in iteration $\ell$.

As established above, when $\budgetf>\rela/(8(1-\alpha))$, once an inactive vertex starts receiving residue, it continues to receive residue in subsequent iterations until it becomes active. Therefore, the cost of identifying newly activated vertices in each iteration can also be charged to the forward-push operations performed in that iteration.

When $\budgetf\le \rela/(8(1-\alpha))=O(1)$, the number of inactive vertices in each iteration is at most $\budgetf=O(1)$. Thus, we can simply scan all inactive vertices to identify those that should be activated, at a cost of $O(\budgetf)$ per iteration. Since there are at most $O(\budgetf)$ iterations, the total cost is $O(\budgetf^2)=O(\budgetf)=O(1)$.

Consequently, the total cost of identifying newly activated vertices over all iterations is $O(\budgetf)$.

\paragraph{RAM model on floating-point numbers.} 
Recall that we work in the RAM model, where arithmetic operations on floating-point numbers with $O(\log n)$-bit exponents and mantissas take constant time. We will show that this precision suffices for all arithmetic operations in \Cref{alg:pushforward_frozen}, ensuring that the accumulated numerical errors do not affect the constant multiplicative factor approximation guarantee.

We first observe that \Cref{alg:pushforward_frozen} does not involve subtraction; it uses only addition, multiplication, and division. More specifically, the arithmetic operations are of two types: multiplying a variable by $(1-\alpha)/d$ for some vertex degree $d$ in the graph, and adding two nonnegative variables. Since the total number of arithmetic operations in~\Cref{alg:pushforward_frozen} is bounded by $\budgetf$ and $d\le n$, every nonzero value arising from these operations is at least $(1-\alpha)^{\budgetf}/n^{\budgetf}$. 
As we shall show later in~\Cref{sec:worst-case}, $\budgetf=O(m\log n)$. Thus, every nonzero value is at least $1/n^{O(n^2\log n)}$. Consequently, the maximum absolute exponent is $O(n^2\log^2 n)$, which can be represented using $O(\log n)$ bits.

Next, we analyze the accumulated numerical errors introduced by the floating-point representation. Suppose each floating-point operation introduces a relative error of at most $\rela$. This implies that each multiplication or division by an exact value increases the relative error by at most $\rela$. Moreover, since all additions involve nonnegative values, the relative error of their sum is at most $\rela$ larger than the maximum relative error of the two operands. Therefore, after $\budgetf$ arithmetic operations, the maximum accumulated relative error is $O(\budgetf\rela)$.

Since $\budgetf=O(m\log n)=O(n^2\log n)$, setting $\rela=1/n^4$ gives a maximum accumulated relative error of
$O\left(\frac{n^2\log n}{n^4}\right)
=o(1)$. 
We also note that using $O(\log n)$ bits for the mantissa suffices to ensure that each floating-point operation introduces a relative error of at most $1/n^4$.
Therefore, $O(\log n)$ bits suffice for both the exponent and the mantissa, ensuring that the accumulated numerical errors do not affect the constant-factor approximation guarantee.

Combining all results above establishes the $O(\budgetf)$ running time of \Cref{alg:pushforward_frozen}. 
\end{proof}

\subsection{Worst-Case Complexity}\label{sec:worst-case}

By incorporating the vertex-freezing technique into the forward-push process as shown in \Cref{alg:pushforward_frozen}, we can already improve the worst-case complexity of single-pair PPR estimation.

Recall that the previously best worst-case complexity is $O(\min\{1/\delta, m\log(1/\delta)\})$, achieved by combining Monte Carlo sampling and the power method. The adaptive variant of \Cref{alg:pushforward_frozen} improves this bound to $O(\min\{1/\delta, m\log n\})$, replacing the $\log(1/\delta)$ factor, which can be as large as $\Theta(n\log n)$, with $\log n$. This matches the lower bound of $\Omega(\min\{1/\delta,m\})$ up to an $O(\log n)$ factor.

This improved complexity bound also extends to the single-source problem (computing constant-factor multiplicative approximations of $\pi(s,t)$ for a given source $s$ and all $t\in V$) and the single-target problem (computing constant-factor multiplicative approximations of $\pi(s,t)$ for a given target $t$ and all $s\in V$), as stated below. For both problems, this improves upon the previously best upper bound of $O(m\log(1/\delta))$.

\begin{theorem}
\label{thm:worst-case-time}
The single-pair, single-source and single-target problems can all be solved in $O(m\log n)$ time deterministically.
\end{theorem}

\begin{proof}
\Cref{alg:pushforward_frozen} takes a running-time budget $\budgetf$ as input. We repeatedly run \Cref{alg:pushforward_frozen} with an increasing budget, initially set to $\budgetf=2$. Whenever the budget is exhausted while some vertices remain unfrozen, we double $\budgetf$ and continue the push process. The process terminates once all vertices have been frozen.

Note that every time we increase the budget, we do not restart the push process from scratch, but rather that we continue the pushing from previous rounds. So the sets of frozen vertices $\frozenf$ and $\frozenb$ are never reset, and vertices frozen in earlier rounds remain frozen in all subsequent rounds. Similarly, the push counters $\cntf(u)$ and $\cntb(u)$ are never reset to zero and can only increase over time. Note, however, that the freezing thresholds increase across rounds, allowing more push operations per vertex in later rounds. So a vertex is less likely to be frozen in a later round than in an earlier one. This choice of increasing freezing thresholds is essential for bounding the approximation error, specified as below.

\begin{lemma}\label{lem:final_approx_worst}
Let $\ifrozenf^{(i)}$ be the freezing threshold in round $i$, and let $\frozenf$ be the set of frozen sets at the end. 
Setting $\ifrozenf^{(i)}=\lceil \log_{1-\alpha/2} \frac{\rela}{2^{3i+4}} \rceil$ for round $i$ ensures that 
\begin{align*}
\sum_{u\in \frozenf}\rf(u)\pi(u,t)\le (\rela/8) \pi(s,t). 
\end{align*}
when the algorithm terminates.  
\end{lemma}

\begin{proof}
Let $\frozenf^{(i)}$ denote the set of frozen vertices in the beginning of round $i$. 
By~\Cref{lem:freeze_f}, setting $\ifrozenf^{(i)}=\lceil \log_{1-\alpha/2} \frac{\rela}{2^{3i+4}} \rceil$ for round $i$ ensures 
\begin{align*}
\sum_{u\in \frozenf^{(i)}/\frozenf^{(i-1)}}\rf(u)\pi(u,t)\le \frac{\rela}{16 \cdot 2^i} \pi(s,t). 
\end{align*}
Consequently, 
\begin{align*}
\sum_{u\in \frozenf}\rf(u)\pi(u,t)= \sum_{i=1}^{\infty}\sum_{u\in \frozenf^{(i)}/\frozenf^{(i-1)}}\rf(u)\pi(u,t)\le (\rela/8)\pi(s,t), 
\end{align*}
completing the proof. 
\end{proof}

We note that when $\budgetf$ exceeds $m\ifrozenf$, all vertices must have been frozen, since forward push is performed on each vertex at most $\ifrozenf$ times before it is frozen. By~\Cref{lem:final_approx_worst}, $\ifrozenf=O(\log\budgetf)$, so the algorithm terminates when $\budgetf=O(m\log n)$. We further recall from \Cref{lem:runtime_freezing} that each execution of \Cref{alg:pushforward_frozen} with budget $\budgetf$ takes $O(\budgetf)$ time. Consequently, the total running time across all executions, up to the final budget of $O(m\log n)$, is $O(m\log n)$. 

\Cref{alg:pushforward_frozen} can be directly extended to solve the single-source PPR estimation problem by using $\pf(u)$ as an estimate of $\pi(s,u)$ for each $u\in V$. For the single-target PPR estimation problem, we will describe a backward version of the algorithm in the next subsection, whose resulting backward reserve $\pb(u)$ can be used as an estimate of $\pi(u,t)$.
\end{proof}

\subsection{Backward Push with Vertex Freezing}

We have shown above how to incorporate the vertex-freezing technique into the forward-push process. Similarly, this technique can also be incorporated into the backward-push process, as illustrated in~\Cref{alg:powermethod_backward}. 

\begin{algorithm}[H]\label{alg:init}
\DontPrintSemicolon
\caption{$\textup{\texttt{BackwardPush-Init}}(t)$}
\KwIn{target vertex $t$}
$\frozenb \gets \emptyset$ \textcolor{gray!60}{\tcp*[r]{set of frozen vertices}}
$\candb\gets \{t\}$\textcolor{gray!60}{\tcp*[r]{set of active vertices}}
$\cntb(v)\gets 0$ for each $v\in V$\textcolor{gray!60}{\tcp*[r]{count the number of backward pushes on $u$}} 
$\rb(v)\gets 0$ for each $v\neq t$, and  $\rb(t)\gets 1$\; 
$\pb(v)\gets 0$ for each $v\in V$\;  
\end{algorithm}

\begin{algorithm}[h]
\DontPrintSemicolon
\caption{$\frozenpushback(t, \budgetb, \ifrozenb)$}\label{alg:pushbackward_frozen}
\KwIn{target vertex $t$, budget $\budgetb$, freezing threshold $\ifrozenb$
}
$\budgetf'\gets \budgetf$ \textcolor{gray!60}{\tcp*[r]{remaining budget}}
\For{$\ell=1,2,\ldots$}{
    $\rmaxb \gets \max_{v\in \{V \setminus \frozenb\}}\rb(v)$  \textcolor{gray!60}{\tcp*[r]{maximum residue over all non-frozen vertices}}
    $\candb \gets \left\{v\in V \setminus \frozenb \mid \rb(v)\ge \alpha \rmaxb/2 \right\}\bigcup \{\candb \setminus \frozenb\}$\; 
    \textbf{if} $\candb$ is empty, \textbf{or} $\sum_{v\in \candb}\din(v)>\budgetb$ \textbf{then break}\;
    
Run $\pushback(v)$ synchronously on every $v\in \candb$ \textcolor{gray!60}{\tcp*[r]{invoke \Cref{alg:backwardpush}}}
$\budgetb \gets \budgetb-\sum_{v\in \candb^{(\ell)}}\din(v)$\;
Increase $\cntb(v)$ for each vertex $v\in \candb$ by $1$ and add all $v$ with $\cntb(v)\ge \ifrozenb$ into $\frozenb$
}
\end{algorithm}

Initially, we set the backward residue $\rb(v)$ and backward reserve $\pb(v)$ to $0$ for all $v\in V$, except that $\rb(t)=1$. 
Let $\rmaxb$ denote the maximum backward residue among all non-frozen vertices, i.e., $\rmaxb=\max_{v\in V\setminus\frozenb}\rb(v)$.
A vertex $v$ is activated for the first time when $\rb(v)\ge \alpha\rmaxb/2$. Once activated, a vertex remains in the candidate set for the next $\ifrozenb$ iterations, after which it is frozen.

Analogously to the forward version, we proceed iteratively for $\ell=1,2,\ldots$. In each iteration $\ell$, we maintain a candidate set $\candb^{(\ell)}$, defined as
\begin{align*}
\candb^{(\ell)} = \left\{v\in V \setminus \frozenb^{(\ell)} \mid \rb^{(\ell)}(v)\ge \alpha \rmaxb^{(\ell)}/2 \right\}\cup \left(\candb^{(\ell-1)}\setminus \frozenb^{(\ell)}\right), 
\end{align*}
where $\frozenb$ denotes the set of frozen vertices, and the superscript $(\ell)$ indicates the value of each variable at the beginning of iteration $\ell$. 
We perform backward push synchronously on all vertices in $\candb^{(\ell)}$ for every $\ell\ge 1$. 

 The algorithm also takes a running-time budget $\budgetb$ as input and terminates if the anticipated cost of the backward-push operations in the next iteration exceeds the remaining budget.

\begin{lemma}\label{lem:decay_backward}
For each integer $\ell\ge 0$, performing backward push synchronously on all vertices in $\candb^{(\ell)}$ ensures that 
\begin{align*}
\rmaxb^{(\ell)}=\max_{v\in \big(V\setminus \frozenb^{(\ell)}\big)}\rb^{(\ell)}(u)\le (1-\alpha/2)^\ell, \quad \text{ and } \quad \rmaxb^{(\ell+1)}\le (1-\alpha/2)\rmaxb^{(\ell)}. 
\end{align*} 
\end{lemma}

\begin{proof}
We prove the lemma by induction on $\ell$. For $\ell=0$, $\frozenb^{(0)}$ is an empty set, so 
\begin{align*}
\rmaxb^{(0)}=\max_{v\in \big(V\setminus \frozenb^{(0)}\big)}\rb^{(0)}(v)=\rb^{(0)}(t)=1 = (1-\alpha/2)^0. 
\end{align*}
We now assume that, for a fixed $\ell$, the following holds:
\begin{align*}
\rmaxb^{(\ell)}=\max_{v\in V\setminus \frozenb^{(\ell)}}\rb^{(\ell)}(v)\le (1-\alpha/2)^{\ell}.
\end{align*}
For each $u\in V$, if it has not yet been activated, then $\rb^{(\ell)}(u)\le \alpha \rmaxb^{(\ell)}/2$, and thus
\begin{align*}
\rb^{(\ell+1)}(u)
\le \rb^{(\ell)}(u)+\sum_{v\in \candb^{(\ell)}}\frac{(1-\alpha)\rb^{(\ell)}(v)}{\dout(u)} \le \alpha \rmaxb^{(\ell)}/2+(1-\alpha)\rmaxb^{(\ell)}
=(1-\alpha/2)\rmaxb^{(\ell)}.
\end{align*}
Otherwise, $u$ must have been pushed in iteration $\ell$, so $\rb^{(\ell)}(u)=0$, and
\begin{align*}
\rb^{(\ell+1)}(u)=\sum_{v\in \candb^{(\ell)}}\frac{(1-\alpha)\rb^{(\ell)}(v)}{\dout(u)}\le (1-\alpha)\rmaxb^{(\ell)}. 
\end{align*}
Recall that we assume $\rmaxb^{(\ell)}\le (1-\alpha/2)^{\ell}$. We thus have $\rb^{(\ell+1)}(u)\le (1-\alpha/2)^{\ell+1}$ for each $u\in V$, completing the inductive step.  
\end{proof}

\begin{lemma}\label{lem:freeze_b}
Activating each vertex $v$ for the first time when $\rb(v)\ge \alpha\rmaxb/2$ and freezing it after it has remained activated for more than $\ifrozenb = \lceil \log_{1-\alpha/2}\frac{\rela \alpha^2}{32 \budgetb} \rceil$ rounds ensures that 
\begin{align*}
\sum_{v\in F_b}\pi(s,v)\rb(v)\le\rela\pi(s,t)/8    
\end{align*}
at the end of \Cref{alg:pushbackward_frozen}. 
\end{lemma}

\begin{proof}
For each $v\in \frozenb$, suppose that we activate $v$ in iteration $\ell_v$, so $\rb^{(\ell_v)}(v)\ge \alpha \rmaxb^{(\ell_v)}/2$, and
\begin{align*}
\pi(s,v)\rb^{(\ell_v)}(v)\le \pi(s,t). 
\end{align*}
This further implies
\begin{align}\label{eqn:upper_pi_back}
\pi(s,v)\le \frac{\pi(s,t)}{ \rb^{(\ell_v)}(v)} \le \frac{2 \pi(s,t)}{\alpha \rmaxb^{(\ell_v)}}. 
\end{align}

Let $\rb(v)$ denote its value at the end of \Cref{alg:pushbackward_frozen}. Since $v$ is activated in iteration $\ell_v$, it is frozen at the end of iteration $(\ell_v+\ifrozenb-1)$.
Afterwards, it receive residue mass from its non-frozen out-neighbors, but never push these residue back. Therefore
\begin{equation*}
\begin{aligned}
\rb(v)
&\le \rb^{(\ell_v+\ifrozenb)}(v)+\sum_{\ell=\ell_v+\ifrozenb}^{\infty}\sum_{w\in \{\Nout(v) \bigcap \candb^{(\ell)}\}}\frac{(1-\alpha)\rb^{(\ell)}(w)}{\dout(v)}\\
&\le \rmaxb^{(\ell_v+\ifrozenb)}+\sum_{\ell=\ell_v+\ifrozenb}^{\infty}\sum_{w\in \{\Nout(v)\bigcap \candb^{(\ell)}\}}\frac{(1-\alpha)\rmaxb^{(\ell)}}{\dout(v)}. 
\end{aligned}
\end{equation*}
Recall from~\Cref{lem:decay_backward} that $\rmaxb^{(\ell+1)}\le (1-\alpha/2)\rmaxb^{(\ell)}$ for each $\ell$. We then have 
\begin{equation}\label{eqn:upper_rb}
\begin{aligned}
\rb(v)
&\le (1-\alpha/2)^{\ifrozenb}\rmaxb^{(\ell_v)}+\sum_{\ell=\ell_v+\ifrozenb}^{\infty} (1-\alpha)\rmaxb^{(\ell)}
\le (1-\alpha/2)^{\ifrozenb}\rmaxb^{(\ell_v)}+\sum_{\ell=\ifrozenb}^{\infty} (1-\alpha)(1-\alpha/2)^{\ell}\rmaxb^{(\ell_v)}\\
&=\rmaxb^{(\ell_v)}(1-\alpha/2)^{\ifrozenb} (1+2(1-\alpha)/\alpha)\le \rmaxb^{(\ell_v)}(1-\alpha/2)^{\ifrozenb} (2/\alpha). 
\end{aligned}
\end{equation}
Combining \Cref{eqn:upper_pi_back} and \Cref{eqn:upper_rb} gives
\begin{align*}
\pi(s,v)\rb(v)\le \frac{2\pi(s,t)}{\alpha \rmaxb^{(\ell_v)}}\cdot \frac{2 \rmaxb^{(\ell_v)} (1-\alpha/2)^{\ifrozenb}}{\alpha}=\frac{4 (1-\alpha/2)^{\ifrozenb} \pi(s,t) }{\alpha^2}. 
\end{align*}
This further implies: 
\begin{align*}
\sum_{v\in \frozenb} \pi(s,v)\rb(v)\le |\frozenb| \frac{4 (1-\alpha/2)^{\ifrozenb} \pi(s,t) }{\alpha^2}. 
\end{align*}
Recall that the total running time of backward push never exceed the required budget  $\budgetb$. 
Each vertex must be pushed $\ifrozenb$ time before being frozen, and each push incurs at least unit cost. The number of frozen vertices therefore satisfies $|\frozenb|\le \budgetb$. As a result, setting $\ifrozenb = \lceil \log_{1-\alpha/2}\frac{\rela \alpha^2}{32 \budgetb} \rceil$ ensures that $\frac{4 |\frozenb| (1-\alpha/2)^{\ifrozenb}}{\alpha^2} \le \frac{4 \budgetb (1-\alpha/2)^{\ifrozenb}}{\alpha^2}\le \frac{\rela}{8}$. 
\end{proof}

\section{Our $\TriPPR$ Algorithm}\label{sec:upperbound}

In this section, we present our $\TriPPR$ algorithm, and show that its expected running time is $O(\Tstar\log n)$, where 
\begin{equation}\label{eqn:Tstar}
\begin{aligned}
\Tstar=\max_{r\in(0,1]}\min\{\TA,\TB\}, & \quad \text{ with }\\
\TA=\sum_{u\in V,\pi(s,u)>0}\min\left\{\dout(u),\frac{\pi(s,u)r}{\pi(s,t)}\right\}, \quad \text{ and }\quad &\TB=\sum_{v\in V,\pi(v,t)\ge r}(1+\din(v)).
\end{aligned}
\end{equation}
For simplicity, we assume $\delta=0$ here. In \Cref{sec:reduction-delta}, we will show how to extend it to a general $\delta$.
We recall that $\din(v)$ and $\dout(v)$ denote the in- and out-degree of vertex $v$, respectively, and  recall that for simplicity we assume $\dout(v)>0$ for each $v\in V$ throughout the paper.

\begin{theorem}\label{thm:upper_bound}
Without prior knowledge of $n$, the estimate $\epi(s,t)$ for $\pi(s,t)$ obtained by the $\TriPPR$ algorithm satisfies
\begin{align*}
\Pr\bigl\{|\epi(s,t)-\pi(s,t)|\le \rela\pi(s,t)\bigr\}\ge 3/4. 
\end{align*}
The expected running time of $\TriPPR$ is $\widetilde O(\Tstar)$.
\end{theorem}

\subsection{Three-Component Structure}
Our $\TriPPR$ algorithm consists of three components. We first run $\frozenpushforward$ from $s$, as described in~\Cref{alg:pushforward_frozen}, followed by $\frozenpushback$ from $t$, as described in~\Cref{alg:pushbackward_frozen}, and finally perform Monte Carlo sampling in the graph. 

Specifically, in the Monte Carlo sampling phase, we generate independent discounted random walks from each vertex $u$ that has not been frozen in the forward push process. Since we only need to control the variance, pairwise independence among these walks suffices.
Let $\nr$ denote the total number of discounted random walks generated. The number of random walks generated from each vertex $u$ is set to $\nr(u)=\lceil \nr\cdot\rf(u)/\rsumf \rceil$, where  $\rf(u)$ denotes the forward residue of $u$, and $\rsumf=\sum_{u\in V\setminus\frozenf}\rf(u)$ is the total forward residue over all non-frozen vertices at the end of the forward-push process.

Let $\indicator_k(u,v)$ denote the indicator variable that equals $1$ if the $k$-th random walk generated from $u$ terminates at vertex $v$, and $0$ otherwise. Therefore,  $\indicator_k(u,v)$ is an unbiased estimate for $\pi(u,v)$ by definition. 
The final estimate $X$ of $\pi(s,t)$ produced by $\TriPPR$ is obtained by combining the quantities computed by the three components, as follows.
\begin{align}\label{eqn:def_X}
X
=\pb(s)
+\sum_{v\notin\frozenb}\pf(v)\rb(v)
+\sum_{u\notin\frozenf}
\sum_{k=1}^{\nr(u)}
\sum_{v\notin\frozenb}\rf(u)\rb(v)\indicator_k(u,v)/\nr(u), 
\end{align}
where $\rf(u), \pf(u), \rb(u), \pb(u), \frozenf, \frozenb$ denote their respective values at the end of the forward- and backward-push processes. The pseudocode for the Monte Carlo sampling component is provided in~\Cref{alg:mc}.

\begin{algorithm}[h]
\DontPrintSemicolon
\small
\caption{$\threephase_R(\nr)$}
\label{alg:mc}
\KwIn{the number of random walks $\nr$, a random seed $R$ for the $\nr$ random walks which should be 2-independent of each other}
\KwOut{estimate $X$ of $\pi(s,t)$}
$X\gets \pb(s)+\sum_{v\notin \frozenb}\pf(v)\rb(v)$\;
\textbf{if} $\rsumf \rmaxb=0$ \textbf{then} \Return{$X$}\;
\ForEach{$u \in (V\setminus \frozenf)$ \textup{with} $\rf(u)>0$}{
 $\nr(u)\gets\lceil\nr \cdot \rf(u)/\rsumf\rceil$\;
 \For{$k=1,\ldots,\nr(u)$}{
  Generate a discounted random walk from $u$; let $v$ be its endpoint\;
  \If{$v\notin \frozenb$}{$X\gets X+\rf(u)\rb(v)/\nr(u)$\;}
 }
}
\Return{$X$}\;
\end{algorithm}

\begin{lemma}\label{lem:sample}\label{lem:variance}
The variance of $X$ derived in~\Cref{alg:mc} satisfies:  $\Var[X]\le \rsumf\rmaxb\pi(s,t)/\nr$. 
\end{lemma}
\begin{proof}

Both the forward and backward processes are deterministic, so the randomness in $X$, as defined in~\Cref{eqn:def_X}, arises solely from Monte Carlo sampling. Therefore, 
\begin{equation*}
\begin{aligned}
\Var[X]
&=\Var\left[\sum_{u\notin\frozenf}
\sum_{k=1}^{\nr(u)}
\sum_{v\notin\frozenb}\rf(u)\rb(v)\indicator_k(u,v)/\nr(u)\right]
\end{aligned}
\end{equation*}
All random walks are generated independently. For any fixed $u$ and $k$, the random variables $\indicator_k(u,v)$, indexed by $v\in V$, are negatively correlated, and $\Var[\indicator_k(u,v)]\le \pi(u,v)$. Therefore,
\begin{align*}
    \Var[X]\le \sum_{u\notin \frozenf}\sum_{v\notin \frozenb}\rf^2(u)\rb^2(v)\pi(u,v)/\nr(u). 
\end{align*}
Since $\rb(v)\le \rmaxb$ and $\nr(u)=\lceil \nr \cdot \rf(u)/\rsumf \rceil$, we further have
\begin{align}\label{eqn:var_upper_mid}
 \Var[X]\le \sum_{u\notin \frozenf}\sum_{v\notin \frozenb}\rf^2(u)\rb(v) \rmaxb\pi(u,v)\Big/ \left\lceil\frac{\nr \rf(u)}{\rsumf}\right\rceil \le \frac{\rsumf \rmaxb}{\nr} \sum_{u\notin \frozenf}\sum_{v\notin \frozenb}\rf(u)\rb(v) \pi(u,v). 
\end{align}

Recall from the backward push invariant~\eqref{eq:back-inv} that
\begin{align*}
\pi(s,t)=\pb(s)+\sum_{v\in V}\pi(s,v)\rb(v)=\pb(s)+\sum_{v\in \frozenb}\pi(s,v)\rb(v)+\sum_{v \notin \frozenb}\pi(s,v)\rb(v).
\end{align*}
Substituting $\pi(s,v)$ in the last term using the forward-push invariant~\eqref{eq:forward-inv}, which states that
$\pi(s,v)=\pf(v)+\sum_{u\in V}\rf(u)\pi(u,v)$, yields
\begin{equation}\label{eqn:3-invariant}
\begin{aligned}
&\pi(s,t)
=\pb(s)+\sum_{v\in \frozenb}\pi(s,v)\rb(v)+\sum_{v\notin \frozenb}\pf(v)\rb(v)+\sum_{v\notin \frozenb} \sum_{u\in V}\rf(u)\pi(u,v) \rb(v). 
\end{aligned}    
\end{equation}
This implies
\[
\sum_{u\notin\frozenf}\sum_{v\notin\frozenb}
\rf(u)\rb(v)\pi(u,v)\le \pi(s,t).
\]
Substituting this into \Cref{eqn:var_upper_mid} gives
$\Var[X]\le \frac{\rsumf\rmaxb}{\nr}\pi(s,t)$, 
as claimed.
\end{proof}

\subsection{Adaptive Rounds}
We have not specified how to set the time budgets $\budgetf$ and $\budgetb$ for the $\frozenpushforward(s, \budgetf)$ and $\frozenpushback(t, \budgetb)$ components, respectively, or the number of random walks $\nr$ for the Monte Carlo sampling component. The ideal choices of these parameters depend on the value of $\pi(s,t)$, which is unknown in advance. We therefore determine them adaptively.

We allocate the same running-time budget $\budget$ to each of the three components. 
Initially, we set the budget $\budget$ to $2$ and adaptively double it in subsequent rounds. 
In each round, we construct three independent estimators $X_1$, $X_2$, and $X_3$. When 
\begin{align}\label{eqn:stopping}
\max\{X_1, X_2\} \ge \frac{1024 \rsumf \rmaxb}{\rela^2 \budget}, 
\end{align}
we terminate the algorithm and return $X_3$ as the final estimate of $\pi(s,t)$. Here we recall that $\rsumf=\sum_{u\in V\setminus\frozenf}\rf(u)$ and $\rmaxb=\max_{v\in V\setminus\frozenb}\rb(v)$, where $\rf(u)$, $\pf(u)$, $\rb(u)$, $\pb(u)$, $\frozenf$, and $\frozenb$ denote their respective values at the end of $\frozenpushforward$ and $\frozenpushback$ in the current round. Since the forward- and backward-push processes are deterministic, in each round, we only need to perform the Monte Carlo sampling independently three times to obtain $X_1$, $X_2$, and $X_3$. The pseudocode of $\TriPPR$ is presented in~\Cref{alg:triPPR}.

\begin{algorithm}[h]
\DontPrintSemicolon
\small
\caption{\TriPPR$(s,t,\rela)$}
\label{alg:triPPR}
\KwIn{source vertex $s$, target vertex $t$, relative error $\rela$}
\KwOut{estimate $\epi(s,t)$ of $\pi(s,t)$}
\textbf{declare global} $\rf()$, $\pf()$, $\cntf()$, $\frozenf$, $\rsumf$, $\rb()$, $\pb()$, $\cntb()$, $\frozenb$, $\rmaxb$\;
$\textup{\texttt{ForwardPush-Init}}(s)$\;
$\textup{\texttt{BackwardPush-Init}}(t)$\;
\For{$i=1,2,\ldots$}{
$\frozenpushforward(s,2^{i-1}, \lceil \log_{1-\alpha/2} \frac{\rela}{2^{3i+4}} \rceil)$\;
$\frozenpushback(t,2^{i-1}, \lceil \log_{1-\alpha/2}\frac{\rela \alpha^2}{2^{2i+5}} \rceil)$\;
\For{$\gamma=1,2,3$ \textup{independently}}{
$X_\gamma \gets \threephase_{R_\gamma}(2^i)$\;
}
$\tau \gets 1024\rsumf \rmaxb/(\rela^2 2^i)$\;
\textbf{if} $\max\{X_1,X_2\}\ge\tau$ \textbf{then return} $X_3$
}
\end{algorithm}

To ensure that both $\rsumf$ and $\rmaxb$ are non-increasing over time, it is important that we do not restart the push operations from scratch in each round, but instead continue from where we left off in previous rounds.
In particular, the sets of frozen vertices $\frozenf$ and $\frozenb$ are never reset, so vertices frozen in earlier rounds remain frozen in all subsequent rounds. Similarly, the push counters $\cntf(u)$ and $\cntb(u)$ are never reset to zero and can only increase over time. Note, however, that the freezing thresholds increase across rounds, allowing more push operations per vertex in later rounds.

A similar idea has been mentioned in~\Cref{sec:worst-case} to achieve the $O(m\log n)$ worst-case running time for estimating $\pi(s,t)$. Here, we adopt analogous freezing thresholds, setting $\ifrozenf^{(i)}=\lceil \log_{1-\alpha/2} \frac{\rela}{2^{3i+4}} \rceil$ and $\ifrozenb^{(i)}=\lceil \log_{1-\alpha/2}\frac{\rela \alpha^2}{2^{2i+5}} \rceil$ accordingly for each round $i$.




\subsubsection{Setting Up the Round Analysis}

For each round $i=0,1,2\ldots$ and each $\gamma\in \{1,2,3\}$, we define $X^{(i)}_{\gamma}$, $\rsumf^{(i)}$, $\rmaxb^{(i)}$, and $\tau^{(i)}$ as the values of $X_\gamma$, $\rsumf$, $\rmaxb$, and $\tau$ in the beginning of round $i$, respectively. 
Additionally, we define $i^*$ to be the smallest integer $i$ such that 
\begin{align}\label{eqn:istar}
\tau^{(i)}=\frac{1024 \rsumf^{(i)}\rmaxb^{(i)}}{\rela^2 2^i}\le 32 \pi(s,t). 
\end{align}
Since $\rsumf^{(i)}$ and $\rmaxb^{(i)}$ are nonincreasing in $i$, so is $\tau^{(i)}$.

We note that $\pi(s,t)\ge 1/n^{O(n)}$, and by~\Cref{thm:worst-case-time}, $\pb(s)\ge (1-\rela)\pi(s,t)$ when the budget $\budget=\omega(m \log{n})$. Therefore, the stopping criterion in \Cref{eqn:stopping} is guaranteed to be satisfied when $\budget=n^{\omega(n)}$. Since the budget doubles in each round, the algorithm is therefore guaranteed to terminate after at most $i^T=O(n\log n)$ adaptive rounds.

To analyze the approximation error and expected runtime of our $\TriPPR$ algorithm, we consider a variant of our algorithm that executes through the entire $i^T$ rounds, regardless of whether the stopping condition is satisfied before reaching round $i^T$. 

With this variant, we will reach any round $i\leq i^T$ regardless of what happens in previous rounds. We also have

\begin{lemma}
For each $i\in [1, i^T]$ and $\gamma \in \{1,2,3\}$, $(1-\rela/4)\pi(s,t)\le \E[X_{\gamma}^{(i)}]\le \pi(s,t)$. 
\end{lemma}

\begin{proof}
Recall that 
\begin{align*}
X_{\gamma}^{(i)}
=\pb^{(i)}(s)
+\sum_{v\notin\frozenb^{(i)}}\pf^{(i)}(v)\rb^{(i)}(v)
+\sum_{u\notin\frozenf^{(i)}}
\sum_{k=1}^{\nr(u)}
\sum_{v\notin\frozenb^{(i)}}\rf^{(i)}(u)\rb^{(i)}(v)\indicator^{(i)}_k(u,v)/\nr(u), 
\end{align*}
where the superscript $(i)$ denotes the corresponding value in the beginning of round $i$. 

Since $\E\left[\indicator^{(i)}_k(u,v)\right]=\pi(u,v)$ for each $i, k, u, v$, we have
\begin{align*}
\E[X^{(i)}_\gamma]=\pb^{(i)}(s)+\sum_{v\notin \frozenb^{(i)}}\pf^{(i)}(v)\rb^{(i)}(v)+\sum_{u\notin \frozenf^{(i)}}\sum_{v\notin \frozenb^{(i)}}\rf^{(i)}(u)\rb^{(i)}(v)\pi(u,v). 
\end{align*}

We therefore have $\E[X^{(i)}_\gamma]\le \pi(s,t)$. Moreover, 
\begin{equation*}
\begin{aligned}
\pi(s,t)-\E[X^{(i)}_\gamma]
&=\sum_{v\in \frozenb^{(i)}}\pi(s,v)\rb^{(i)}(v)+\sum_{v\notin \frozenb^{(i)}} \sum_{u\in \frozenf^{(i)}}\rf^{(i)}(u)\pi(u,v) \rb^{(i)}(v)\\
&\le \sum_{v\in \frozenb^{(i)}}\pi(s,v)\rb^{(i)}(v)+\sum_{u\in \frozenf^{(i)}}\rf^{(i)}(u)\sum_{v \in V} \pi(u,v) \rb^{(i)}(v). 
\end{aligned}
\end{equation*}
Since $\sum_{v \in V} \pi(u,v) \rb^{(i)}(v)\le \pi(u,t)$ by the backward push invariant~\eqref{eq:back-inv}, we then have
\begin{align*}
\pi(s,t)-\E[X^{(i)}_\gamma]\le \sum_{v\in \frozenb^{(i)}}\pi(s,v)\rb^{(i)}(v)+\sum_{u\in \frozenf^{(i)}}\rf^{(i)}(u)\pi(u,t). 
\end{align*}
Recall from~\Cref{lem:final_approx_worst} and~\Cref{lem:freeze_b} that
\begin{align*}
\sum_{v\in \frozenb^{(i)}}\pi(s,v)\rb^{(i)}(v)\le \rela\pi(s,t)/8, \text{ and } \sum_{u\in \frozenf^{(i)}}\rf^{(i)}(u)\pi(u,t)\le \rela\pi(s,t)/8. 
\end{align*}
We therefore have 
$\pi(s,t)-\E[X^{(i)}_\gamma]\le \rela\pi(s,t)/4$, as claimed. 
\end{proof}

Therefore, by~\Cref{lem:variance}, for each $i\in [0, i^T]$ and $\gamma \in \{1, 2, 3\}$, we have
\begin{align}\label{eqn:var2}
\E[X^{(i)}_\gamma] \le\pi(s,t), \text{ and } \Var[X^{(i)}_\gamma]\le  \frac{\rsumf^{(i)}\rmaxb^{(i)}\pi(s,t)}{2^i}. 
\end{align}

\subsubsection{Goodness}
Suppose that the stopping criterion in~\Cref{eqn:stopping} is satisfied for the first time in round $i^R$. So \Cref{alg:triPPR} outputs $X_3^{(i^R)}$ as its final estimate of $\pi(s,t)$. We now prove that $X_3^{(i^R)}$ is a good estimate of $\pi(s,t)$. 
\begin{lemma}\label{lem:goodness}
$\Pr\left\{|X_3^{(i^R)}-\pi(s,t)|\ge \rela \pi(s,t)\right\}\le 1/4$.  
\end{lemma}

\begin{proof}
Recall from \Cref{lem:variance} that 
\begin{align*}
(1-\rela/4)\pi(s,t)\le \E[X_{\gamma}^{(i^R)}]\le \pi(s,t), \quad \text{deterministically}.
\end{align*}
So we only need to ensure
\begin{align*}
\Pr\left\{|X_3^{(i^R)}-\E[X_3^{(i^R)}]|\ge (3\rela/4) \pi(s,t)\right\}\le 1/4. 
\end{align*}

Recall that $i^*$ is the smallest integer $i$ such that
\begin{align*}
\tau^{(i)}=\frac{1024 \rsumf^{(i)}\rmaxb^{(i)}}{\rela^2 2^i}\le 32\pi(s,t) \iff \frac{ \rsumf^{(i)}\rmaxb^{(i)}}{\rela^2 2^i}\le \frac{\pi(s,t)}{32}.
\end{align*}
For each $i\ge i^*$, applying Chebyshev’s inequality together with \Cref{eqn:var2} and the above inequality yields
\begin{align}
\Pr\left\{|X_3^{(i)}-\E[X_{\gamma}^{(i)}]|\ge (3\rela/4) \pi(s,t)\right\}
&\le \frac{2\Var[X_3^{(i)}]}{\rela^2 \pi^2(s,t)}
\le \frac{2\rsumf^{(i)}\rmaxb^{(i)}}{\rela^2 2^i \pi(s,t)}
\le \frac{1}{16 \cdot 2^{i-i^*}}.
\end{align}
By the union bound, with probability at least $1-1/8$, the estimate $X_3^{(i)}$ is within a multiplicative factor of $(1\pm\rela)$ of $\pi(s,t)$ simultaneously in all rounds $i\ge i^{*}$.

For each $i<i^*$, Markov's inequality gives
\begin{align}
\Pr\left\{X_{\gamma}^{(i)}\ge \tau^{(i)}\right\}
&\le \frac{\E[X_{\gamma}^{(i)}]}{\tau^{(i)}}
\le \frac{\pi(s,t)}{\frac{1024\rsumf^{(i)}\rmaxb^{(i)}}{\rela^2 2^i}}
\le \frac{1}{32 \cdot 2^{i^*-i-1}}.
\end{align}
So the probability that the stopping criterion of~\Cref{alg:triPPR} is satisfied in round $i$ is at most
\begin{align*}
\Pr\left\{\max\{X_1^{(i)}, X_2^{(i)}\}\ge \tau^{(i)}\right\}\le \frac{1}{16 \cdot 2^{i^*-i-1}}.
\end{align*}
Applying the union bound over all rounds $i<i^{}$, we conclude that the probability of termination before round $i^{}$ is at most $1/8$. Equivalently, $i^R\ge i^*$ with probability at least $1-1/8$.

Combining these two bounds establishes the claimed results.
\end{proof}

\subsubsection{Efficiency}
We now prove that the expected running time of $\TriPPR$ is $O(\Tstar\log n)$. 
\begin{lemma}\label{lem:efficiency}
The expected time of~\Cref{alg:triPPR} is $O(2^{i^*})$. 
\end{lemma}

\begin{proof}
By~\Cref{lem:runtime_freezing}, both $\frozenpushforward$ and $\frozenpushback$ run in $O(\budget)$ time with budget parameter $\budget$. Thus, in round $i$, these two components take $O(2^i)$ time in total. The Monte Carlo sampling phase generates $2^i$ discounted random walks and also takes $O(2^i)$ time. Therefore, if we get to round $i$, the total time cost we spend is $O(2^i)$. As a result, the expected time of \Cref{alg:triPPR} is bounded by 
\begin{align}\label{eqn:timebound_adaptive}
\sum_{i=1}^{i^T}2^i \Pr\{\max\{X^{(i)}_1, X^{(i)}_2\}< \tau^{(i)}\},    
\end{align}
where $i^T$ denotes the largest possible round the algorithm can reach.

To bound the expected running time of~\Cref{alg:triPPR} by $O(2^{i^{*}})$, it suffices to analyze rounds $i>i^{*}+5$, since the total running time of all earlier rounds is $O(2^{i^{*}})$. Recall from~\Cref{eqn:istar} that
\[
\tau^{(i^{*})}
=\frac{1024\rsumf^{(i^{*})}\rmaxb^{(i^{*})}}{\rela^2 2^{i^{*}}}
\le 32\pi(s,t).
\]
Since $\rsumf^{(i)}$ and $\rmaxb^{(i)}$ are nonincreasing in $i$, for every $i>i^{*}+5$ we have
\begin{align*}
\tau^{(i)}
\le \frac{\tau^{(i^{*})}}{2^{i-i^{*}}}
\le \frac{32\pi(s,t)}{2^{i-i^{*}}}
\le \frac{\pi(s,t)}{2}.
\end{align*}
Together with $(1-\rela/4)\pi(s,t)\le \E[X_{\gamma}^{(i)}]\le\pi(s,t)$ from~\Cref{lem:variance}, this gives
\begin{align*}
\Pr{X_{\gamma}^{(i)}<\tau^{(i)}}
&\le \Pr{\pi(s,t)-X_{\gamma}^{(i)}>\pi(s,t)/2}\le \Pr{\E[X_{\gamma}^{(i)}]-X_{\gamma}^{(i)}>\pi(s,t)/4}.
\end{align*}
Applying Chebyshev’s inequality together with the variance bound $\Var[X_\gamma^{(i)}]\le \rsumf \rmaxb \pi(s,t)/2^i$ from~\Cref{eqn:var2}, we obtain
\begin{align*}
\Pr{\E[X_{\gamma}^{(i)}]-X_{\gamma}^{(i)}>\pi(s,t)/4}\le \frac{16\Var[X_{\gamma}^{(i)}]}{\pi^2(s,t)}\le \frac{16 \frac{\rsumf \rmaxb}{2^i} \pi(s,t) }{\pi^2(s,t)} \le \frac{1}{64 \cdot 2^{i-i^*-5}}. 
\end{align*}
By the independence of $X_1^{(i)}$ and $X_2^{(i)}$ for each $i$, we further have 
\begin{align*}
\Pr\{\max\{X_1^{(i)}, X_1^{(2)}\}<\tau^{(i)}\}\le \frac{1}{64^2 \cdot 4^{i-i^*-5}}. 
\end{align*}
Substituting into \Cref{eqn:timebound_adaptive} gives the time cost upper bound as
\begin{align*}
2^{i^*+5}+\sum_{i=i^*+5}^{i^T} \frac{2^i}{64^2 \cdot 4^{i-i^*-5}}=O(2^{i^*}),  
\end{align*}
as claimed. 
\end{proof}

\begin{lemma}\label{lem:optimality_upper}
$2^{i^*}=O(T^* \log{n})$. 
\end{lemma}

\begin{proof}
Recall from~\Cref{eqn:istar} that $i^{*}$ is the smallest integer $i$ such that
\begin{align*}
\tau^{(i)}=\frac{1024 \rsumf^{(i)}\rmaxb^{(i)}}{\rela^2 2^i}\le 32\pi(s,t)
\iff
\rsumf^{(i)}\rmaxb^{(i)}\le \frac{\rela^2 2^i}{32}\pi(s,t).
\end{align*}
Thus, for any round $i<i^{*}$, we have
\begin{equation}\label{eq:failed-round}
\rsumf^{(i)}\rmaxb^{(i)}>
\frac{\rela^2 2^i}{32}\pi(s,t).
\end{equation}
Fix such a round $i$.

\paragraph{Relating $\boldsymbol{\TB}$ to the running time of round $\boldsymbol{i}$.}
Let $\totalactivateb^{(i)}$ denote the set of vertices activated during the backward-push phase of round $i$, namely, during $\frozenpushback(t,2^i)$, and write
\[
D_b^{(i)}=\sum_{v\in\totalactivateb^{(i)}}(1+\din(v)).
\]
Since backward push is performed on each activated vertex at most $\ifrozenb$ times, the total cost of the backward-push operations already performed is at most $\ifrozenb D_b^{(i)}$.

Since $i<i^{*}$, \Cref{eq:failed-round} implies that $\rmaxb^{(i)}>0$. Thus, the backward-push procedure terminates in round $i$ only when the anticipated cost of the next iteration exceeds the remaining budget. This next iteration, which is not executed, would perform at most one backward push on each vertex in a subset of $\totalactivateb^{(i)}$, so its anticipated cost is at most $D_b^{(i)}$. Hence, the cost already incurred plus the anticipated cost of the next iteration exceeds $2^i$, giving
\[
2^i<\ifrozenb D_b^{(i)}+D_b^{(i)}
=(\ifrozenb+1)D_b^{(i)}.
\]
Consequently,
\begin{equation}\label{eqn:din_lower}
D_b^{(i)}=\sum_{v\in\totalactivateb^{(i)}}(1+\din(v))
>\frac{2^i}{\ifrozenb+1}.
\end{equation}

Let $\rmaxb^{(i)}$ denote the value of $\rmaxb$ at the end of round $i$. By~\Cref{lem:freeze_b}, this is its minimum value over all iterations of $\frozenpushback(t,2^i)$. The activation condition $\rb(v)\ge\alpha\rmaxb/2$ in~\Cref{alg:pushbackward_frozen} therefore ensures that every $v\in\totalactivateb^{(i)}$ has backward residue at least $\alpha\rmaxb^{(i)}/2$ upon activation. 
By the backward invariant~\eqref{eq:back-inv}, we have $\pi(v,t)\ge \rb(v)\pi(v,v)$. By definition, $\pi(v,v)\ge\alpha$, since a discounted random walk starting at $v$ terminates immediately with probability $\alpha$. It follows 
\[
\pi(v,t)\ge \alpha^2\rmaxb^{(i)}/2
\qquad\text{for every }v\in\totalactivateb^{(i)}.
\]

Recall from~\Cref{eqn:Tstar} that
\[
T_b(r)=\sum_{v\in V:\,\pi(v,t)\ge r}(1+\din(v)).
\]
Combining this definition with~\Cref{eqn:din_lower}, we obtain
\begin{equation}\label{eq:backward-lower-bound}
T_b\left(\alpha^2\rmaxb^{(i)}/2\right)
\ge \sum_{v\in\totalactivateb^{(i)}}(1+\din(v))
>\frac{2^i}{\ifrozenb+1}.
\end{equation}

\paragraph{Relating $\boldsymbol{\TA}$ to the running time of round $\boldsymbol{i}$.}
In the $\frozenpushforward(s,2^i)$ algorithm as shown in~\Cref{alg:pushforward_frozen}, we may terminate because the
anticipated cost of the next iteration exceeds the remaining budget, or the total
residue of the candidate set is less than $\rsumf^{(i)}/2$.
For the former, 
let $\totalactivatef^{(i)}$ denote the set of vertices activated during the forward-push phase of round $i$, namely, during $\frozenpushforward(s,2^i)$. Then an analogous budget argument gives
\begin{equation}\label{eqn:forward_cost}
\sum_{u\in\totalactivatef^{(i)}}\dout(u)
>\frac{2^i}{\ifrozenf+1}.
\end{equation}

Let $\rsumf^{(i)}$ denote the value of $\rsumf$ at the end of round $i$. By~\Cref{lem:freeze_f}, this is its minimum value over all iterations of $\frozenpushforward(s,2^i)$. The activation condition $\rf(u)\ge\rsumf\dout(u)/2^i$ in~\Cref{alg:pushforward_frozen} therefore ensures that every $u\in\totalactivatef^{(i)}$ has forward residue at least $\rsumf^{(i)}\dout(u)/2^i$ upon activation. Analogously,
by the forward push invariant~\eqref{eq:forward-inv}, we have
\[
\pi(s,u)\ge\frac{\alpha\rsumf^{(i)}\dout(u)}{2^i}
\qquad\text{for every }u\in\totalactivatef^{(i)}. 
\]
Recall from~\Cref{eq:failed-round} that $\rsumf^{(i)}\rmaxb^{(i)}>
\frac{\rela^2 2^i}{32}\pi(s,t)$. It follows
\[
\pi(s,u)
\ge\frac{\alpha\rsumf^{(i)}\dout(u)}{2^i}
\ge\frac{\alpha\rela^2\pi(s,t)\dout(u)}{32\rmaxb^{(i)}}.
\]
This further implies
\[
\frac{\pi(s,u)}{\pi(s,t)}
\cdot\frac{\alpha^2\rmaxb^{(i)}}{2}
\ge\frac{\alpha^3\rela^2}{64}\dout(u).
\]

Recall from~\Cref{eqn:Tstar} that
\[
T_f(r)=\sum_{u\in V:\,\pi(s,u)>0}
\min\left\{\dout(u),\frac{\pi(s,u)r}{\pi(s,t)}\right\}.
\]
Since $\alpha^3\rela^2/64<1$, the preceding bound together with~\Cref{eqn:forward_cost} gives
\begin{equation}\label{eq:forward-budget-lower-bound}
T_f\left(\alpha^2\rmaxb^{(i)}/2\right)
\ge\frac{\alpha^3\rela^2}{64}
\sum_{u\in\totalactivatef^{(i)}}\dout(u)
>\frac{\alpha^3\rela^2 2^i}{64(\ifrozenf+1)}.
\end{equation}

As mentioned above, the forward-push phase may also terminate because the total
residue of vertices in the candidate set $\candf^{(i)}$ is less than $\rsumf^{(i)}/2$. 
That is, when $\sum_{u\in \candf^{(i)}}\rf^{(i)}(u)< \rsumf^{(i)}/2$, 
where we recall that $\rsumf^{(i)}=\sum_{u\in V\setminus \frozenf^{(i)}}\rf^{(i)}(u)$. 
This implies
\begin{align}\label{eqn:rf_u}
\sum_{u\in (V\setminus\frozenf^{(i)})\setminus\candf^{(i)}}\rf^{(i)}(u)> \rsumf^{(i)}/2.     
\end{align}
We also note that $\dout(u)>
2^i\rf^{(i)}(u)/\rsumf^{(i)}$ for every $u\in (V\setminus\frozenf^{(i)})\setminus\candf^{(i)}$, given that the candidate set $\candf^{(i)}$ is defined to contain all vertices $v$ with $\rf^{(i)}(u)/\dout(v)\ge \rsumf^{(i)}/2^i$. Combining this with \Cref{eqn:rf_u} gives
\begin{align*}
\sum_{u\in (V\setminus\frozenf^{(i)})\setminus\candf^{(i)}}\dout(u)>\sum_{u\in (V\setminus\frozenf^{(i)})\setminus\candf^{(i)}} \frac{2^i \rf^{(i)}(u)}{\rsumf^{(i)}}>2^i/2. 
\end{align*}

Additionally, we have
$\pi(s,u)\ge\alpha\rf^{(i)}(u)$ by the forward push invariant~\eqref{eq:forward-inv}. 
We also recall from~\Cref{eq:failed-round} that 
\begin{align*}
\rsumf^{(i)}\rmaxb^{(i)}>
\frac{\rela^2 2^i}{32}\pi(s,t).
\end{align*}
Consequently, 
\[
\frac{\pi(s,u)}{\pi(s,t)}
\cdot\frac{\alpha^2\rmaxb^{(i)}}{2}
\ge
\frac{\alpha^3\rf^{(i)}(u)\rmaxb^{(i)}}{2\pi(s,t)}
\ge
\frac{\alpha^3\rela^2 2^i}{64\rsumf^{(i)}}\rf^{(i)}(u).
\]
Combining this with \Cref{eqn:rf_u} gives
\begin{equation}\label{eq:forward-mass-lower-bound}
\begin{aligned}
T_f\left(\alpha^2\rmaxb^{(i)}/2\right)
\ge
\frac{\alpha^3\rela^2 2^i}{64\rsumf^{(i)}}
\sum_{u\in (V\setminus\frozenf^{(i)})\setminus\candf^{(i)}}\rf^{(i)}(u)>
\frac{\alpha^3\rela^2 2^i}{128}
\ge
\frac{\alpha^3\rela^2 2^i}{64(\ifrozenf+1)},
\end{aligned}
\end{equation}
where the last inequality uses $\ifrozenf\ge1$.

\paragraph{Combining $\boldsymbol{\TA}$ and $\boldsymbol{\TB}$.}
Combining~\Cref{eq:backward-lower-bound,eq:forward-budget-lower-bound,eq:forward-mass-lower-bound}, we obtain
\[
\Tstar
\ge\min\left\{
T_f\left(\alpha^2\rmaxb^{(i)}/2\right),
T_b\left(\alpha^2\rmaxb^{(i)}/2\right)
\right\}
>\frac{\alpha^3\rela^2 2^i}
{64\max\{\ifrozenb+1,\ifrozenf+1\}}.
\]
By~\Cref{lem:freeze_b,lem:freeze_f}, both $\ifrozenb$ and $\ifrozenf$ are $O(i+1)$. Thus,
\begin{equation}\label{eq:failed-round-budget}
2^i=O(\Tstar(i+1)).
\end{equation}
Setting $i=i^{*}-1$ gives $2^{i^{*}}=O(\Tstar i^{*})$, which implies $i^{*}=O(\log(2+\Tstar))$. Consequently,
\[
2^{i^{*}}
=O(\Tstar\log(2+\Tstar))
=O(\Tstar\log n),
\]
completing the proof.
\end{proof}

Combining~\Cref{lem:goodness,lem:efficiency,lem:optimality_upper} establishes \Cref{thm:upper_bound}.

\section{Lower Bound When $n$ is Unknown}
\label{sec:lowerbound_withoutn}

In this section, we prove a lower bound of $\Omega(\Tstar)$ for any good algorithm when $n$ is unknown, showing that our $\TriPPR$ algorithm is instance optimal in this case.
For simplicity, we also assume $\delta=0$ here. In \cref{sec:reduction-delta}, we will show how to extend it to a general $\delta$.
We start with notations and some general technical lemmas about random walk probabilities.

\paragraph{Notations.}
For any edge $e=(u,v)$, let $\pi(w,e)=\pi(w,u)/\dout(u)$.

For any vertices $u,v$, let $H(u,v)$ denote the probability that an $\alpha$-discounted random walk starting from $u$ visits $v$ before termination.
For any edge $e=(u,v)$, let $H(w,e)$ denote the probability that an $\alpha$-discounted random walk starting from $w$ visits $e$.

For any set of edges $F$, let $\pi(u,F)=\sum_{e\in F}\pi(u,e)$.
Let $\vpi(u,F,v)$ be the probability that an $\alpha$-discounted random walk starting from $u$ visits at least one edge in $F$ and eventually terminates at $v$. Conversely, $\vpi(u,\overline{F},v)$ denotes the probability that the walk terminates at $v$ without visiting any edge in $F$.

Let $V_s=\{u\in V \mid \pi(s,u)>0\}$.
Recall that \[\Tstar=\max_{r\in(0,1]}\min\{\TA,\TB\},\]
where 
\begin{align*}
    \TA=\sum_{u\in V_s}\min\left\{\dout(u),\frac{\pi(s,u)r}{\pi(s,t)}\right\} \quad \text{ and }\quad \TB=\sum_{v\in V,\pi(v,t)\ge r}(1+\din(v)).
\end{align*}

\begin{lemma}
\label{lem:pi-uvw}
    For any vertices $u,v,w$, $\pi(u,v)\ge\pi(u,w)\pi(w,v)$.
\end{lemma}

\begin{proof}
    By definition, we have
    \begin{align*}
        \pi(u,w)=H(u,w)\pi(w,w) \quad\text{and}\quad
        \pi(u,v)\ge H(u,w)\pi(w,v).
    \end{align*}
    Then 
    \begin{align*}
        \pi(u,v)\ge \frac{\pi(u,w)\pi(w,v)}{\pi(w,w)} \ge \pi(u,w)\pi(w,v).
    \end{align*}
\end{proof}

\begin{lemma}
\label{lem:E'}
    For any vertices $u$ and $v$ and any set of edges $E'$,
    \begin{align}\label{eqn:E'}
        \max\left\{\pi(u, \notedgesub, v), \pi(\mu(\edgesub,v), \notedgesub, v)\right\}=\Omega(\pi(u,v)), 
    \end{align}
    where $\mu(\edgesub,v)=\arg \max_{w: \exists u, (u,w)\in \edgesub} \pi(w, \notedgesub, v)$. 
\end{lemma}

\begin{proof}
    By definition, we have 
    \begin{align*}
    \pi(u,v)=\pi(u, \edgesub, v)+\pi(u, \notedgesub, v). 
    \end{align*}
    If $\pi(u, \notedgesub, v)\ge \pi(u,v)/2$, then~\Cref{eqn:E'} follows. So in the following we only consider the case where $\pi(u, \edgesub, v)\ge \pi(u,v)/2$. 
    
    Let $\tp^{(i)}(u, v)$ denote the probability that a standard (non-terminating) random walk starting at $u$ visits $v$ at its $i$-th step, that is, $\vpi(u, v)=\sum_{i=0}^{\infty}\alpha (1-\alpha)^i \vpi^{(i)}(u,v)$.
    Then we have
    \begin{equation*}
        \begin{aligned}
            \pi(u, \edgesub, v) 
            &\le \sum_{w: \exists u, (u,w)\in \edgesub}\sum_{\ell=0}^\infty (1-\alpha)^{\ell}\tp^{(\ell)}(u, w)\pi(w, \notedgesub, v)\\
            &\le \sum_{w: \exists u, (u,w)\in \edgesub}\sum_{\ell=0}^\infty (1-\alpha)^{\ell}\tp^{(\ell)}(u, w)\pi(\mu(\edgesub,v), \notedgesub, v)
        \end{aligned}
    \end{equation*}
    Since $\sum_{w: \exists u, (u,w)\in \edgesub}\tp^{(\ell)}(u, w)\le \sum_{w\in V}\tp^{(\ell)}(u, w)=1$, this implies
    \begin{align*}
        \pi(u, \edgesub, v) \le \sum_{\ell=0}^{\infty}(1-\alpha)^\ell \pi(\mu(\edgesub,v), \notedgesub, v)=\pi(\mu(\edgesub,v), \notedgesub, v)/\alpha. 
    \end{align*}
    Recall that $\pi(u, \edgesub, v)\ge \pi(u,v)/2$, so \Cref{eqn:E'} follows. 
\end{proof}

The next lemma has been proven by \cite{thorup2025instance}.

\begin{lemma}[\cite{thorup2025instance} Lemma 3.2]
\label{lem:subdivide}
    For any edge $(u,v)$, if we replace it with edges $(u,w)$ and $(w,v)$ where $w$ is a new vertex, then for any vertices $u',v'$, $\pi(u',v')$ can only be decreased by a factor of at most $(2-\alpha)/(1-\alpha)$.
\end{lemma}

\begin{theorem}
\label{thm:not-knowing-n}
    For any graph $G$, any good algorithm must spend $\Omega(\Tstar)$ time when $n$ is unknown.
\end{theorem}

\begin{proof}
    Consider any graph $G$, any good algorithm $A$ and any $r\in(0,1]$.
    For any edge $(u,v)$, let $\pr(u,v)$ denote the probability that $A$ visits $(u,v)$. For any vertex $v$, let $\pr(v)$ denote the probability that $A$ queries the degree of $v$. Let 
    \begin{align*}
    \qr(v)=\pr(v)+\sum_{u\in\Nin}\pr(u,v)+\sum_{w\in\Nout}\pr(v,w). 
    \end{align*}
    For any set of edges $F$, let $\pr(F)=\sum_{e\in F}\pr(e)$.
    For any set of vertices $U$, let $\qr(U)=\sum_{v\in U}\qr(v)$.
    Note that for any vertex $v$, the probability that $A$ visits $v$ is at most $\qr(v)$, and $A$ spends $\Omega(\qr(V))$ time in total.

    Let $C$ be some large enough constant to be determined later. Suppose for contradiction that 
    \begin{align*}
        \qr(V) < \min\{\TA,\TB\}/C.
    \end{align*}
    We will construct a graph $G'$ from $G$ such that $\pi_{G'}(s,t) \gg\pi_G(s,t)$ and 
    \begin{equation}
    \label{eqn:total-prob}
        \pr(E')+\qr(V') \ll 1,
    \end{equation}
    where $E'$ is the set of edges deleted from $G$ and $V'$ is the set of vertices in $G$ whose degrees are changed. Then $A$ cannot distinguish $G$ and $G'$ with constant probability, which is a contradiction.

    We first have
    \begin{equation}
    \label{eqn:r}
        r>C\pi(s,t)/10,
    \end{equation}
    since otherwise $\qr(V) < \TA/C \le 0.1$, which means that $A$ does not query anything with probability at least $0.9$.

    Since $\qr(V) < \TB/C$, there is a vertex $y$ such that $\pi(y,t)\ge r$ and $\qr(y) < (1+\din(y))/C$.
    By \cref{lem:pi-uvw}, we have
    \begin{equation}
    \label{eqn:pisy}
        \pi(s,y)\le \pi(s,t)/r.
    \end{equation}

    \begin{lemma}
    \label{lem:U}
        At least one of the following is true:
        \begin{itemize}
            \item There is a vertex $u\neq y$ such that $\pi(s,u)/\dout(u)>(C/50)\pi(s,t)/r$ and $\qr(u)<\dout(u)/20$;
            \item There is a vertex $u\neq y$ such that $\pi(s,u)/\dout(u)\le (C/50)\pi(s,t)/r$, $\pi(s,u)>(C/50)\pi(s,t)/r$ and $\qr(u)<(50/C)\pi(s,u)r/20\pi(s,t)$;
            \item There is a set of vertices $U\not\owns y$ such that $\pi(s,u)\le (C/50)\pi(s,t)/r$ for all $u\in U$, $\pi(s,U)>(C/50)\pi(s,t)/r$ and $\qr(U)<1/20$.
        \end{itemize} 
    \end{lemma}

    \begin{proof}
        Let $V_0=\{u\in V_s\setminus\{y\} \mid \frac{\pi(s,u)}{\dout(u)} > \frac{C\pi(s,t)}{50r}\}$, $V_1=\{u\in V_s\setminus(V_0\cup\{y\}) \mid \pi(s,u) > \frac{C\pi(s,t)}{50r}\}$ and $V_2=V_s\setminus(V_0\cup V_1\cup\{y\})$. Then
        \[\qr(V) < \left(\sum_{u\in V_0}\dout(u)+\frac{\pi(s,V_1)r}{\pi(s,t)}+\frac{\pi(s,V_2)r}{\pi(s,t)} + \frac{\pi(s,y)r}{\pi(s,t)}\right)/C.\]
        If there is a vertex $u\in V_0$ such that $\qr(u)<\dout(u)/20$, or there is a vertex $u\in V_1$ such that $\qr(u)<(50/C)\pi(s,u)r/20\pi(s,t)$, then we are done.
        Otherwise, we have
        \[\qr(V_2) < \left((1-C/20)|V_0|-1.5(C/50)|V_1|+\frac{\pi(s,V_2)r}{\pi(s,t)}+1\right)/C.\]
        
        We claim that $\pi(s,V_2)>(C/50)\pi(s,t)/r$. When $V_0\cup V_1\neq\emptyset$, this is true for large enough $C$ because
        \[\frac{\pi(s,V_2)r}{\pi(s,t)}+1-(C/20-1)|V_0|-1.5(C/50)|V_1|>0.\]
        When $V_0=V_1=\emptyset$, we have $V_2=V\setminus\{y\}$, so
        \[\pi(s,V_2)=1-\pi(s,y)>(C/10-1)\pi(s,t)/r\]
        by \cref{eqn:r,eqn:pisy}.

        We sort the vertices in $V_2$ in the decreasing order of $\pi(s,u)/\qr(u)$, and let $U$ be the shortest prefix such that $\pi(s,U)>(C/50)\pi(s,t)/r$. By definition, we have \[\pi(s,U)\le(C/25)\pi(s,t)/r\] and \[\frac{\pi(s,U)}{\qr(U)}\ge\frac{\pi(s,V_2)}{\qr(V_2)}.\]
        So we have
        \[\qr(U) \le \frac{\qr(V_2)}{\pi(s,V_2)}\pi(s,U) < \frac{1}{C}\left(\frac{r}{\pi(s,t)}+\frac{1}{\pi(s,V_2)}\right)\pi(s,U) < \frac{1+50/C}{C}\frac{r}{\pi(s,t)}\pi(s,U) \le \frac{1}{20}. \]
    \end{proof}

    \begin{lemma}
    \label{lem:F}
        There is a set of edges $F$ such that $\pi(s,F)>(C/50)\pi(s,t)/r$ and $\pr(F)<1/10$.
        Let $U=\{u\in V \mid \exists v\ (u,v)\in F\}$, we also have $y\notin U$.
    \end{lemma}

    \begin{proof}
        Consider the $3$ cases in \cref{lem:U}.
        \begin{itemize}
            \item \textbf{Case 1.} There is a vertex $u\neq y$ such that $\pi(s,u)/\dout(u)>(C/50)\pi(s,t)/r$ and $\qr(u)<\dout(u)/20$. Since $\qr(u)<\dout(u)/20$, there is an edge $(u,v)$ such that $\pr(u,v)<1/20$. Let $F=\{(u,v)\}$ and $U=\{u\}$.
            \item \textbf{Case 2.} There is a vertex $u\neq y$ such that $\pi(s,u)/\dout(u)\le(C/50)\pi(s,t)/r$, $\pi(s,u)>(C/50)\pi(s,t)/r$ and $\qr(u)<(50/C)\pi(s,u)r/20\pi(s,t)$.
            Sort the out-edges of $u$ in the increasing order of $\pr(u,v)$. Let $F$ be the shortest prefix such that $\pi(s,F)>(C/50)\pi(s,t)/r$.
            By definition, we have
            \[\pi(s,F)\le(C/25)\pi(s,t)/r\]
            and
            \[\frac{\pr(F)}{|F|}\le\frac{\qr(u)}{\dout(u)}.\]
            So
            \[\pr(F)\le \frac{Q(u)|F|}{\dout(u)} = \frac{Q(u)\pi(s,F)}{\pi(s,u)} < \frac{1}{10}.\]
            \item \textbf{Case 3.} There is a set of vertices $U\not\owns y$ such that $\pi(s,U)>(C/50)\pi(s,t)/r$ and $\qr(U)<1/20$. Let $F$ be the set of all out-edges of vertices in $U$.
        \end{itemize}
    \end{proof}

    We create a new vertex $w$.
    For every edge $(u,v)\in F$, we create two new vertex $w_{uv}$ and $w'_{uv}$, and replace $(u,v)$ by $(u,w_{uv})$, $(w_{uv},w)$ and $(w'_{uv},v)$.
    Call this new graph $G'$.
    Then we have $\pi_{G'}(s,w)=\Omega(\pi_G(s,t)/r)$ by the following lemma.

    \begin{lemma}
    \label{lem:pi-sw-G'}
        $\pi_{G'}(s,w)=\Omega(\pi_G(s,F))$.
    \end{lemma}

    \begin{proof}
        The probability that a random walk in $G$ stops at $F$ (which means the last edge of the random walk is in $F$) is $\Omega(\pi_G(s,F))$.
        Let $F'=\{(u,w_{uv}) \mid (u,v)\in F\}$.
        Since we only deleted $F$ and the out-degrees of $U$ are unchanged, a random walk in $G'$ will behave as if in $G$ until it visits $F'$.
        So, the probability that a random walk in $G'$ visits $F'$ is at least the probability that a random walk in $G$ visits $F$, which is at least $\Omega(\pi_G(s,F))$.
        Given that a random walk in $G'$ visits $F'$, the conditional probability that it stops at $w$ is $(1-\alpha)\alpha=\Omega(1)$, so the lemma follows.
    \end{proof}

    Since $\pi(y,t)=\Omega(r)$, by \cref{lem:E'}, there exists $y'\in\{y,\mu(F,t)\}$ such that $\pi_G(y',\overline{F},t)=\Omega(r)$. Since we only deleted $F$ and the out-degrees of $U$ are unchanged, $\pi_{G'}(y',t)\ge\pi_{G'}(y',\overline{F},t)=\pi_G(y',\overline{F},t)=\Omega(r)$.
    The only remaining thing is to connect $w$ with $y'$. We do it as follows:
    \begin{itemize}
        \item If $y'=y$ and $\din(y)>0$, since $\qr(y)<(1+\din(y))/C$, there exists an edge $(x,y)$ such that $\pr(x,y)<2/C$. We replace $(x,y)$ with $(x,w)$ and $(w,y)$.
        \item If $y'=y$ and $\din(y)=0$, then $\qr(y)<(1+\din(y))/C=1/C$. We simply add an edge $(w,y)$.
        \item If $y'=\mu(F,t)$, then there exists an edge $(x,y')\in F$. Recall that this edge has been replaced by $(w'_{xy'},y')$ (and something else). We further replace $(w'_{xy'},y')$ with $(w,y')$.
    \end{itemize}

    Call the new graph $G''$.
    
    \begin{lemma}
    \label{lem:pi-st-G''}
        $\pi_{G''}(s,t) > \Omega(C\pi_G(s,t))$.
    \end{lemma}

    \begin{proof}
        Since $(w,y)$ is the only out-edge of $w$, we have $\pi_{G''}(s,t)=\Omega(\pi_{G''}(s,w)\pi_{G''}(y',t))$.
        We claim that $\pi_{G''}(s,w)=\Omega(\pi_{G'}(s,w))$ and $\pi_{G''}(y',t)=\Omega(\pi_{G'}(y',t))$. This is because, the only things we may do from $G'$ to $G''$ are:
        \begin{itemize}
            \item Adding an edge $(w,y')$ where this edge becomes the only out-edge of $w$. This will not decrease any $\pi(\cdot,\cdot)$.
            \item Deleting an edge $(w'_{xy'},y')$ where $\din(w'_{xy'})=0$. This will not decrease any $\pi(u,\cdot)$ for $u\neq w'_{xy'}$.
            \item Replacing $(x,y')$ with $(x,w)$ where the only out-edge of $w$ is $(w,y')$. This is equivalent to subdividing $(x,y')$ and then merging the new vertex with $w$. By \cref{lem:subdivide}, this can only decrease any $\pi(\cdot,\cdot)$ by a constant factor.
        \end{itemize}
        
        By \cref{lem:pi-sw-G',lem:F}, $\pi_{G'}(s,w)>\Omega(C\pi(s,t)/r)$. By definition, $\pi_{G'}(y',t)=\Omega(r)$. So $\pi_{G''}(s,t)>\Omega(C\pi_G(s,t))$.
    \end{proof}
    
    We remark that, by \cref{lem:pi-st-G''}, we can choose a large enough $C$ so that, e.g. $\pi_{G''}(s,t) > 10\pi_G(s,t)$.
    It is straightforward to check that our modification from $G$ to $G''$ satisfies \cref{eqn:total-prob}.
\end{proof}

Combining \cref{thm:not-knowing-n,thm:upper_bound} proves the instance-optimality of $\TriPPR$.
\section{Instance-Non-Optimality on Non-Sparse Graphs When $n$ is Known}\label{sec:instance-non-optimality}

We show that when $n$ is known, every good algorithm must spend the usual worst-case time $\Omega(\min\{m, 1/\pi(s,t)\})$ on an instance which can be solved in time $O(n)$ by some correct algorithm.
For simplicity, we also assume $\delta=0$ here. In \cref{sec:reduction-delta}, we will show how to extend it to a general $\delta$.

The proof is similar to that of an analogous result for bidirectional Dijkstra's algorithm on simple graphs~\cite[Theorem 5.1]{dijkstra-simple}.
It exploits that if $n$ is known, and we observe that a vertex $u$ has e.g.\ out-degree $n-1$, then we know that every other vertex has an in-edge from $u$, without explicitly querying any in-edges.

\begin{theorem}\label{thm:n-known-best-possible}
    When $n$ is known, every algorithm returning an estimate $\epi(s,t)$ of $\pi(s,t)$ satisfying 
    \[\Pr\bigl\{|\epi(s,t)-\pi(s,t)|\le \rela\pi(s,t)\bigr\}\ge 2/3,\]
    has instance-optimality ratio
    \[\Omega\left(\frac{1}{n}\min\left\{m, \frac{1}{\pi(s,t)}\right\}\right).\]

    More precisely, for every $n$, $m$, and $p$ with $n \le m \le n^2$ and $0 < p \leq 1$, and for every correct algorithm, there exists an instance with $\Theta(n)$ vertices, $\Theta(m)$ edges, and $\pi(s,t) = \Theta(\max\{1/m, p\})$, on which the algorithm queries $\Omega(\min\{m, 1/p\}) = \Omega(\min\{m, 1/\pi(s,t)\})$ edges in expectation, but some correct algorithm spends $O(n)$ queries and time.
\end{theorem}
\begin{proof}
    We can assume that $n$ is at least a sufficiently large constant for the following construction to go through.
    Let $k = \sqrt{\min\{m, 1/p\}}$.
    We construct a graph $G$ with vertices partitioned into eight subsets: $V = \{s,t\} \cup V_1 \cup V_2 \cup \cdots \cup V_8$, where $\abs{V_i}=\Theta(k)$ for $i \in \{1,\ldots,6\}$, $\abs{V_8} = \Theta(m/n)$ and $\abs{V_7} = n - \Theta(k)-\Theta(m/n)$, such that $\abs{V} = n$.
    The source $s$ has an edge to every vertex in $V_1$.
    Every vertex in $V_2$ has an edge to $t$.
    Every vertex in $V_1 \cup V_3 \cup V_6$ has an edge to every vertex in $V_5$.
    Every vertex in $V_6$ has an edge to every vertex in $V_2 \cup V_4 \cup V_5$ (including those from $V_6$ to $V_5$ already mentioned).
    There is a perfect matching from $V_1$ to $V_4$ and from $V_3$ to $V_2$.
    The vertices in $V_7$ form a path, and there is an edge from $t$ to the first vertex on the path.
    Finally, every vertex in $V_8$ has an edge to every other vertex in the graph, i.e.\ full out-degree $n-1$.
    Note that the graph has $\Theta(m)$ edges.

    We also construct a modified graph $G'$, as follows.
    Pick edges $(u_1, v_4) \in V_1 \times V_4$ and $(u_3, v_2) \in V_3 \times V_2$ uniformly at random from the matchings in $G$.
    Then remove these edges and insert new edges $(u_1, v_2)$ and $(u_3, v_4)$ in their place in the respective adjacency lists.
    Note that this preserves the in-degree and out-degree of every vertex.
    See~\Cref{fig:graph-best-possible} for an illustration of $G$ and $G'$, without $V_8$.

    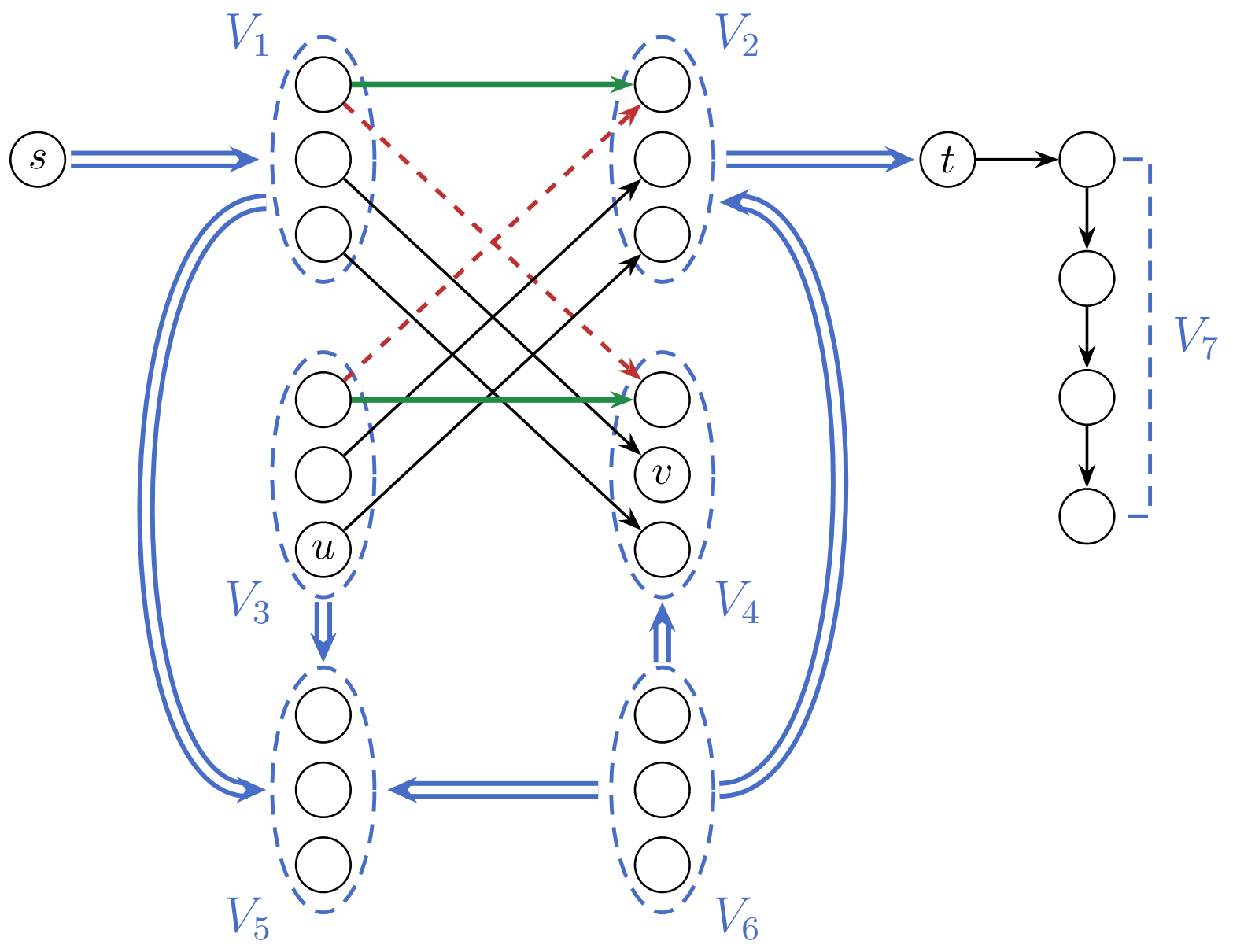
\begin{figure}[t]
      \centering

\definecolor{edgeblue}{RGB}{71,109,199}
\definecolor{edgegreen}{RGB}{35,139,72}
\definecolor{edgered}{RGB}{191,49,49}

\begin{tikzpicture}[
  vertex/.style={circle, draw=black, fill=white,
    minimum size=4.6mm, inner sep=0pt, line width=0.5pt},
  cluster/.style={draw=edgeblue, dashed, dash pattern=on 4.7pt off 3pt,
    line width=0.92pt},
  dbl/.style={draw=edgeblue, double=white, double distance=1.95pt,
    line width=1.12pt, -{Stealth[length=2.5mm,width=2.2mm]}},
  single/.style={draw=black, line width=0.68pt,
    -{Stealth[length=2.0mm,width=1.45mm]}},
  rededge/.style={single, draw=edgered, dashed, line width=1.05pt},
  greenedge/.style={single, draw=edgegreen, line width=1.42pt},
  every node/.style={font=\small}
]

\foreach \k/\x/\y in {
  1/0/4.40, 2/2.85/4.40,
  3/0/1.75, 4/2.85/1.75,
  5/0/-0.90,6/2.85/-0.90
} {
  \foreach \i/\dy in {1/0.63,2/0,3/-0.63} {
    \node[vertex] (v\k-\i) at (\x,{\y+\dy}) {};
  }
  \draw[cluster] (\x,\y) ellipse [x radius=0.43cm, y radius=1.03cm];
}

\foreach \k/\x/\y in {
  1/-0.63/5.45, 2/3.48/5.45,
  3/-0.63/0.68, 4/3.48/0.68,
  5/-0.63/-1.98,6/3.48/-1.98
} {
  \node[text=edgeblue, font=\large] at (\x,\y) {$V_{\k}$};
}

\node[vertex] (s) at (-2.40,4.40) {$s$};
\node[vertex] (t) at (5.25,4.40) {$t$};
\foreach \i/\y in {1/4.40,2/3.40,3/2.40,4/1.40} {
  \node[vertex] (w\i) at (6.42,\y) {};
}
\draw[cluster] (6.72,4.40) -- (6.95,4.40) --
               (6.95,1.40) -- (6.72,1.40);
\node[right, text=edgeblue, font=\large] at (7.00,2.90) {$V_7$};

\draw[dbl, shorten <=0.4mm, shorten >=3.0mm] (s) -- (v1-2);
\draw[dbl, shorten <=3.0mm, shorten >=0.4mm] (v2-2) -- (t);

\draw[dbl] (-0.48,4.04)
  .. controls (-1.80,4.04) and (-1.80,-0.90) .. (-0.48,-0.90);
\draw[dbl] (3.33,-0.90)
  .. controls (4.65,-0.90) and (4.65,4.04) .. (3.33,4.04);

\draw[dbl, shorten <=3.0mm, shorten >=3.0mm] (v6-2) -- (v5-2);
\draw[dbl] (0,0.68) -- (0,0.17);        
\draw[dbl] (2.85,0.17) -- (2.85,0.68);  

\foreach \i in {1,2,3} {
  \ifnum\i=1
    \draw[rededge] (v3-\i) -- (v2-\i);
    \draw[rededge] (v1-\i) -- (v4-\i);
  \else
    \draw[single] (v3-\i) -- (v2-\i);
    \draw[single] (v1-\i) -- (v4-\i);
  \fi
}

\draw[greenedge] (v1-1) -- (v2-1);
\draw[greenedge] (v3-1) -- (v4-1);

\draw[single] (t) -- (w1);
\draw[single] (w1) -- (w2);
\draw[single] (w2) -- (w3);
\draw[single] (w3) -- (w4);

\node at (v3-3) {$u$};
\node at (v4-2) {$v$};
\end{tikzpicture}
      \caption{Graph construciton for the proof of~\Cref{thm:n-known-best-possible}. Blue double edges indicate that every edge from a subset to another is present. The dashed red edges are only present in $G$. The thick green edges are only present in $G'$. Note that $t$ is reachable from $s$ in $G'$ but not in $G$.}
      \label{fig:graph-best-possible}
    \end{figure}

    Note that in $G$ we have $\pi(s,t)=0$, while in $G'$ we have $\pi(s,t) = \frac{1-\alpha}{\abs{V_1}}\cdot\frac{1-\alpha}{\abs{V_5}+1}(1-\alpha)\alpha = \Theta(1/k^2) = \Theta(\max\{1/m, p\})$.
    Every correct algorithm must thus distinguish $G$ and $G'$ with constant probability.

    We first show that if adjacency lists are given in uniformly random orders, every algorithm must spend $\Omega(k^2)$ time to distinguish $G'$ from $G$ with constant probability.
    This implies that every algorithm requires $\Omega(k^2)$ time (in expectation over the internal randomness of the algorithm) on some (algorithm-specific) ordering.
    Since all in-degrees and out-degrees are the same in $G$ and $G'$, the only way for the algorithm to deviate on the two graphs is to query the edge $(u_1, v_2)$ from $u_1$ or $v_2$, or query the edge $(u_3, v_4)$ from $u_3$ or $v_4$.
    Running the algorithm on $G$, each matching edge in $G$ is hidden among $\Omega(k)$ neighbors at both endpoints, so $q$ (adaptive) queries discover $O(q/k)$ matching edges in expectation.
    Since the two selected matching edges are chosen uniformly at random among $\Omega(k)$ options, the probability of discovering either is $O(q/k^2)$.
    Coupling the orders of $G$ and $G'$, this shows that the algorithm must spend $\Omega(k^2)$ queries to distinguish the two graphs.

    Next, we show that on any fixed adjacency-list ordering, it can be certified that the given graph is $G'$ in $O(n)$ time.
    Let us describe a certification algorithm.
    First, the algorithm visits all out-neighbors of $s$ and all in-neighbors of $t$, identifying $V_1$, $V_2$, and $V_8$.
    Here, $V_8$ is distinguished as the subset of neighbors with in-degree zero.
    By observing that every vertex in $V_8$ has out-degree $n-1$, the algorithm certifies that every vertex in $V_8$ has an edge to every other vertex in the graph, using that $n$ is known.
    Next, querying all edges of $u_1$ and $v_2$, which can be guessed correctly because we fixed the adjacency-list orderings, identifies $V_5$ and $V_6$.
    Querying all edges of a vertex in $V_5$ and a vertex in $V_6$ identifies $V_3$ and $V_4$.
    Next, the algorithm queries all matching edges, including the replaced ones, using hardwired adjacency-order guesses.
    The algorithm also traverses the path in $V_7$ from $t$.
    Since $n$ is known, the algorithm knows that there are no more vertices.
    The algorithm now observes that no vertex in $\{s,t\} \cup V_2 \cup V_4 \cup V_7$ has an unqueried out-edge, and in particular no edge to $V_5$.
    So since each vertex in $V_5$ can be seen to have in-degree $\abs{V_1 \cup V_3 \cup V_6 \cup V_8}$, it is certified that every vertex in $V_1 \cup V_3 \cup V_6$ (and as already observed $V_8$) has an edge to every vertex in $V_5$.
    It is similarly certified that $V_6$ has an edge to every vertex of $V_2 \cup V_4 \cup V_5$.
    Summing all vertex degrees, it is certified, that there are no more edges.
    This certifies that the graph is $G'$.
\end{proof}
\section{Lower Bound When $n$ is Known}\label{sec:lowerbound_withn}

In this section, we will prove the following theorem.
\begin{theorem}\label{thm:lower-bound-h}
    Fix any constant $c\in(0,1)$.
    Every algorithm that, on every valid instance, outputs an estimate $\hat{\pi}(s,t)$ satisfying
    $|\hat{\pi}(s,t)-\pi(s,t)|\leq c\pi(s,t)$ with probability at least $2/3$,
    has expected running time
    \[
        T=\Omega\p{\varepsilon\min\left\{\frac{T^*}{h+1},\frac{n}{\log n}\right\}}.
    \]
\end{theorem}
This implies the following bound on the instance-optimality ratio of $\TriPPR$.

\begin{corollary}
    $\TriPPR$ is $\tilde O\left((1/\varepsilon)\max\left\{h+1,\frac{\Tstar}{n}\right\}\right)=\tilde O\left((1/\varepsilon)\max\left\{h+1,\frac{1}{n}\min\left\{m,\frac{1}{\pi(s,t)}\right\}\right\}\right)$-instance-optimal.
\end{corollary}
\begin{proof}
    Since $\TriPPR$ has expected running time $\tilde O(\Tstar)$, \Cref{thm:lower-bound-h} implies an instance-optimality ratio of
    \[
        \tilde O\left(\frac{\Tstar}{\varepsilon\min\{\Tstar/(h+1),n\}}\right)
        =\tilde O\left(\frac{1}{\varepsilon}\max\left\{h+1,\frac{\Tstar}{n}\right\}\right).
    \]
    Moreover, for every $r\in[0,1]$,
    \[
        \TA\leq\min\left\{m,\frac{r}{\pi(s,t)}\sum_{u\in V}\pi(s,u)\right\}
        \leq\min\left\{m,\frac{1}{\pi(s,t)}\right\}.
    \]
    Hence $\Tstar\leq\min\{m,1/\pi(s,t)\}$, proving the corollary.
\end{proof}
We may also want to parameterize the instance-optimality factor in terms of $\delta$. Recall that $\delta$ is the approximation threshold, so we only need to approximate $\pi(s,t)$ when $\pi(s,t)\ge\delta$. By the reduction in \Cref{sec:reduction-delta}, we may therefore replace $1/\pi(s,t)$ by $1/\delta$ in the instance-optimality factor.

To prove~\Cref{thm:lower-bound-h}, we wish to show that for every good algorithm $A$ and graph $G$, the running time of $A$ on $G$ is $\Omega(T^*)$.
To do so, we assume for the sake of contradiction, that the running time of $A$ on $G$ is $o(T^*)$, and show that there exists a modified graph $G'$, which $A$ does not distinguish from $G$, but where $(1-c)\pi_{G'}(s,t) > (1+c)\pi_G(s,t)$, contradicting correctness of $A$.
To construct $G'$, we first find a vertex $y$, that $A$ examines with low probability, such that the backward probability mass $\pi(y,t)$ is large, even if we only count random walks inside some local subgraph around $t$.
This is formalized in~\Cref{lem:hidden-target-mass} and can be thought of as the lower-bound counter-part to the backward pushing in our \TriPPR{} algorithm.
Next, we modify $G$, in a way that avoids the mentioned local subgraph around $t$, and touches only edges examined with low probability by $A$, such that a large forward probability mass $\pi(s,W)=\sum_{w\in W}\pi(s,w)$ is concentrated in some set $W$, which is examined with low probability, and satisfies that all low-degree vertices have many non-edges to $W$.
This is formalized in~\Cref{lem:hidden-source-mass} and can be thought of as the lower-bound counter-part to the forward pushing and Monte Carlo phase in our \TriPPR{} algorithm. 
Finally, we obtain $(1-c)\pi_{G'}(s,t) > (1+c)\pi(s,t)$, by funneling the forward probability source in $W$ into the backward probability sink in $y$, through modified edges, that are unlikely to be examined.
This is formalized in the proof of~\Cref{thm:lower-bound-h}, and can be thought of as the lower-bound counter-part to the Monte Carlo sampling in our \TriPPR{} algorithm. 

Fix a graph $G$, $r\in (0,1]$, and $\varepsilon\in (0,1)$.
Let $A$ be any algorithm satisfying \Cref{thm:lower-bound-h}, and let $T$ denote its expected running time on $G$.
Assume
\begin{equation}\label{ineq:T-small}
    T \leq c_0 \varepsilon\min\left\{\frac{\min\{\TA,\TB\}}{h+1},\frac{n}{\log n}\right\}    .
\end{equation}
where
\begin{align*}
    c_0 = \frac{(1-c)^2\alpha^4(1-\alpha)^3}{320000(1+c)^2}.
\end{align*}
In particular, we assume (as $T$ is trivially greater than $1$)
\begin{equation}\label{eq:assumptions}
    \TA,\TB \geq \frac{(h+1)T}{c_0\varepsilon} \geq \frac{(h+1)}{c_0}.
\end{equation}
We now formalize the two ingredients described above in the following lemmas.
We use the same definitions as in \Cref{sec:lowerbound_withoutn}.
Figure~\Cref{fig:sketch-lb-1} illustrates how the two lemmas combine in the proof of \Cref{thm:lower-bound-h}.

\begin{figure}[h]
    \centering
    \includegraphics[width=0.5\linewidth]{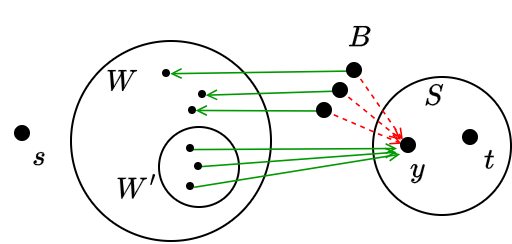}
    \caption{
Sketch of \Cref{thm:lower-bound-h}. We find a set $S$ containing $t$ and a vertex $y$ with large backward probability mass to $t$, chosen so that this mass remains large as long as the internal structure of $S$ is preserved. We then find a set $W$ that is unlikely to be examined by the algorithm, together with $W'\subseteq W$ carrying substantial forward probability mass from $s$. Finally, we funnel the probability mass from $W'$ to $y$ by replacing edges from $B$ to $y$ with edges from $W'$ to $y$, while adding replacement edges from $B$ to $W$ to preserve the out-degrees of vertices in $B$. These modifications are unlikely to be detected by the algorithm, but increase $\pi(s,t)$ enough to contradict its correctness.
}
    \label{fig:sketch-lb-1}
\end{figure}

\begin{lemma}[Hidden Backward Probability Mass]\label{lem:hidden-target-mass}
    There exist $S\subseteq V$, $y\in S$, and $B\subseteq\Nin(y)\setminus \Hout$ satisfying each of the following.
    \begin{itemize}
        \item $|S\cup B|\leq \frac{\varepsilon n}{5}$.
        \item Either $B=\emptyset$ and $\qr(y)\le\frac{1}{10}$, or $|B|=h+1$ and $\pr(B\times \{y\})\le\frac{1}{10}$.
        \item $\pi_{G'}(y,t)\geq \alpha^2 r$ where $G'$ is any modified graph preserving the out-degrees of vertices in $S$ and leaving all edges within $S$, except possibly those in $B\times \{y\}$.
    \end{itemize}
\end{lemma}

We say that a deterministic algorithm $A$ \emph{distinguishes} two graphs $G$ and $G'$ if its query transcript on $G$ differs from its query transcript on $G'$, where the query transcript is the ordered sequence of queries made by $A$ together with the corresponding answers. Thus, differing outputs alone do not constitute distinguishing the two graphs. For a randomized algorithm, we say that $A$ distinguishes $G$ and $G'$ for a given random seed if the corresponding deterministic executions distinguish them, and the probability of distinguishing is taken over the internal randomness of $A$.

\begin{lemma}[Hidden Forward Probability Mass]\label{lem:hidden-source-mass}
    Let $S$ and $B$ be as in~\Cref{lem:hidden-target-mass}.
    There exists a modified graph $G'$ of $G$ preserving the vertex set, the out-degrees of vertices in $S$, and keeping all internal edges of $S$, possibly apart from edges in $B\times\{y\}$, such that $A$ distinguishes $G'$ from $G$ with probability at most $\frac{1}{10}$, and such that there exists $W\subseteq V\setminus (S\cup B)$ and $W' \subseteq W$ satisfying each of the following.
    \begin{enumerate}
        \item\label{item:concentrated-large-hit-prob} $H_{G'}(s,\overline{S},W') > \frac{1+c}{1-c}\pi_G(s,t)\frac{h+1}{r(1-\alpha)\alpha^2}$.
        \item\label{item:concentrated-unlikely-visited} $\qr(W)\le\frac{1}{10}$ in $G$.
        \item\label{item:concentrated-outdeg-small} In $G'$, all $w' \in W'$ have $\dout(w') \leq h$.
        \item\label{item:concentrated-b-non-edge} For every $b \in B$ there exists a $w \in W$ with $(b,w) \notin E(G')$.
        \item\label{item:concentrated-w-size} $1 \le |W'| \le h+1$.
    \end{enumerate}
\end{lemma}

\subsection{Proof of~\Cref{thm:lower-bound-h}}

We now wish to obtain $(1-c)\pi_{G'}(s,t) > (1+c)\pi_G(s,t)$, by funneling the forward probability source in $W'$, obtained in~\Cref{lem:hidden-source-mass}, into the backward probability sink in $y$, obtained in~\Cref{lem:hidden-target-mass}, through modified edges, that are unlikely to be examined.

\begin{proof}[Proof of \Cref{thm:lower-bound-h}]
    We will show that our assumption~\labelcref{ineq:T-small} leads to a contradiction using our lemmas.
    Choose $S$, $y$ and $B$ as in~\Cref{lem:hidden-target-mass} and choose matching $G'$, $W$, and $W'$ as in~\Cref{lem:hidden-source-mass}.
    We will construct a further modified graph $G''$ of $G'$, which $A$ distinguishes from $G$ with probability less than $\frac{1}{3}$, where the forward probability mass in $W'$ is routed into $y$, ensuring $(1-c)\pi_{G''}(s,t) > (1+c)\pi_G(s,t)$.
    This will contradict $A$ having success probability at least $2/3$, thereby showing that the assumed~\labelcref{ineq:T-small} is false, completing the proof.

    Our construction $G''$ will preserve the out-degrees of vertices in $S$ and leave all edges within $S$, except possibly those in $B \times \{y\}$, so $\pi_{G''}(y,t) \geq \alpha^2 r$.
    Moreover, our construction will not decrease the hitting probability of $W'$ of walks avoiding $S$, so $H_{G''}(s,\overline S, W') \ge H_{G'}(s,\overline S,W') > \frac{1+c}{1-c}\pi_G(s,t)\frac{h+1}{r(1-\alpha)\alpha^2}$.
    Finally, our construction will contain every possible edge from $W'$ to $\{y\}$, while increasing the out-degree of each vertex in $W'$ by at most one, that is, increasing them to at most $h+1$.
    We will thus have $(1-c)\pi_{G''}(s,t)\ge (1-c)H_{G''}(s,\overline S,W')\frac{1-\alpha}{h+1}\pi_{G''}(y,t) > (1+c)\pi_G(s,t)$, as wished.

    Consider first the case where $B = \emptyset$ and $\qr(y) \le \frac{1}{10}$.
    We construct $G''$ by inserting all missing edges $W' \times \{y\}$ into $G'$.
    This graph $G''$ is not distinguished from $G'$ unless a vertex in $W'\cup\{y\}$ is queried in $G'$.
    In particular, not unless $G$ is distinguished from $G'$ or a vertex in $W'\cup\{y\}$ is queried in $G$.
    Recall from~\Cref{lem:hidden-source-mass} that $G$ is distinguished from $G'$ with probability at most $\frac{1}{10}$ and $\qr(W') \le \qr(W) \le \frac{1}{10}$.
    Recall from~\Cref{lem:hidden-target-mass} that $\qr(y) \le \frac{1}{10}$.
    The probability that $G$ is distinguished from $G''$ is thus at most $\frac{3}{10} < \frac{1}{3}$, as wished.
    Moreover, the construction satisfied the above promised properties giving $(1-c)\pi_{G''}(s,t) > (1+c)\pi_G(s,t)$, finishing the case.

    Next, consider the remaining case where $\abs{B}=h+1$ and $\pr(B\times\{y\})\le\frac{1}{10}$.
    Choose injective maps $f\colon W' \to B$ and $g\colon B \to W$ such that $(b, g(b)) \notin E(G')$ for all $b\in B$.
    To construct $G''$, consider each $w' \in W'$ for which $(w',y)$ is not an edge in $G'$.
    Insert $(w',y)$ and $(f(w'),g(f(w')))$ as edges in $G''$ and remove $(f(w'), y)$.
    Note that this modification is well-defined, and the degrees of vertices in $V\setminus W$ are preserved.
    More specifically, the edge $(w',y)$ is inserted in the inward adjacency list of $y$ in place of $(f(w'),y)$, and $(f(w'),g(f(w')))$ is inserted in the outward adjacency list of $f(w')$ in place of $(f(w'),y)$.
    This graph $G''$ is not distinguished from $G'$ unless a vertex in $W$ or an edge in $B\times\{y\}$ is queried in $G'$.
    In particular, not unless $G$ is distinguished from $G'$ or a vertex in $W$ or an edge in $B\times\{y\}$ is queried in $G$.
    Again, the lemmas tell us that this happens with probability at most $\frac{1}{10}+\frac{1}{10}+\frac{1}{10}<\frac{1}{3}$, as wished.
    Moreover, the construction satisfied the above promised properties giving $(1-c)\pi_{G''}(s,t) > (1+c)\pi_G(s,t)$, since replacing an edge $(f(w'),y)$ by $(f(w'), g(f(w')))$ cannot decrease $h(s,\overline S,W')$ since $y \in S$.
    This completes the case.
\end{proof}

\subsection{Proof of \Cref{lem:hidden-target-mass}}

\begin{proof}[Proof of \Cref{lem:hidden-target-mass}]
    Let $V_r=\{v \in V \mid \pi(v,t)\geq r\}$.
    Let $S\subseteq V_r$ be the $\min\{\abs{V_r}, 10T\}$ vertices $v \in V_r$ with largest $\pi(v,t)$.
    Note that $|S|\leq 10T \leq \frac{\varepsilon n}{10}$.

    Suppose first that $S\subsetneq V_r$.
    Then $\abs{S} = 10T$.
    Since $Q(S) \leq T$, there must exist a $y\in S$ with $Q(y)\le T/\abs{S}=1/10$.
    In this case, choose $B=\emptyset$.

    Otherwise, $S=V_r$. 
    Then $|S|\leq 10T \leq \frac{\varepsilon \TB}{10(h+1)}$, and the number of in-edges of $S$ is
    \begin{align*}
        \TB - |V_r|
        \geq \TB - \frac{\varepsilon \TB}{(h+1) 4} 
        \geq \frac{3\TB}{4}.
    \end{align*}
    Consider the subgraph constructed by removing all edges $(u,v)$ with $u \in \Hout$, then removing all edges $(u,v)$ with $\din(v) \le h$, and finally removing all vertices with in-degree zero.
    This removes at most $2\abs{S}h < \frac{\varepsilon\TB}{4}$ edges, leaving us with at least $\TB/2$ edges.
    In this subgraph, there must exist a vertex $y$ with $\frac{\pr(\Nin(y)\times y)}{\din(y)} \leq \frac{2T}{\TB}$.
    Take such a $y$ and choose $B\subseteq \Nin(y)$ to be the $h+1$ vertices $v\in\Nin(y)$ with smallest $\pr(v,y)$, implying that $\pr(B\times \{y\}) \leq \frac{(h+1)2T}{\TB} \leq \frac{1}{50}$.
    Furthermore, in this case
    \begin{align*}
        \TB \leq n |S| \leq 10nT \leq \frac{n\varepsilon \TB}{10(h+1)},
    \end{align*}
    so $|B|=h+1 \leq \frac{\varepsilon n}{10}$, and $|S\cup B| \leq \frac{\varepsilon n}{5}$.

    Let $p_S$ be the probability that a discounted random walk starting in $y$ terminates in $t$, without ever walking outside $S$.
    The probability of moving outside $S$ is at most $1-\alpha$, since it requires at least one step.
    Furthermore, the probability of terminating at $t$ when starting outside $S$ is at most $\max_{u \in V\setminus S}\pi(u,t) \leq \pi(y,t)$, using the definition of $S$.
    Hence, the the probability of moving outside $S$ and later terminating at $t$ is at most $(1-\alpha)\pi(y,t)$.
    Thus $p_S \geq \pi(y,t)-(1-\alpha)\pi(y,t) = \alpha\pi(y,t) \geq \alpha r$.

    The probability that a random walk starting at $y$ traverses an edge in $B\times \{y\}$ is also at most $1-\alpha$, since it requires at least one step.
    After traversing such an edge, the walk continues from $y$, so from that point on, it terminates at $t$ without leaving $S$ with probability $p_S$.
    Thus, the probability that a random walk starting in $y$ terminates in $t$ without leaving $S$ or traversing an edge in $B\times\{y\}$ is at least $p_S-(1-\alpha)p_S=\alpha^2 r$.
    Finally, note that is a lower bound on $\pi_{G'}(y,t)$ for any modified graph $G'$ preserving the out-degrees of vertices in $S$ and leaving all edges within $S$, except possibly those in $B\times\{y\}$.
\end{proof}

\subsection{Proof of \Cref{lem:hidden-source-mass}}

We start by sketching the proof.
First, we find a suitable vertex set $W\subseteq V\setminus(S\cup B)$ that is unlikely to be examined by $A$. We then analyze the forward probability mass from $s$ and show:
either sufficiently much of this mass already reaches $W$, or we identify a set of edges $F$, that $A$ is unlikely to examine, carrying sufficiently large forward mass that can be redirected into $W$.
In either case, we construct a modified graph in which $W$ receives the required amount of forward probability mass while remaining unlikely to be examined.
Finally, we concentrate this mass on a subset $W'\subseteq W$ of at most $(h+1)$ low-out-degree vertices, yielding the properties claimed in \Cref{lem:hidden-source-mass}.

We start by finding the suitable vertex set $W$.

\begin{lemma}\label{lem:unlikely}
    Sample a subset $W \subseteq V$ of vertices by sampling each vertex in $V'=V\setminus (S\cup B)$ independently with probability $1/(100T)$.
    With probability at least $3/5$, each of the following properties hold in $G$.
    \begin{enumerate}
        \item (Unlikely visited) $Q(W) \leq \frac{1}{16}$,
        \item (Large) $\frac{n}{200T}\leq\abs{W} \leq \frac{n}{50T}$ and $\varepsilon|W|\geq 16$,
        \item (Many non-neighbors) $\forall v\notin \Hout\colon |W\setminus \Nout(v)| \geq \varepsilon|W|/4$ and $\forall v\notin \Hin\colon |W\setminus \Nin(v)| \geq \varepsilon|W|/4$,
    \end{enumerate}
\end{lemma}
\begin{proof}
    All properties hold in expectation, (with $\varepsilon|W|\ge16$ following from~\eqref{ineq:T-small}) so the remaining part of the proof is to establish concentration using Markov's inequality and Chernoff bounds.
    
    We will prove that these properties hold simultaneously by a union bound.
    By the first property of $\Cref{lem:hidden-target-mass}$, $|V'|\geq n - \varepsilon n/5 \geq 3n/4$.

    \emph{Unlikelt visited}.
    Recall $Q(V) \leq 2 T$, and thus $\Ex{Q(W)} \leq \frac{1}{50}$.
    By Markov's inequality, $Q(W) > \frac{1}{16}$ with probability at most $8/25$.

    \emph{Large}.
    Since $|S\cup B|\leq \varepsilon n/4$, we have
    $
        \Ex{|W|} 
        \geq \p{1-\frac{\varepsilon}{4}} \frac{n}{100T} 
        \geq \frac{3n}{400T}.
    $
    Chernoff bounds give $|W|<n/(200T)$ or $|W|>n/(50T)$ with probability at most $1/25$.
    On the complement of this event, $\varepsilon|W|\geq \varepsilon n/(200T)\geq \log n/(200c_0)\geq 16$, assuming $n$ is sufficiently large.

    \emph{Many non-neighbors}.
    Fix $v\notin \Hout$.
    By definition of $\Hout$, $v$ has at least $\varepsilon n$ non-out-neighbors, of which at least $\varepsilon n-|S\cup B|\geq 4\varepsilon n/5$ lie in $V'$.
    We have $\Ex{\abs{W\setminus \Nout(v)}} \geq \frac{\varepsilon n}{125 T}$.
    By Chernoff's bound and a union bound, the probability that there exists a vertex \(v\notin\Hout\) with \(|W\setminus\Nout(v)|<\varepsilon n/(200T)\) is at most \(1/50\).
    If $|W|\leq n/(50T)$, then $\varepsilon n /(200T)\geq \varepsilon|W|/4$.    
    By symmetry, we can show the same for vertices in $\Hin$

    All properties hold simultaneously with probability at least $1-8/25-1/25-1/50-1/50 = 3/5$.
\end{proof}

For a vertex $v\in V$, let $E(v)=\{(a,b)\in E \mid v=a\lor v=b\}$ be the edges incident to $v$.
For a vertex set $A\subseteq V$, let $E(A) = \cup_{v\in A} E(v)$.
For an edge, we define the $z$-mass of an edge $(u,v)$ to be $z(u,v)=\min\{1, \pi(s,u)r/(\pi(s,t)\dout(u))\}$.
One can vaguely think of this as the amount of times the edge should be sampled in expectation by the forward push or Monte Carlo section of~\TriPPR{}.
For an edge set $R\subseteq E$, let $z(R)=\sum_{e\in R}z(e)$.
Let $z(v) = z(E(v))$ and $z(A)=z(E(A))$.
Note $z(V)=z(E)=\TA$.

\begin{fact}\label{fact:S}
    $H(s,S) \leq \frac{\pi(s,t)}{r}$.
\end{fact}
\begin{proof}
    Since every $u\in s$ satisfies $\pi(u,t)\geq r$, it follows that $\pi(s,t)\geq H(s,S)r$.
\end{proof}

\begin{lemma}\label{lem:z-mass-prop}
    For any set $A\subseteq V$, $\frac{(1-\alpha)\pi(s,t)}{2r}z(A) \leq \pi(s,A)\leq H(s,A)$.
\end{lemma}
\begin{proof}
    From the basic identity
    \begin{align*}
        \pi(s,v)&=\alpha\mathbf{1}_{\{v=s\}}+(1-\alpha)\sum_{u\in \Nin(v)}\frac{\pi(s,u)}{\dout(u)},
    \end{align*}
    we get
    \begin{align*}
        \sum_{u\in \Nin(v)}\frac{\pi(s,u)}{\dout(u)} \leq  \frac{\pi(s,v)}{(1-\alpha)}.
    \end{align*}
    We then have
    \begin{align*}
        z(A)
        &\leq \sum_{v\in A} \p{\sum_{u\in \Nout(v)}z(v,u)+\sum_{u\in \Nin(v)}z(u,v)}\\
        &\leq \frac r{\pi(s,t)}\p{\sum_{v\in A}\pi(s,v)+\sum_{v\in A}\sum_{u\in \Nin(v)}\frac{\pi(s,u)}{\dout(u)}}\\
        &\leq \frac r{\pi(s,t)} \left(1+\frac1{1-\alpha}\right)\pi(s,A)\\
        &\leq \frac {2r \pi(s,A)}{(1-\alpha)\pi(s,t)} .
    \end{align*}
\end{proof}

Let $E_H=\{(u,v)\in E:u\in\Hout\ \lor\ v\in\Hin\}$ be the set of edges coming from high-out-degree vertices or going to high-in-degree vertices.

\begin{lemma}\label{lem:E-H}
    If $z(E_H) \geq \TA/4$, there exists a subset $W \subseteq V\setminus(S\cup B)$ satisfying the properties of \Cref{lem:unlikely} and
    \begin{align*}
        H_G(s,W) > \p{1+\frac{32(1+c)(h+1)}{\varepsilon(1-c)(1-\alpha)^2\alpha^2}}\frac{\pi_G(s,t)}{r}
    \end{align*}
\end{lemma}
\begin{proof}
    We show that the $W$ sampled in~\Cref{lem:unlikely} satisfies these properties with positive probability.
    Let us first present a short sketch.
    To show the large hitting-probability, it suffices by~\Cref{lem:z-mass-prop} to show that $z(W)$ is large.
    This is true in expectation, because $z(V)$ is large, and $W$ contains a $\Theta(1/T)$-fraction of the mass of $V$ in expectation.
    To get small variance, we have to be more careful.
    For a vertex set $A\subseteq V$, we define $z_H(A)=Z(E(A)\cap E(\Hin \cup \Hout))$.
    Focusing on $z_H$ will allow us to bound the variance by a quantity involving a factor $h+1$ which would otherwise have been $n$.
    
    Note that $z_H(A)\leq z(\Hin \cup \Hout)$ and $z_H(A)\leq z(A)$.
    Let $V'=V\setminus (S\cup B)$ denote the support of the sampling process.
    Every edge incident to $\Hin\cup\Hout$ is also incident to $V'$, unless both its endpoints belong to $S\cup B$.
    Consequently, 
    \begin{align*}
        z_H(V')
        &=z(\Hin\cup\Hout)
          -z(E(\Hin\cup\Hout)\cap (S\cup B)^2)\\
        &\geq z(\Hin\cup\Hout)-z(E\cap (S\cup B)^2)
    \end{align*}
    We continue to bound $z(E\cap (S\cup B)^2)$.
    Note that $E(S)\cup (E\cap B^2) \supseteq E\cap (S\cup B)^2$.
    We have, by \cref{lem:z-mass-prop} and \cref{fact:S},
    \begin{align*}
        z(S) \leq \frac{2rH(s,S)}{(1-\alpha)\pi(s,t)} \leq \frac{2}{1-\alpha}.
    \end{align*}
    Also note, since every out-edge of a vertex $b$ has the same $z$-value,
    \begin{align*}
        z(E\cap B^2) 
        \leq |B| \sum_{b \in B} z(b,y) 
        \leq \frac{|B| r}{\pi(s,t)} \sum_{b\in B}\frac{\pi(s,b)}{\dout(b)}
        \leq \frac{|B| r \pi(s,y)}{(1-\alpha)\pi(s,t)}
        \leq \frac{h+1}{1-\alpha}.
    \end{align*}
    Therefore $z(E\cap (S\cup B)^2)\leq \frac{1}{1-\alpha}(2+h+1) \leq \frac{3(h+1)}{1-\alpha} \leq \frac{\TA}{8}$, where the last inequality uses~\eqref{eq:assumptions}. 
    At last, $z_H(V') \geq \TA/8$, since $z(\Hin \cup \Hout) \ge z(E_H) \ge \TA/4$ using our assumption.

    We will show that $\Ex{z_H(W)}\geq z_H(V')/T$.
    However, we also want to show that this holds with constant probability.
    Therefore, we will do a careful variance analysis.
    
    Assign every edge $(u,v)\in E(V')\cap E(\Hin \cup\Hout)$ to one of its endpoints in $V'$ as follows:
    If $(u,v)\in (V')^2$, assign the edge to an endpoint whose other endpoint belongs to $\Hin\cup\Hout$ (e.g. if $u\in\Hin\cup\Hout$ assign it to $v$), breaking ties arbitrarily.
    Otherwise, assign the edge to its unique endpoint in $V'$.
    For each $v\in V'$, let $b(v)$ be the total $z$-weight assigned to $v$.
    Then $\sum b(v) = z_H(V')$ and $b(v)\leq 2h+2|B|+z(S)$, since $v$ can be assigned at most $2h$ edges from $(V')^2$, and at most $2|B|$ edges incident to $B$, and all $z$-mass from $S$. 
    Thus, $b(v)\leq 2h+ 2(h+1)+\frac{2}{1-\alpha}\leq \frac{4(h+1)}{1-\alpha}$.
    Let $X=\sum_{v\in W}b(v)$.
    We then have
    \begin{align*}
        \Ex{X}
        = \frac{1}{100T} \sum_{v\in V'}b(v)
        = \frac{z_H(V')}{100T}
        \geq \frac{\TA}{800 T}
    \end{align*}
    and
    \begin{align*}
        \Vari{X} \leq \frac{4(h+1)}{1-\alpha}\Ex{X}.
    \end{align*}
    By Chebyshev's inequality, 
    \begin{align*}
        \P{X\leq \frac{1}{2}\Ex{X}}
        \leq \P{\abs{X-\Ex{X}}\geq \frac{1}{2}\Ex{X}}
        \leq 4\cdot \frac{4(h+1)}{1-\alpha}\frac{800T}{\TA}
        \leq \frac{1}{10}.
    \end{align*}
    Thus, with probability at least $9/10$,
    \[
    z(W)
    \geq X
    \geq \frac{\TA}{1600 T}
    \geq \frac{h+1}{1600c_0\varepsilon}
    > \p{1+\frac{32(1+c)(h+1)}{\varepsilon(1-c)(1-\alpha)^2\alpha^2}}\frac{2}{1-\alpha}.
    \]
    Consequently, by \Cref{lem:z-mass-prop},
    \begin{align*}
        H(s,W)
        \geq \frac{(1-\alpha)}{2}z(W)\frac{\pi(s,t)}{r} \geq \p{1+\frac{32(1+c)(h+1)}{\varepsilon(1-c)(1-\alpha)^2\alpha^2}}\frac{\pi(s,t)}{r}.
    \end{align*}

    The remaining properties follow by a union bound with~\Cref{lem:unlikely}.
\end{proof}

For $U\subseteq V$, let $\pi(s, \bar U, t)$ denote the probability of termination at $t$ without visiting $U$.
Let $z_{\bar U}(u,v) = \min\{1,r\pi(s,\bar U,u)/(\pi(s,t)\dout(u))\}$ be our analogous modification of $z(u,v)$.
Let $H(s,\bar U,u)$ be the probability of visiting $u$ no later than $U$.
For an edge set $F \subseteq E$, let $\Delta(F)$ be the maximum number of edges in $F$ going out of a single vertex.

\begin{lemma}\label{lem:constructing-F}
    Fix a $W$ satisfying the properties of \Cref{lem:unlikely}.
    Let $E_0 =E\setminus (E(S\cup W)\cup E_H)$.
    If $z_{\overline{S\cup W}}(E_0) \geq \TA/2$, then there exists $F\subseteq E_0$ satisfying each of the following.
    \begin{itemize}
        \item $z_{\overline{S\cup W}}(F) \geq \frac{2(1+c)(h+1)}{(1-c)(1-\alpha)^3\alpha^2}$. 
        \item $P(F) \leq \frac{1}{32}$.
        \item $\Delta(F) \leq \varepsilon|W|/8$.
    \end{itemize}
\end{lemma}
\begin{proof}
    Set $C_F= \frac{2(1+c)}{(1-c)(1-\alpha)^3\alpha^2} = \Theta(1)$.
    Let $E'$ be a bipartite graph with a left and right copy of $V$, and edges from $G$ placed from left to right.
    Partition these edges into at most $n$ matchings $P_1,\ldots,P_\ell\subseteq E'$ by bipartite edge coloring.
    For each $P_i$, let $Z_i=z_{\overline{S\cup W}}(P_i)$ and $Q_i=\pr(P_i)$.

    Discard matchings with $Z_i<\TA/(4n)$, removing total $z_{\overline{S\cup W}}$-weight at most $\TA/4$.
    The remaining matchings have total weight at least $\TA/4$.
    Order them by increasing $Q_i/Z_i$, and take matchings in this order until the cumulative $z_{\overline{S\cup W}}$-weight of the edges reaches $C_F(h+1)$.
    In the final of these matchings, we more precisely order the edges by increasing $\pr(e)/z_{\overline{S\cup W}}(e)$ and take the shortest prefix making the total weight at least $C_F(h+1)$.
    Let $F$ be the set of edges selected in this matching, together with all edges from the previous matchings.
    Since $z_{\overline{S\cup W}}(e)\leq1$, we have $C_F(h+1)\leq z_{\overline{S\cup W}}(F)\leq C_F(h+1)+1$.
    Since $\TA/4 \geq C_F(h+1)$, such an $F$ always exists.

    We next bound $\pr(F)$.
    Since we picked the matchings with the lowest $Q_i/Z_i$, we have
    \begin{equation}\label{eq:forward-ratio}
        \frac{\pr(F)}{z_{\overline{S\cup W}}(F)}
        \leq\frac{\sum_{P_i\in\mathcal P}Q_i}{\sum_{P_i\in\mathcal P}Z_i},
    \end{equation}
    where $\mathcal P$ are the non-discarded matchings.
    This can be seen e.g.\ by writing each side as weighted averages of the individual matching ratios.
    Consequently,
    \[
        \pr(F)\leq\frac{2T}{\TA/4}(C_F(h+1)+1) \leq \frac{1}{32}.
    \]
    Finally, every fully selected matching has weight at least $\TA/(4n)$, and their total weight is less than $C_F(h+1)$.
    Hence,
    \begin{align*}
        \Delta(F) \leq 1+\frac{4C_F(h+1)n}{\TA}
        \leq 1+ \frac{4C_Fc_0\varepsilon n}{T} 
        \leq \varepsilon|W|/8.
    \end{align*}
\end{proof}

Let $\mathcal G$ be the set of modified graphs $G'$ of $G$ preserving the vertex set, the out-degrees of vertices in $S$, and keeping all internal edges of $S$, possibly apart from edges in $B\times\{y\}$, such that $A$ distinguishes $G'$ from $G$ with probability at most $\frac{1}{10}$.

\begin{lemma}\label{lem:swap-F}
    Fix a $W$ satisfying the properties of \Cref{lem:unlikely}, and let $F$ satisfy the properties of \Cref{lem:constructing-F}.
    Let $W'\subseteq W$ with $|W'| = \max\{1,\Delta(F)\}$.
    Then there exists a graph $G'\in \mathcal G$ satisfying $Q(W)\leq 1/16$ and each of the following.
    \begin{itemize}
        \item $H_{G'}(s,\overline{S\cup W},W')\geq (1-\alpha)\frac{\pi_G(s,t)}{r}z_{\overline{S\cup W}}(F)$.
        \item For every $v\notin \Hin, |W\setminus\Nin(v)|\geq \varepsilon|W|/8$ in $G'$.
        \item For every $b\in B$, there exists $w\in W$ with $(b,w)\notin E(G')$.
    \end{itemize}
\end{lemma}
\begin{proof}
    Start by deleting edges in $W\times W$.
    Next we isolate $W'$ in the following way:

    A vertex $u\notin \Hout \cup W$ has, by \Cref{lem:unlikely}, at least $|W\setminus \Nout(v)|\geq \varepsilon|W|/4$ non-out-neighbors in $W$.
    And since $|W'|\leq \varepsilon|W|/8$, we can move all out-edges from $u$ to $W'$ to $W\setminus W'$.
    Similarly, we can do the same for $v\notin \Hin \cup W$.
    After this isolation, $W'$ only has out-edges from $\Hout$ and in-edges from $\Hin$.

    We will swap the edges of $F$ into $W'$.
    We do this by deleting $F$, and now restoring all the degrees of vertices.    
    As $F$ is not in $E_H\cup W$, the only vertices with out-edges in $F$ is a subset of $V\setminus W\cup \Hout$.
    Consider such a vertex $u\notin W\cup \Hout$.
    It has lost at most $\Delta(F)\leq \varepsilon|W|/8$ edges, and since $|W'|\geq \varepsilon|W|/8$, we can restore the degree of $u$ by adding the lost edges to $W'$.
    We can do the exact same for the in-edges.

    Let $G'$ denote the modified graph.
    We now verify the claimed properties.
    For every $v\notin\Hin\cup W$, the isolation preserves the number of in-neighbors in $W$, and swapping $F$ increases it by at most $\Delta(F)$.
    Hence
    \begin{align*}
        |W\setminus\Nin(v)|
        &\geq \frac{\varepsilon|W|}{4}-\Delta(F)
        \geq \frac{\varepsilon|W|}{8}.
    \end{align*}
    For $v\in W$, the same bound holds since $v$ has at most one in-neighbor in $W$.
    Similarly, every $b\in B$ has at least $\varepsilon|W|/8$ non-out-neighbors in $W$, since $B\cap\Hout=\emptyset$.
    
    Consider the probability of visiting an edge of $F$ no later than visiting $S\cup W$ in $G$,
    \begin{align*}
        H_G(s,\overline{S\cup W}, F)
        \geq (1-\alpha)\sum_{(u,v)\in F}\frac{\pi(s,\overline{S\cup W},u)}{\dout(u)}
        \geq (1-\alpha)\frac{\pi_G(s,t)}{r}z_{\overline{S\cup W}}(F).
    \end{align*}
    Since $F$ is swapped into $W'\subseteq W$, we have
    \begin{align*}
        H_G(s,\overline{S\cup W},F) = H_{G'}(s,\overline{S\cup W},F) \leq H_{G'}(s, \overline{S\cup W}, W').
    \end{align*}

    Finally, $G$ and $G'$ cannot be distinguished unless $A$ touches a vertex in $W$ or examines an edge of $F$ in $G$.
    Indeed, all degrees outside $W$ are preserved, and we can order their adjacency lists so that all unchanged edges retain their original positions.
    Consequently, the probability that $A$ distinguishes $G'$ from $G$ is at most
    $
        \qr(W)+\pr(F)
        \leq\frac{1}{16}+\frac{1}{32}
        =\frac{3}{32}<\frac{1}{10}.
    $
    Hence $G'\in\mathcal G$, completing the proof.    
\end{proof}

\begin{lemma}\label{lem:bound-z-U}
    Let $U\subseteq V$.
    Then $z_{\bar U}(E\setminus E(U))\geq \TA - \frac{2r H(s,U)}{(1-\alpha)\pi(s,t)}$.
\end{lemma}
\begin{proof}
    First of all,
    \begin{align*}
        z(E) -z_{\bar U}(E)
        &=\sum_e(z(e)-z_U(e)) \\
        &\leq \frac{r}{\pi(s,t)}\sum_u(\pi(s,u)-\pi(s,\bar U,u)) \\
        &= \frac{r}{\pi(s,t)}(1- \pi(s,\bar U, V)) \\
        &\le \frac{r}{\pi(s,t)}H(s,U),
    \end{align*}
    where the first inequality uses the fact that choosing the second term of $z(\cdot)$ maximizes the difference.
    Furthermore, $\pi(s,\bar U,u)=0$ for every $u\in U$.
    Thus
    \begin{align*}
        z_{\bar U}(E(U)) \leq \frac{r}{\pi(s,t)}\sum_{(u,v)\in E: u\notin U,v\in U} \frac{\pi(s,\bar U,u)}{\dout(u)}
        \leq \frac{r}{(1-\alpha) \pi(s,t)}H(s,U).
    \end{align*}
    Combining both inequalities gives
    \begin{align*}
        z_{\bar U}(E\setminus E(U))
        &= z_{\bar U}(E)-z_{\bar U}(E(U))\\
        &= \TA - (z(E) - z_{\bar U}(E)) - z_{\bar U}(E(U))\\
        &\ge \TA -  \p{1+\frac{1}{1-\alpha}}\frac{r}{\pi(s,t)}H(s,U)\\
        &\ge  \TA -\frac{2r}{(1-\alpha)\pi(s,t)}H(s,U).
    \end{align*}
\end{proof}

\begin{lemma}[Forward Probability Mass Dichotomy]\label{lem:forward-dichotomy}
    There exists $W \subseteq V\setminus(S\cup B)$ satisfying~\Cref{lem:unlikely} such that at least one of the following holds.
    \begin{enumerate}
        \item\label{item:dichotomy-large} $H_G(s,W) > \p{1+\frac{32(1+c)(h+1)}{\varepsilon(1-c)(1-\alpha)^2\alpha^2}}\frac{\pi_G(s,t)}{r}$.
        \item\label{item:dichotomy-F} There exists $F \subseteq E_0 = E\setminus(E(S\cup W)\cup E_H)$ satisfying the properties of~\Cref{lem:constructing-F}.
    \end{enumerate}
\end{lemma}
\begin{proof}
    We can assume that $z(E_H) < \TA/4$, as otherwise we are done by~\Cref{lem:E-H}.
    Choose any $W$ satisfying~\Cref{lem:unlikely}.
    We can assume $H_G(s,W) \le \p{1+\frac{32(1+c)(h+1)}{\varepsilon(1-c)(1-\alpha)^2\alpha^2}}\frac{\pi_G(s,t)}{r}$.
    Together with~\Cref{fact:S} we get
    \[
    H_G(s,S\cup W)
    \le H_G(s,S)+H_G(s,W)
    \le \p{2+\frac{32(1+c)(h+1)}{\varepsilon(1-c)(1-\alpha)^2\alpha^2}}\frac{\pi_G(s,t)}{r}.
    \]
    Using these facts together with~\Cref{lem:bound-z-U},
    \begin{align*}
        z_{\overline{S\cup W}}(E_0) 
        &= z_{\overline{S\cup W} }(E\setminus (E(S\cup W) \cup E_H)) \\
        &\geq z_{\overline{S\cup W} }(E\setminus (E(S\cup W)) - z_{\overline{S\cup W} }(E_H) \\
        &> \TA - \frac{2r H_G(s,S\cup W)}{(1-\alpha)\pi(s,t)} - \frac{\TA}{4} \\
        &\geq \frac{\TA}{2},
    \end{align*}    
    where the last inequality uses that $\TA >\!\!> (h+1)/\varepsilon$ by the inequality on $T$ we assumed for the sake of contradiction.
    Now~\Cref{lem:constructing-F} completes the proof.
\end{proof}

\begin{lemma}\label{lem:pre-hidden-source-mass}
    Let $S$ and $B$ be as in~\Cref{lem:hidden-target-mass}.
    There exists a modified graph $G'\in \mathcal G$ and sets $W\subseteq V\setminus (S\cup B)$ and $W' \subseteq W$ satisfying each of the following.
    \begin{enumerate}
        \item\label{item:large-hit-prob} $H_{G'}(s,\overline{S\cup W},W') > 2\frac{1+c}{1-c}\pi_G(s,t)\frac{h+1}{r(1-\alpha)^2\alpha^2}$.
        \item\label{item:unlikely-visited} $\qr(W)\leq\frac{1}{10}$ in $G$.
        \item\label{item:in-nonneighbors} $\forall v\notin \Hin\colon |W\setminus \Nin(v)| \geq \varepsilon|W|/8$ in $G'$.
        \item\label{item:b-non-edge} For every $b \in B$ there exists a $w \in W$ with $(b,w) \notin E(G')$.
        \item\label{item:w-large} $1 \le \abs{W'} \le \varepsilon |W|/16$.
    \end{enumerate}
\end{lemma}
\begin{proof}
    Let $W$ be as in~\Cref{lem:forward-dichotomy}.
    Assume first that~\Cref{item:dichotomy-large} of~\Cref{lem:forward-dichotomy} holds.
    We choose $G' = G$.
    For each vertex $w\in W$, we consider the probability that a walk from $s$ first enters $W$ at $w$ without visiting $S$ before entering $W$.
    This is exactly the probability $H(s,\overline{S\cup W},w)$.
    Note that since the events are disjoint, $H(s,\bar S,W)=\sum_{w\in W} H(s,\overline{S\cup W},w)$. 
    We choose $W'$ to consist of the $\floor{\varepsilon \abs{W}/8}$ vertices with the largest such probabilities. 
    Since $\varepsilon|W|\geq 16$, we have $|W'|\geq \varepsilon|W|/16$ and therefore, 
    \begin{align*}
        H(s,\overline{S\cup W}, W')\geq  \frac{|W'|}{|W|} H(s,\overline{S\cup W}, W') \geq \frac{\varepsilon}{16}\p{H(s,W')-H(s,S)}.
    \end{align*}
    Applying \Cref{fact:S} with the above inequality,
    \begin{align*}
        H(s,\overline{S\cup W}, W') 
        &\geq \frac{\varepsilon}{16}\p{\p{1+\frac{32(1+c)(h+1)}{\varepsilon(1-c)(1-\alpha)^2\alpha^2}}\frac{\pi_G(s,t)}{r}-\frac{\pi_G(s,t)}{r}}\\
        &\geq 2\frac{1+c}{1-c}\pi_G(s,t)\frac{h+1}{r(1-\alpha)^2\alpha^2}.
    \end{align*}
    This finishes the proof in this case.

    Assume instead that~\Cref{item:dichotomy-F} of~\Cref{lem:forward-dichotomy} holds.
    Let $F$ be as in the \lcnamecref{lem:forward-dichotomy}, that is, satisfying the properties of~\Cref{lem:constructing-F}.
    Choose $W' \subseteq W$ to be any subset of size $\max\{1,\Delta(F)\}$.
    Now~\Cref{lem:swap-F} completes the proof.
\end{proof}

Now,~\Cref{lem:hidden-source-mass} follows from~\Cref{lem:pre-hidden-source-mass} by concentrating the forward mass into a smaller subset, and moving out-edges away from this subset.

\begin{proof}[Proof of~\Cref{lem:hidden-source-mass}]
    The graph claimed to exist will be constructed under the name $G''$ in the proof, and it will be a modification of the graph $G'$ given by~\Cref{lem:pre-hidden-source-mass}.
    The subset of vertices claimed to exist will be constructed under the name $W''$ in the proof, and it will be a subset of the set $W'$ given by~\Cref{lem:pre-hidden-source-mass}, which itself is a subset of the set $W$ given by that~\lcnamecref{lem:pre-hidden-source-mass}.

    To construct $G''$ from $G'$, first clean the slate by removing all edges in $W \times W$.
    
    Our next step is to reduce the out-degree of vertices in $W'$ to at most $h$.
    Consider each edge $(w', v)$ with $w' \in W'$ and $v \notin \Hin$.
    Since $\abs{W\setminus\Nin(v)} \ge \varepsilon\abs{W}/8 \ge \abs{W'}$, there exists for each such edge a distinct non-edge $(w,v) \notin E(G')$ with $w \in W\setminus W'$.
    Replace $(w',v)$ by $(w,v)$ in $G''$.
    Now $\Nout(w') \subseteq \Hin$ for each $w' \in W'$ and in particular $\dout(w') \le h$.

    We now move on to concentrating the forward probability mass of $W'$ into a subset $W''$ of size at most $h+1$.
    Choose $W''$ to be any subset of $W'$ of size $\min\{\abs{W'},h+1\}$.
    Recall that at this point, we have no edges inside $W$.
    Insert all edges from the possibly empty complement $W'\setminus W''$ to $W''$.
    Any vertex $u \in W'\setminus W''$ will have exactly $h+1$ edges to $W''$ and at most $2h+1$ out-edges in total.
    The probability that a random walk standing in $W'\setminus W''$ moves to $W''$ is thus at least $\frac{1-\alpha}{2}$.
    Since $G''$ only differs from $G'$ in out-edges of $W$, the hitting probability guaranteed by~\Cref{lem:pre-hidden-source-mass} is maintained.
    Therefore, $H_{G''}(s,\overline{S}, W'') \ge H_{G'}(s,\overline{S\cup W})\frac{1-\alpha}{2} \ge \frac{1+c}{1-c}\pi_G(s,t)\frac{h+1}{r(1-\alpha)\alpha^2}$, as wished.

    The remaining properties follow from~\Cref{lem:pre-hidden-source-mass}.
\end{proof}

\appendix
\crefalias{section}{appendix}
\section{Extension to a General $\delta$}
\label{sec:reduction-delta}

In this section, we show a simple modification of our algorithm so that it works for the variant that we are only required to estimate $\pi(s,t)$ when it is at least $\delta$. Equivalently speaking, we are allowed to have an additional additive error of $\delta$.

Let \[\Td=\max_{r\in(0,1]}\min\{\TAd,\TB\},\]
where 
\begin{align*}
    \TAd=\sum_{u\in V_s}\min\left\{\dout(u),\frac{\pi(s,u)r}{\max\{\pi(s,t),\delta\}}\right\}.
\end{align*}
We will show that our modified algorithm spends $O(\Td)$ time, and our lower bounds still hold after replacing $\Tstar$ with $\Td$.

For the algorithm, we just imagine that in the beginning, with probability $\delta$, we just magically fly to $t$ and stop.
Practically, the only modification is that, right after Line 5 of \cref{alg:triPPR}, we further do $X_{\gamma}\gets\delta+(1-\delta)X_{\gamma}$. Note that this only introduces an additional additive error of at most $\delta(1-X_{\gamma})\le \delta$. It is straightforward to check that, since we always have $X_{\gamma}\ge\delta$, the time complexity will become $O(\Td)$.

For the lower bounds, the modification is also simple. After replacing $\Tstar$ with $\Td$, if a good algorithm $A$ spends too little time, then by the same construction, we are able to construct a graph $G'$ such that $\pi_{G'}(s,t)\gg\max\{\pi_G(s,t),\delta\}$ and $A$ cannot distinguish $G$ and $G'$.
For example, to prove the $\delta$-variant of \cref{thm:not-knowing-n}, we only need to replace all $\pi(s,t)$ in the proof with $\max\{\pi(s,t),\delta\}$.

%


\bibliographystyle{alphaurl}
\bibliography{paper}

\end{document}